# Virtual Element based formulations for computational materials micro-mechanics and homogenization

Ph.D. thesis submitted to the University of Palermo

by

*Marco Lo Cascio*

Tutor

*Prof. Alberto Milazzo*

*Prof. Ivano Benedetti*

Dipartimento di Ingegneria
Università degli Studi di Palermo
Viale delle Scienze, Ed. 8 - 90128 Palermo

Marco Lo Cascio
Palermo, April 2020
e-mail:marco.locascio01@unipa.it

# Preface

In this thesis, a computational framework for microstructural modelling of transverse behaviour of heterogeneous materials is presented. The context of this research is part of the broad and active field of Computational Micromechanics, which has emerged as an effective tool both to understand the influence of complex microstructure on the macro-mechanical response of engineering materials and to tailor-design innovative materials for specific applications through a proper modification of their microstructure.

While the classical continuum approximation does not account for microstructural details within the material, computational micromechanics allows detailed modelling of a heterogeneous material's internal structural arrangement by treating each constituent as a continuum. Such an approach requires modelling a certain material microstructure by considering most of the microstructure's morphological features.

The most common numerical technique used in computational micromechanics analysis is the Finite Element Method (FEM). Its use has been driven by the development of mesh generation programs, which lead to the quasi-automatic discretisation of the artificial microstructure domain and the possibility of implementing appropriate constitutive equations for the different phases and their interfaces. In FEM's applications to computational micromechanics, the phase arrangements are discretised using continuum elements. The mesh is created so that element boundaries and, wherever required, special interface elements are located at all interfaces between material's constituents. This approach can be effective in modelling many microstructures, and it is readily available in commercial codes. However, the need to accurately resolve the kinematic and stress fields related to complex material behaviours may lead to very large models that may need prohibitive processing time despite the increasing modern computers' performance. When rather complex microstructure's morphologies are considered, the quasi-automatic discretisation process stated before might fail to generate high-quality meshes. Time-consuming mesh regularisation techniques, both automatic and operator-driven, may be needed to obtain accurate numeric results. Indeed, the preparation of high-quality meshes is today one of the steps requiring more attention, and time, from the analyst. In this respect, the development of computational techniques to deal with complex and evolving geometries and meshes with accuracy, effectiveness, and robustness attracts relevant interest.

The computational framework presented in this thesis is based on the Virtual Element Method (VEM), a recently developed numerical technique that has proven to provide robust numerical results even with highly-distorted mesh. These peculiar features have been exploited to analyse two-dimensional representations of heterogeneous materials' microstructures. Ad-hoc polygonal multi-domain meshing strategies have been developed and tested to exploit the discretisation freedom that VEM allows. To further simplify the preprocessing stage of the analysis and reduce the total computational cost, a novel hybrid formulation for analysing multi-domain problems has been developed by combining the Virtual Element Method with the well-known Boundary Element Method (BEM). The hybrid approach has been used to study both composite material's transverse behaviour in the presence of inclusions with complex geometries and damage and crack propagation in the matrix phase. Numerical results are presented that demonstrate the potential of the developed framework.

# Contents

# Chapter 1

# Introduction

Modern advanced structural components' behaviour strongly relies on the tailored behaviour of their constituent material. Since the properties of a material at a certain scale depend on the features of and mutual interactions among the material constituents at lower scales [136], an effective way to obtain a desired macroscopically response is to enhance a base material's properties by altering its microstructure. Therefore, materials with highly complex microstructures are frequently developed for many modern engineering structures.

Concurrently, the design of structures for advanced engineering applications requires a full understanding of a component response to different operational and environmental conditions. Usually, the knowledge of their failure mechanisms is of paramount importance. Component level phenomenological models may not always predict complex materials behaviours, especially if damage initiation and evolution are of concern. Today, it is widely recognised that these aspects may be better understood if the material microstructure features are considered and brought into the modelling framework. The link between microstructure and material macroscopic properties is technologically interesting as it may provide valuable information for the design of enhanced materials. The ability to understand, explain and predict macroscopic material properties from a suitable description of the micro-scale is of relevant technological interest, especially in connection with the contemporary availability of manufacturing technologies offering a tighter control on the microstructure of the material.

In the ambitious paradigm known as *materials-by-design*, the much sought-after capability of modelling materials *ab initio*, i.e. starting from the smallest nano-scales, exploiting first principles, such as quantum mechanics, should enable the design of materials with properties tailored to specific applications.

In the last few decades, remarkable developments in experimental materials characterisation have contributed to the development of the *materials by design* paradigm [5], which aims at developing novel and sustainable materials with desired optimal features by combining elementary constituents in a bottom-up approach using a variety of production techniques.

A pillar of such a paradigm is provided by the capability of *multi-scale materials characterisation and modelling* [167] that, by embodying deeper and richer layers of information about the materials hierarchical organisation, often spanning several different scales, contribute to the understanding of complex material/structural behaviours and to the design of novel high-performance applications, with apparent technological benefits. In such a context, the possibility of modelling, with acceptable fidelity, the microstructure of a considered material, and the complex interactions between its building blocks, plays a fundamental role. However, the inclusion of deeper layers of fidelity requires the ability to robustly address several kinds of modelling complexities, including those arising, for example, from the need to represent involved material morphological details, which may also present statistical variability.

To reduce the costs of experimental characterisation techniques, prediction of new material's behaviour can be performed by numerical simulation, with the primary goal being to accelerate trial and error experimental testing [186]. The recent substantial increase in computational power available for mathematical modelling and simulation raises the possibility that modern numerical methods can play a significant role in analysing heterogeneous microstructures with enhanced efficiency.

## 1.1 Numerical methods in micromechanics

Computational micromechanics has emerged as a consistent framework supporting the understanding of the link between the material microstructure and its macroscopic properties, i.e. the *structure-property relationship*. The field has enormously benefited from the rapid advancements of experimental techniques for materials microscopic characterisation and reconstruction, able to provide a wealth of useful processable information, and from the increased affordability of high-performance computing (HPC), which provide the complementary ability to combine and to process such information towards a better understanding, prediction and manipulation raising the possibility that modern numerical methods can play a significant role in analysing heterogeneous microstructures with enhanced efficiency. In this respect, the development of computational techniques to deal with complex and evolving geometries and meshes with accuracy, effectiveness, and robustness attracts relevant interest.

The Finite Element Method (FEM) is the most popular technique used in computational micromechanics analysis and has been extensively used to investigate several kinds of materials, including polycrystalline [23, 72, 101, 154] and composite materials [91, 130, 131]. Its extensive use has been driven by the development of mesh generation programs, which lead to the quasi-automatic discretisation of the artificial microstructure domain.

In FEM's applications to computational micromechanics, the phase arrangements are discretised using continuum elements and the mesh is created so that element boundaries and, wherever required, special interface elements are located at all interfaces between material's constituents. This approach can be effective in modelling many microgeometries, and it is available in commercial codes.

FEM is also popular because of its capabilities in nonlinear analyses where its flexibility and capability of supporting a wide range of constitutive descriptions for the constituents and the interfaces between can be exploited trough the possibility of implementing appropriate constitutive equations for the different phases and their interfaces. Constitutive models for constituents used in FEM-based micromechanics have included a wide range of elasto-plastic, viscoelastic, viscoelastoplastic and continuum damage mechanics descriptions as well as crystal plasticity models [127], and nonlocal models [25]. Besides, FEM has supported a range of modelling options for interfaces between phases.

However, the need to accurately resolve the kinematic and stress fields related to complex material behaviours with a conformal discretisation of the domain that retains proper topological features may lead to very large models that may need prohibitive processing time despite the increasing modern computers' performance.

## 1.2 A key issue in computational micromechanics

One of the key aspects in the effective modelling of materials micromechanics is the availability of a suitable representation of the material micromorphology. This can be based either the computer reconstruction of real microstruc-tures or on the generation of artificial models embodying the relevant statistical features of the microstructural aggregate, which may exhibit involved shapes. The potential presence of complex morphological features has a direct effect on the complexity of the numerical grid, or mesh, used to discretise the considered boundary value problem. Examples of actual complex morphologies for two types of composite materials are shown in Fig.(1.1) and Fig.(1.2). The quality of the mesh, in turn, may have an important effect on the accuracy of the numerical reconstruction of the mechanical fields.

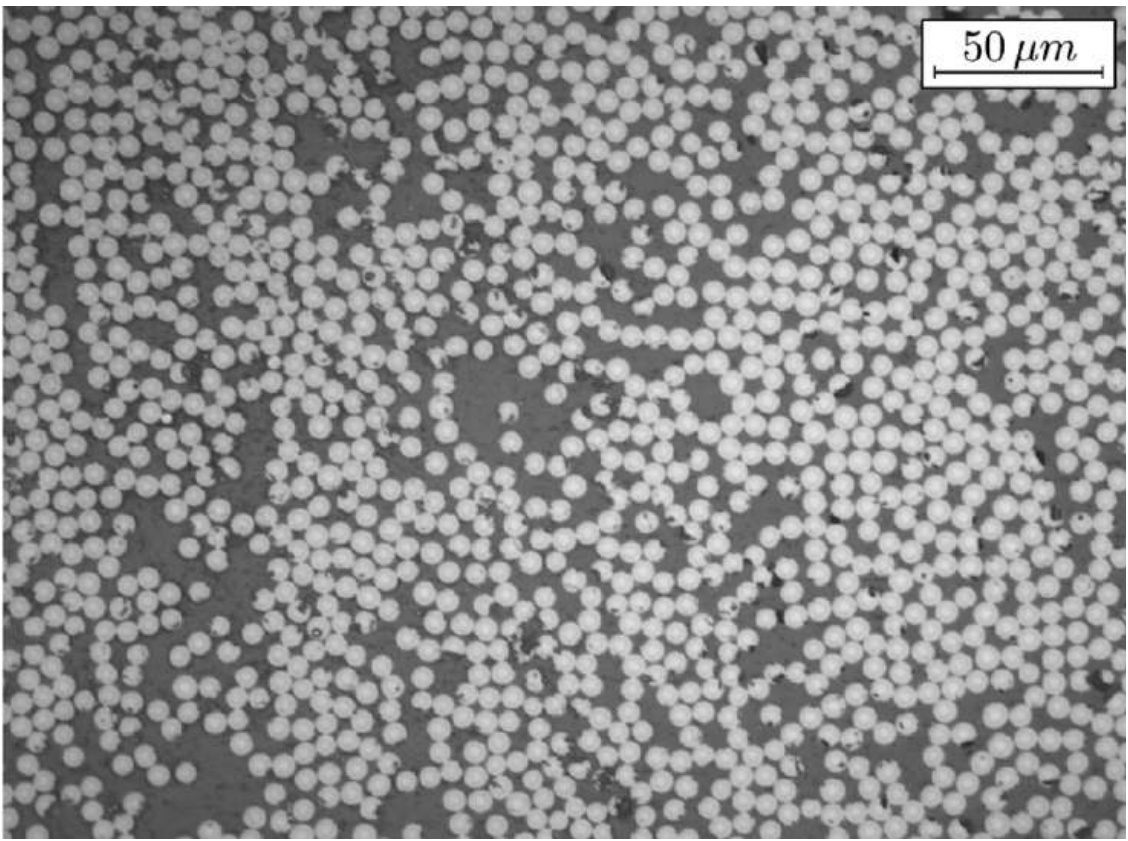

Figure 1.1: Cross section micrograph of a UD AS4/8552 composite [88].

When rather complex microstructure's morphologies are considered, the quasi-automatic discretisation process described before might fail to generate high-quality meshes. Time-consuming mesh regularisation techniques, both automatic and operator-driven, may be needed to obtain accurate numeric results. Indeed, the preparation of high-quality meshes is today one of the steps requiring more attention, and time, from the analyst [149, 150].

Besides FEM, many numerical techniques have been employed for con-tinuum micromechanical studies of discrete microstructures, including Finite Difference (FD) [2] and Finite Volume algorithms [22, 147] and spring lattice models [141]. FEM derived techniques such the extended finite element method (X-FEM) [164, 163] and the scaled boundary finite element method (SBFEM)[38] have also been employed. Other popular numerical approaches rely on the use of meshfree methods [67, 176] and the Boundary Element Method. The latter has been extensively used for micromechanical studies on composite materials [1, 69, 113, 114] and polycrystalline materials [79, 84, 49, 47] because it allows reducing the problem dimensionality as a direct consequence of the underlying integral formulation, with the consequent reduction in the number of degrees of freedom required in the analysis simplification and such a feature may result particularly appealing when materials morphologies with high statistical variability have to be automatically generated, meshed and analysed [46].

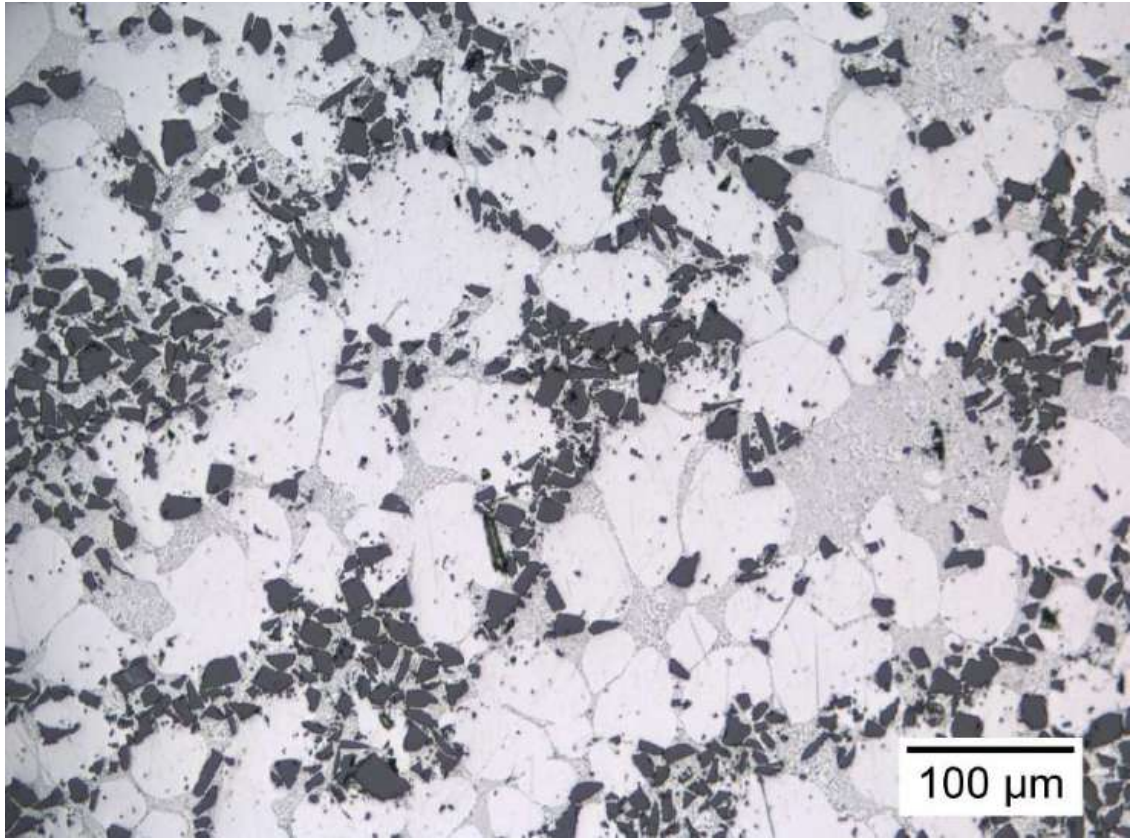

Figure 1.2: Microstructure of a A359/SiCp metal–matrix composite (MMC) [66].

Given the considerations outlined above, the development of computational techniques capable of dealing with complex and evolving geometries and meshes with accuracy, effectiveness, efficiency, and robustness is still of relevant interest.

The computational framework presented in this thesis is based on the Virtual Element Method (VEM), a recently developed numerical technique which has already proven to provide robust numerical results even with highly-distorted mesh.

In this thesis, attention is focused on *continuum* micromechanics, i.e. on the study of materials whose basic building blocks, e.g. individual crystals, fibres or matrix, can be modelled resorting to the continuum idealisation. In particular, the research focus is on polycrystalline materials and fibre-reinforced composites that are two classes of materials widely employed in engineering applications. In polycrystalline materials, the availability of information about the grains' mechanical properties and their inter-granular interfaces, their crystallographic orientation and size distribution can be conveniently exploited to predict the aggregate's properties and suggest potential manufacturing pathways for material optimisation [48]. Analogously, the knowledge of the properties of carbon fibres, epoxy matrix and the characterisation of the fibre-matrix interface can be used to investigate the effectiveness of different fibre arrangements on composite laminates' structural performances [85, 129]. Both materials present morphologies that may become challenging to model given the microstructure's representation's complex shapes.

The hypothesis that VEM's peculiar features could be exploited to analyse two-dimensional representations of heterogeneous materials' microstructures efficiently forms the basis of this research work. To further simplify the pre-processing stage of the analysis and reduce the total computational cost, a novel hybrid formulation for analysing multi-domain problems is developed and employed by combining the Virtual Element Method with the Boundary Element Method. The hybrid approach will be used to study both composite material's transverse behaviour in the presence of inclusions with complex geometries and damage and crack propagation in the matrix phase.

## 1.3 Content of the thesis

The following Chapters of this thesis are organised as follows.

Chapter (2) introduces the main features of the Virtual Element Method (VEM). VEM's lowest-order formulation in the context of two-dimensional linear elasticity is reviewed, setting the framework of this thesis's ensuing developments. Insights of the in-house developed code are given while discussing some specific implementation aspects of the VEM formulation previously introduced.

Chapter (3) presents an application of the Virtual Element Method for computational homogenisation of composite and heterogeneous materials. The reported applications are focused on modelling the transverse mechanical behaviour of polycrystalline and unidirectional fibre-reinforced composite materials exploiting VEM's flexibility in the analysis of randomly generated and meshed microstructures.

In Chapter (4), a novel two-dimensional hybrid virtual-boundary element formulation is presented. Numerical tests are performed to assess its accuracy using a representative test case. The application of such a novel formulation to the computational homogenisation problem of a composite material with randomly distributed inclusions of complex shape is also reported.

Chapter (5) is meant to introduce further applications of the hybrid VEM-BEM formulation for modelling damage phenomena in composite materials.

Eventually, the last Chapter summarises and discusses the research work outcome, providing suggestions for future work on the topic.

# Chapter 2

# The Virtual Element Method

The present Chapter is intended to introduce the displacement-based lowest-order VEM formulation for two-dimensional linear elasticity problems, setting the framework of this thesis's ensuing developments. Section 2.1 is devoted to introduce the core idea of the method and its peculiar features. A selected list of references is also provided to the interested reader that comprise both the method's fundamentals and its most prominent scientific application to date. Section 2.2 recalls both the strong and weak form of the governing equations for two-dimensional linear elasticity. Section 2.3 recalls the lowest-order displacement-based VEM formulation for two-dimensional linear elasticity problems and details the construction of all the VEM approximation scheme components. Section 2.4 introduces VESTA, the virtual element program that has been developed and used to perform all the analysis of the present thesis. In Section 2.5 some aspects of the implementation of the VEM formulation are discussed. Eventually, Section 2.6 presents a numerical example used to assess VEM's accuracy in reproducing an exact linear solution even with heavily distorted mesh.

## 2.1 Introduction

The Virtual Element Method (VEM) [31] is a numerical technique used to find approximate solutions to problems described by partial differential equations. VEM can be considered as a generalisation of the Finite Element Method (FEM), a powerful and well established numerical technique that is used in a large number of engineering applications.

FEM's common approach involves discretising a continuous domain into a set of discrete sub-domains that is usually referred to as a *mesh*. In the case of two and three-dimensional domains, the mesh is usually constructed with triangular or tetrahedral and quadrilateral or hexahedral elements while the use of general polygonal and polyhedral meshes is rather uncommon in industrial applications. Nevertheless, there are some advantages in using polygonal/polyhedral meshes over the simpler ones, the most important of which are a significant simplification of the partitioning of the domain even for very complex geometries and a reduction of the complexity of adaptive mesh refinement and coarsening algorithms as no special treatment of hanging nodes are required to maintain the conformity of the mesh.

While the use of polygonal and polyhedral meshes for the approximate solution of boundary value problems is well established in the literature [165] it turns out that extending FEM to general polygonal/polyhedral meshes leads to very complex and computationally expensive schemes. This is because the construction of the basis functions on elements with a very general shape is a non-trivial and complex task [36].

VEM most appealing feature is its capability to use *polytopal mesh* (polygons in two dimensions or polyhedra in three dimensions). An example of a VEM two-dimensional polygonal discretisation is shown in Fig.(2.2). Despite, its rather recent introduction, VEM has already proven to be able to provide accurate results in several class of engineering problems even in the presence of severe mesh distortions. At the same time, as will be shown later in this Chapter, VEM allows treating irregularly shaped elements with relative simplicity and efficiency.

The possibility of using meshes with elements of arbitrary shape is certainly one of the most evident features of VEM. This features allows VEM to manage meshes which in the standard FEM are considered non-conformal. An example of non-conformal mesh is shown in Fig.(2.1). These meshes are often generated during adaptive refining and may include the presence of so-called hanging-nodes (red nodes in the figure). VEM straightforwardly treats hanging nodes introduced by the refinement of a neighbouring element as new element nodes since the VEM formulation admits adjacent co-planar element interfaces. The adaptive process results, in fact, in a VEM-conforming mesh. A mesh adaptivity process could also involve a coarsening process, frequent in the numerical solution of evolution PDEs where it is standard practice to efficiently track evolving fronts and singularities. Adaptive coarsening could be interpreted as the opposite process of mesh refining and could be exemplified by reading Fig.(2.1) from right to left. In such a context, VEM allows a rather straightforward and inexpensive implementation as node removal does not necessitate any further local mesh modification. In general, VEM's mesh flexibility may have the potential to provide complexity reduction with respect to standard FEM.

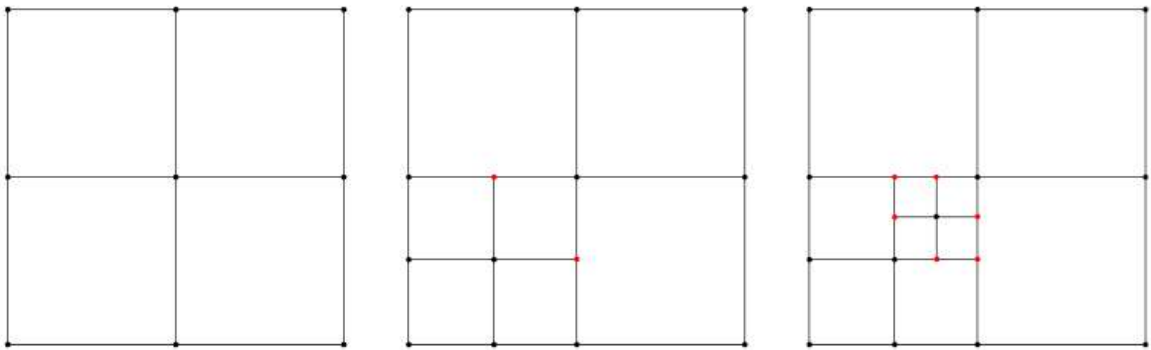

Figure 2.1: Example of an adaptive mesh refinement process that involve the generation of hanging nodes (marked in red).

The VEM has been first introduced as an evolution of the Mimetic Finite Difference (MFD) method [57, 111], inheriting the capability of obtaining robust approximate solutions of boundary value problems even on unstructured meshes whose elements may have very general geometries. By exploiting the aforementioned intrinsic capabilities, the VEM has been successfully applied to general linear elasticity [34, 76, 13], inelastic materials at small strains [37, 14], hyper-elastic materials at finite strains [178, 62], contact mechanics [179], topology optimization [75, 11], magneto-static problems [32, 33], geomechanical simulations of reservoir models [9], damage and fracture analysis [68, 6, 52, 138, 93, 17] and plate bending problems [58, 182, 183, 133, 30]. The possibility of using virtual elements of general shape, also highly distorted, makes VEM particularly interesting for applications where the meshed domain may undergo large deformations [174].

Another interesting application of VEM is to micromechanics problems where the occurrence of problematic morphological features, likely sources of mesh irregularities, may not be *a priori* excluded: a typical example is provided by problems of statistical homogenisation, where a relevant number of random unit cells is generated and analysed with the aim of inferring the emerging material properties through volume averaging techniques. The application of VEM to micromechanics problems are rather recent [12, 16, 177, 124] and represent the core research work reported in the present thesis [116, 117, 118].

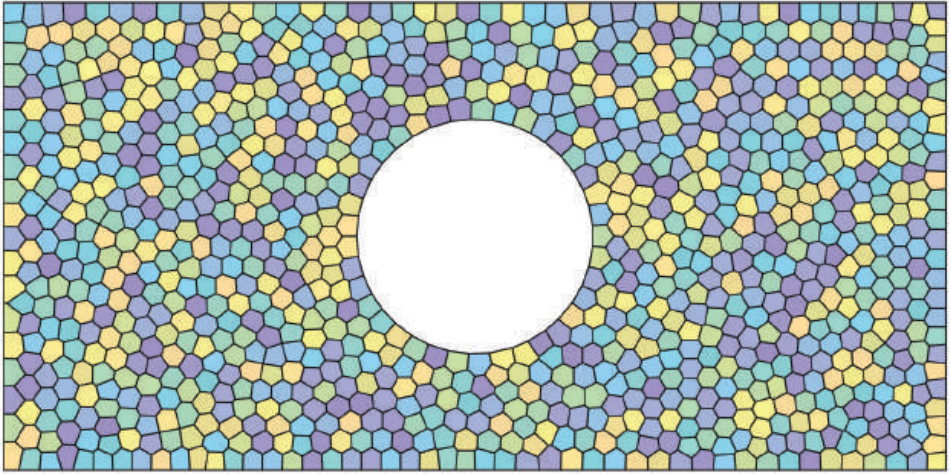

Figure 2.2: An example of a two-dimensional polygonal discretization of a rectangular plate with a circular hole.

## 2.2 Governing equations for 2D linear elasticity

### 2.2.1 Strong form

Let continuous body occupying $\Omega \in \mathbb{R}^2$ a two-dimensional region bounded by the curve $\Gamma \equiv \partial \Omega$ in a reference system $(x, y)$.
The strong formulation of the small strains elastic problem is based on the use of the strain-displacement equations

$$\varepsilon_{ij}(\boldsymbol{u}) = \frac{1}{2}\left(u_{i,j} + u_{j,i}\right), \tag{2.1}$$

of the linear elastic constitutive laws

$$\sigma_{ij} = C_{ijkl}\,\varepsilon_{kl}, \tag{2.2}$$

and of the indefinite equilibrium equations

$$\sigma_{ij,j} + f_i = 0, \tag{2.3}$$

where $\boldsymbol{u} = \{u_x, u_y\}^\mathsf{T}$ represents the displacement vector field, $\varepsilon_{ij}$ are the components of the strain tensor, $\sigma_{ij}$ are the components of the stress tensor, $C_{ijkl}$ are the stiffness tensor components and $f_i$ are the components of volume distributes loads. Eventually, suitable boundary conditions are enforced on the boundary $\Gamma$ of the the considered body, so that

$$\forall i = 1, 2 \quad \begin{cases} u_i = \bar{u}_i & \text{on } \Gamma_{u_i} \\ t_i = \bar{t}_i & \text{on } \Gamma_{t_i} \end{cases} \tag{2.4}$$

and $\Gamma_{u_i} \cap \Gamma_{t_j} \equiv \varnothing$ if $i = j$ and $\Gamma_{u_i} \cup \Gamma_{t_i} \equiv \Gamma$ for $i = x, y$.

It may sometimes be convenient, for the sake of expressivity and compactness, to employ the Voigt notation and express the above sets of equations in matrix form, so that the strain and stress tensor are expressed as the vectors

$$\boldsymbol{\varepsilon} = \{\varepsilon_1 = \varepsilon_{xx}, \varepsilon_2 = \varepsilon_{yy}, \varepsilon_6 = 2\varepsilon_{xy}\}^\mathsf{T}, \tag{2.5}$$

$$\boldsymbol{\sigma} = \{\sigma_1 = \sigma_{xx}, \sigma_2 = \sigma_{yy}, \sigma_6 = \sigma_{xy}\}^\mathsf{T}, \tag{2.6}$$

and the strain-displacements relationships, the constitutive equations and the equilibrium equations read respectively

$$\boldsymbol{\varepsilon} = \mathscr{D}\boldsymbol{u}, \tag{2.7}$$

$$\boldsymbol{\sigma} = \mathbf{C}\boldsymbol{\varepsilon}, \tag{2.8}$$

$$\mathscr{D}^\mathsf{T}\boldsymbol{\sigma} + \boldsymbol{f} = \mathbf{0}, \tag{2.9}$$

with the associated boundary conditions

$$\begin{cases} \boldsymbol{u} = \bar{\boldsymbol{u}} & \text{on } \Gamma_u \\ \boldsymbol{t} = \mathscr{D}_n^\mathsf{T}\boldsymbol{\sigma} = \bar{\boldsymbol{t}} & \text{on } \Gamma_t \end{cases} \tag{2.10}$$

where

$$\mathscr{D} = \begin{bmatrix} \partial_x & 0 \\ 0 & \partial_y \\ \partial_y & \partial_x \end{bmatrix}, \tag{2.11}$$

denotes the small-strains linear differential matrix operator, $\partial_x = \partial(\cdot)/\partial x$ and $\partial_y = \partial(\cdot)/\partial y$, and

$$\mathscr{D}_n = \begin{bmatrix} n_x & 0 \\ 0 & n_y \\ n_y & n_x \end{bmatrix}, \tag{2.12}$$

is the matrix operator built with the components $n_x$ and $n_y$ of the boundary unit normal $\boldsymbol{n}$.

### 2.2.2　Weak form

The weak form for the boundary value problem is provided by the *principle of virtual displacements*, which states that the solution field is given by the displacements $u(\mathbf{x}) \in \mathcal{V}$ satisfying the equality

$$\int_\Omega \varepsilon(v)^\mathsf{T} \mathbf{C}\, \varepsilon(u)\, d\Omega = \int_\Omega v^\mathsf{T} f\, d\Omega \qquad \forall v(\mathbf{x}) \in \mathcal{V}, \tag{2.13}$$

where $\mathcal{V} := \left[H_0^1(\Omega)\right]^2$ is the space of kinematically admissible displacements and $H_0^1(\Omega)$ denotes the first order Sobolev space on $\Omega$, consisting of functions vanishing on $\Gamma$ and square integrable over $\Omega$ together with their first order derivatives. With the sole aim of simplifying the formal introduction of the method, in Eq.(2.13) it has been assumed that the displacements $u$ vanish along the boundary $\Gamma$ of the analysed domain. As it will be recalled later in this Section, this assumption does not affect the generality of the formulation: indeed, due to the choice of the element-wise virtual space of admissible displacements, see Section 2.3.2, either non-homogeneous Dirichlet or Neumann boundary conditions can be implemented following the same procedure as in the standard finite element method [13].

Defining integral operators

$$\mathscr{L}(u,v) := \int_\Omega \varepsilon(v)^\mathsf{T} \mathbf{C}\, \varepsilon(u)\, d\Omega, \tag{2.14}$$

which identifies the virtual strain energy symmetric bilinear form and

$$\mathscr{G}(v) := \int_\Omega v^\mathsf{T} f\, d\Omega, \tag{2.15}$$

which identifies the loads' virtual work linear functional, Eq.(2.13) can be written in the compact form

$$\mathscr{L}(u,v) = \mathscr{G}(v) \quad \forall v \in \mathcal{V}, \tag{2.16}$$

useful in subsequent developments.

## 2.3　VEM formulation

The VEM core idea is rooted in the assumption that the trial and test functions over each mesh element belong to a space containing all the polynomials up to a certain previously selected order *k* plus other additional functions that, in general, are not polynomials and are solutions, within the element, of a suitably defined boundary value problem. Such additional functions are explicitly known only over the element edges while, within each element, they are not explicitly known and never computed, which justifies the adjective *virtual* referred to the method. The implicit definition of virtual functions does not allow the computation of their values or their gradients' values in any element interior point. VEM resorts to a projection operator to obtain an approximate polynomial expression of the strains associated with the virtual functions, thus obtaining a computable approximation of the internal virtual work. Through a particular choice of the element degrees of freedom, such projection is exactly computed as a function of the degrees of freedom themselves, without actually solving the local boundary value problem. In the particular case of the VEM lowest-order formulation, as it will be shown later in this Chapter, the aforementioned VEM approach results in a computational cost that, in terms of generation of the problem's system of equations, is faster than FEM's standard quadrilateral isoparametric elements and it is comparable to FEM's simplicial elements.

### 2.3.1 Domain partition and discrete weak form

This Section introduces the discrete form of the problem and the notations used throughout this thesis when referring to a generic polygonal virtual element. As in standard FEM procedures, the weak form in Eq.(2.16) is employed to build an approximate solution to the elastic boundary-value problem. For this purpose, the domain $\Omega$ is sub-divided into a set $\Omega^h$ of finite non-overlapping elements $E \in \Omega^h$, mutually interconnected at the nodal points lying on their edges. The superscript *h* refers to the association to a discretization of the domain $\Omega$, which is parametrized by a characteristic length scale *h*.

Once the discretisation $\Omega^h$ is identified, a function space $\mathcal{V}^h \subset \mathcal{V}$ constituting a finite-dimensional approximation of $\mathcal{V}$ can be associated to it. The Galerkin approximation of the problem is provided by $\boldsymbol{u}^h \in \mathcal{V}^h$ such that

$$\mathscr{L}(\boldsymbol{u}^h, \boldsymbol{v}^h) = \mathscr{G}(\boldsymbol{v}^h) \quad \forall \boldsymbol{v}^h \in \mathcal{V}^h, \tag{2.17}$$

where the integral operators can be split into elemental contributions $\mathscr{L}_E(\cdot, \cdot)$

and $\mathscr{G}_E(\cdot)$ as

$$\mathscr{L}(\boldsymbol{u}^h, \boldsymbol{v}^h) = \sum_{E \in \Omega^h} \mathscr{L}_E(\boldsymbol{u}^h, \boldsymbol{v}^h) = \sum_{E \in \Omega^h} \int_E \boldsymbol{\varepsilon}(\boldsymbol{v}^h)^\mathsf{T} \mathbf{C}\, \boldsymbol{\varepsilon}(\boldsymbol{u}^h)\, dE, \qquad (2.18)$$

and

$$\mathscr{G}(\boldsymbol{v}^h) = \sum_{E \in \Omega^h} \mathscr{G}_E(\boldsymbol{v}^h) = \sum_{E \in \Omega^h} \int_E \boldsymbol{v}^{h\,\mathsf{T}} \boldsymbol{f}\, dE. \qquad (2.19)$$

An admissible VEM discretization of a two-dimensional domain is allowed to comprise polygonal elements with very general shapes: in particular *polygons with an arbitrary number of edges are allowed and even the requirement of convexity may be waived*. Moreover, two consecutive edges of an element are allowed to form a straight angle.

In the following, for a generic element $E$, $|E|$ will denote the area of the element, $h_E$ its diameter, $\partial E$ the element boundary and $\mathbf{n}_E = \begin{bmatrix} n_x & n_y \end{bmatrix}^\mathsf{T}$ the unit normal vector to $\partial E$. The counter-clockwise ordered vertices $\mathrm{v}_i$, $i = 1, 2 \ldots, m$, have coordinates $\mathbf{x}_i = \{x_i, y_i\}$ and their *local scaled coordinates* are defined by

$$\xi_i = \frac{x_i - x_E}{h_E}, \qquad \eta_i = \frac{y_i - y_E}{h_E}. \qquad (2.20)$$

The symbol $e_i$, $i = 1, 2 \ldots, m$ will refer to the edge having $\mathrm{v}_i$ as its first vertex, see Fig.(2.3).

### 2.3.2 Virtual space and degrees of freedom

This Section introduces the local and global space for the vector valued test VEM functions and the associated degrees of freedom for the general VEM formulation of order $k$. The local discrete *virtual* space of order $k$ of the admissible displacements is defined for the generic element $E$ as [34]

$$\mathcal{V}^h(E) := \{ \boldsymbol{v}^h \in \left[H^1(E)\right]^2 : \Delta \boldsymbol{v}^h \in [\mathbb{P}_{k-2}(E)]^2, \boldsymbol{v}^h_{|e} \in [\mathbb{P}_k(e)]^2 \, \forall e \in \partial E \} \quad (2.21)$$

Expression (2.21) uses the following notations:

- $H^1(E)$ is the first order Sobolev space;
- $\Delta$ denotes the component-wise Laplace operator;
- $k \in \mathbb{N}$ is the degree of accuracy of the method;

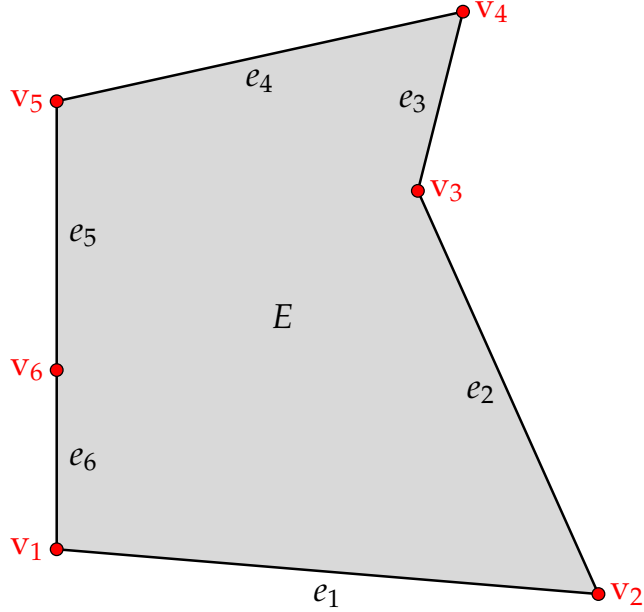

Figure 2.3: Example of a generic VEM element.

- $[\mathbb{P}_k(\bullet)]^2$ denotes the space of vector polynomials of degree up to $k$ defined on $(\bullet)$, with the convention that $\mathbb{P}_{-1} = \{0\}$.

The definition (2.21) states that a function $v^h \in \mathcal{V}^h(E)$ has the following properties:

- $v^h$ is a vector valued continuous functions on $E$;

- $v^h$ is a vector polynomial of degree $\leq k$ on each edge $e$ of $E$;

- $\Delta v^h$ is a vector polynomial of degree $k - 2$ in $E$ when $k > 1$ and it is equal to zero when $k = 1$.

According to the definition in Eq.(2.21), the local space contains all polynomials of degree up to $k$ plus other functions that are well defined but are not explicitly known inside the element but only on the element boundary. The name *virtual* for the space $\mathcal{V}^h(E)$ comes from the fact that, differently from standard FE method, the above definition of the local displacement approximation is not fully explicit.

The global *virtual* element space can be defined based on the local space as

$$\mathcal{V}^h := \{v \in \mathcal{V} : v^h_{|E} \in \mathcal{V}^h(E) \ \forall E \in \Omega^h\}. \tag{2.22}$$

Any function $v^h \in \mathcal{V}^h$ is univocally defined by the following set of degrees of freedom

- the values of $v^h$ at the $m$ vertexes of the polygon $E$;

- for $k \geq 2$, the values of $v^h$ at $k-1$ internal points on every edge $e \in \partial E$, this points can be chosen according to the Gauss-Lobatto quadrature rule with $k+1$ points.

- for $k \geq 2$, the moments $\frac{1}{|E|} \int_E \boldsymbol{p} \cdot v^h dE \quad \forall \boldsymbol{p} \in [\mathbb{P}_{k-2}(E)]^2$.

In the lowest-order VEM formulation ($k = 1$), $\mathcal{V}^h(E)$ is a space of harmonic vector valued functions that are piecewise linear (edge by edge) and continuous on the boundary of the element and the associated degrees of freedom are the the values of $v^h$ at the $m$ vertexes of the polygon $E$ and, since $v^h \in \mathcal{V}^h(E)$ is linear on each edge, the value of $v^h$ at every point along the boundary $\partial E$ can be determined.

### 2.3.3 Projection operator

This Section introduces the definition and describe the computational implementation of the projection operator that plays a fundamental role in the Virtual Element Method.

In the previous section it has been shown that the functions of the local virtual space $\mathcal{V}^h(E)$ are not explicitly known inside the generic element $E$. Consequently, the local discrete bilinear form $\mathscr{L}_E(\cdot, \cdot)$ cannot be computed by standard numerical integration, as usually done in FEM.

For this purpose, the virtual element approximation employs a polynomial *projector* operator $\Pi$, defined on $E$, $\forall v^h \in \mathcal{V}^h(E)$ by the orthogonality condition

$$\int_E \boldsymbol{p}^\intercal \left[ \Pi(v^h) - \boldsymbol{\varepsilon}(v^h) \right] dE = 0 \quad \forall \boldsymbol{p} \in [P_0(E)]^3, \tag{2.23}$$

which identifies $\Pi(v_h)$ as a polynomial *approximation* of the strains field associated to the *unknown* displacements $v_h$, so that the error with respect to the approximated strain field $\boldsymbol{\varepsilon}(v^h)$ has no components along the space of polynomial of order $k - 1$ over $E$. Having $\boldsymbol{p}$ constant components when $k = 1$, Eq.(2.23) simplifies to

$$\Pi(v^h) = \frac{1}{|E|} \int_E \boldsymbol{\varepsilon}(v^h) \, dE. \tag{2.24}$$

In the following Section it will be shown how to compute the right-hand side of Eq.(2.24).

**Projector operator matrix**

To be effectively employed, the projector operator must be computed in a discrete form. To do so, each function $v^h \in \mathcal{V}^h(E)$, whose explicit expression is unknown, can be thought as

$$v^h = \mathbf{N}\,\tilde{v}, \tag{2.25}$$

where

$$\tilde{v} = \begin{bmatrix} \tilde{v}_{x1} & \tilde{v}_{y1} & \ldots & \ldots & \tilde{v}_{xm} & \tilde{v}_{ym} \end{bmatrix}^\mathsf{T}, \tag{2.26}$$

collects the point-wise values of $v^h$ associated with the element vertices. The matrix $\mathbf{N} \in \mathbb{R}^{2 \times 2m}$ collects the *virtual* shape functions $N_i$, associated with each vertex $i$ of the element $E$. It has the following structure

$$\mathbf{N} = \begin{bmatrix} N_1 & 0 & N_2 & 0 & \ldots & \ldots & N_m & 0 \\ 0 & N_1 & 0 & N_2 & \ldots & \ldots & 0 & N_m \end{bmatrix}, \tag{2.27}$$

which remarks its analogy to the shape function matrix operator of a standard displacement-based finite element formulation. It's worth noting that the functional entries of the matrix $\mathbf{N}$ are never analytically known at the interior points of the generic element; only their restriction to the element edges is explicitly known, due to the features of the space $\mathcal{V}^h(E)$ defined in Section 2.3.2. Moreover, each virtual shape function $N_i$ associated to a vertex $v_i$ is endowed with the usual Kronecker delta property of having value 1 at vertex $v_i$ and 0 at any other vertex. Here, along the lines of Ref.[13], the expression in Eq.(2.25) is introduced to allow a formal presentation of the method more suited to readers with an engineering background, thus familiar with the concept of shape functions.

Employing Eq.(2.25) in Eq.(2.24), and considering the strain-displacement relationship in Eq.(2.7), yields

$$\Pi(v_h) = \frac{1}{|E|} \int_E \mathscr{D}\left(\mathbf{N}\tilde{v}\right) dE = \frac{1}{|E|} \left( \int_{\partial E} \mathscr{D}_n \cdot \mathbf{N}\, d\partial E \right) \tilde{v} = \mathbf{\Pi}\,\tilde{v}, \tag{2.28}$$

where the boundary integral appearing in the third term has been obtained applying the Green's theorem to the domain integral and the operator $\mathscr{D}_n$ has been defined in Eq.(2.12).

Eq.(2.28) defines the discrete projector operator $\mathbf{\Pi} \in \mathbb{R}^{3\times 2m}$ as

$$\mathbf{\Pi} = \frac{1}{|E|} \int_{\partial E} \mathscr{D}_n \cdot \mathbf{N} \, d\partial E = \frac{1}{|E|} \sum_{k=1}^{m} \int_{e_k} \mathscr{D}_n \cdot \mathbf{N} \, ds, \tag{2.29}$$

From Eq.(2.29), it follows that the discrete projector operator has the following explicit matrix form

$$\mathbf{\Pi} = \begin{bmatrix} \pi_{1x} & 0 & \pi_{2x} & 0 & \ldots & \ldots & \pi_{mx} & 0 \\ 0 & \pi_{1y} & 0 & \pi_{2y} & \ldots & \ldots & 0 & \pi_{my} \\ \pi_{1y} & \pi_{1x} & \pi_{2y} & \pi_{2x} & \ldots & \ldots & \pi_{my} & \pi_{mx} \end{bmatrix}, \tag{2.30}$$

where the generic coefficient $\pi_{ix}$ can be expressed as

$$\pi_{ix} = \frac{1}{|E|} \sum_{i=1}^{m} n_x^i \int_{e_i} N_i ds. \tag{2.31}$$

A similar expression can be used for the generic coefficient $\pi_{iy}$

$$\pi_{iy} = \frac{1}{|E|} \sum_{i=1}^{m} n_y^i \int_{e_i} N_i ds, \tag{2.32}$$

$n_x^i$ and $n_y^i$ are the components of the unit outward normal vector to the edge $e_i$. Recalling the Kronecker delta property of the virtual shape functions, according to which the integral of the generic virtual shape function $N_i$ is not null only on the edges adjacent to vertex $i$, the sum of the edge integrals appearing in Eqs.(2.31) and (2.32) can be further simplified leading to the following expressions for the coefficients of the projection operator matrix $\mathbf{\Pi}$

$$\begin{aligned} \pi_{ix} &= \frac{1}{|E|} \left[ n_x^{i-1} \int_{e_{i-1}} N_i ds + n_x^i \int_{e_i} N_i ds \right] \\ &= \frac{1}{2|E|} \left[ n_x^{i-1} |e_{i-1}| + n_x^i |e_i| \right], \end{aligned} \tag{2.33}$$

$$\begin{aligned} \pi_{iy} &= \frac{1}{|E|} \left[ n_y^{i-1} \int_{e_{i-1}} N_i ds + n_y^i \int_{e_i} N_i ds \right] \\ &= \frac{1}{2|E|} \left[ n_y^{i-1} |e_{i-1}| + n_y^i |e_i| \right], \end{aligned} \tag{2.34}$$

where $|e_i|$ is the length of the $i$-th edge. It's worth highlighting that Eqs.(2.33) and (2.34) allow the explicit computation of the discrete projector operator $\mathbf{\Pi}$ directly from geometric considerations.

### 2.3.4 Virtual element stiffness matrix

The construction of the virtual element approximation of the local symmetric bilinear form $\mathscr{L}_E(\cdot,\cdot)$ appearing in Eq.(2.18) is based on the following local decomposition [31]

$$\mathscr{L}_E^h\left(u^h,v^h\right) = \int_E \Pi(v^h)^\mathsf{T} \mathbf{C}\, \Pi(u^h)\, dE + s_E\left(u^h,v^h\right), \qquad (2.35)$$

defined $\forall E \in \Omega_h$ and $\forall u^h, v^h \in \mathcal{V}^h(E)$. Eq.(2.35) is the sum of two terms, respectively related to the *consistency* and *stability* property of the method and in Ref.[31, 34]) it has been demonstrated that such decomposition guarantees the convergence of the virtual element method.

The first term on the right-hand side of Eq.(2.35) is the bilinear form associated to the linear displacements and constant strain modes that ensures *consistency*. The approximation $\mathscr{L}_E^h(\cdot,\cdot)$ is said to be linear consistent if the exact bilinear form $\mathscr{L}_E(\cdot,\cdot)$ is recovered when the first entry is a linear polynomial [37]

$$\mathscr{L}_E^h\left(p,v^h\right) = \mathscr{L}_E\left(p,v^h\right) \quad \forall p \in [\mathbb{P}_1(E)]^2,\, \forall v^h \in \mathcal{V}^h(E), \qquad (2.36)$$

In other words, if the solution of the original problem is globally a linear polynomial, then the discrete solution and the exact solution coincide. Thus, the consistency matrix ensures that a virtual element is capable of exactly representing a constant strain state (linear displacement field), passing the *linear patch test* [26].

The consistency term uses the local projector defined in Eq.(2.24) and leads to the definition of the element consistency stiffness matrix $\mathbf{K}_E^c \in \mathbb{R}^{2m \times 2m}$. In fact, using Eq.(2.28), such a term can be approximated as

$$\int_E \Pi(v^h)^\mathsf{T} \mathbf{C}\, \Pi(u^h)\, dE = \tilde{v}^\mathsf{T} \left(\int_E \mathbf{\Pi}^\mathsf{T}\, \mathbf{C}\, \mathbf{\Pi}\, dE\right) \tilde{u} = \tilde{v}^\mathsf{T}\, \mathbf{K}_E^c\, \tilde{u}, \qquad (2.37)$$

which, being the integrand constant over $E$, readily gives the consistency stiffness matrix

$$\mathbf{K}_E^c = |E|\, \mathbf{\Pi}^\mathsf{T} \mathbf{C}\, \mathbf{\Pi}. \qquad (2.38)$$

The second term on the right-hand side of Eq.(2.35) is the bilinear form associated to the non-polynomial terms or higher-order terms and leads to the definition of the element stability stiffness matrix $\mathbf{K}_E^s \in \mathbb{R}^{2m \times 2m}$. The stability term is necessary for the virtual element approximation scheme since the consistency term alone cannot suppress the development of hourglass modes, nonphysical, zero-energy modes of deformation that produce zero strain and no stress and usually develop in under-integrated finite elements [40].

An algebraic interpretation of the hourglass modes' appearing may be given by considering the structure of the consistency stiffness matrix $\mathbf{K}^c$. For a generic virtual element with $m$ vertices, a system of $2m$ equations has to be solved for the $2m$ unknowns given by the vertex values of the displacement components. By using the constitutive equations once per element, 3 independent relations can be supplemented. Thus, the stiffness matrix has $2m - 3$ equations that are not independent. By fixing the 3 zero-energy modes corresponding to the 3 rigid body modes (two translations and one rotation), other 3 independent relations can be supplemented leaving with $2m - 6$ equations that are not independent. When the considered polygonal virtual element is a triangle, the total number of equations is $2m = 6$, the consistency matrix has proper rank and no hourglass instability can arise. However, when considering polygonal virtual elements with 4 or more vertices, $2m = 6$ zero-energy modes can develop, thus requiring a stabilization term that ensures proper rank of the virtual element stiffness matrix.

Within VEM's literature, there exist different approaches to construct the stability term. Whatever the approach adopted, the stability term must correct the consistency term in such a way that (a) the resulting bilinear form is stable, (b) consistency is not affected, (c) the correction is computable using only the degrees of freedom.

In this thesis, the approach introduced in Ref.[34] has been used. The approximated constant strain field computed trough the projector operator matrix using Eq.(2.28) is not compatible with the boundary piece-wise linear approximation of the virtual displacement field unless the number of element vertices $m$ is equal to 3. In fact, such a displacement field could generate deformation modes of a higher order than the constant ones. Those modes are neglected by the consistent term of the virtual approximation of the bilinear form. However, the approximate strain field $\Pi(v_h)$ obtained from the projection operation is compatible with a linear displacement field $\mathbf{u}^c$ that can be expressed as

$$\mathbf{u}^c = \mathbf{P}(\xi, \eta)\, \mathbf{a}, \tag{2.39}$$

where

$$\mathbf{P}(\xi, \eta) = \begin{bmatrix} 1 & 0 & \xi & 0 & \eta & 0 \\ 0 & 1 & 0 & \xi & 0 & \eta \end{bmatrix}, \tag{2.40}$$

is a matrix of polynomial basis functions, $\xi$ and $\eta$ are the local scaled coordinates defined in Eq.(2.20) and

$$\mathbf{a} = \begin{bmatrix} a_1 & a_2 & a_3 & a_4 & a_5 & a_6 \end{bmatrix}^\mathsf{T}, \tag{2.41}$$

is a vector of constant coefficients. The compatible displacements $\mathbf{u}^c$ can be evaluated at the $i$-th vertex of a generic virtual element $E$ as

$$\mathbf{u}_i^c = \mathbf{P}\left(\xi_i, \eta_i\right) \mathbf{a}, \tag{2.42}$$

The components of the compatible displacements at all the element vertices can be collected in the vector

$$\tilde{\mathbf{u}}^c = \mathbf{D}\,\mathbf{a}. \tag{2.43}$$

where $\mathbf{D} \in \mathbb{R}^{2m \times 6}$ is the matrix obtained assembling the matrices $\mathbf{P}\left(\xi_i, \eta_i\right)$ and whose explicit expression is

$$\mathbf{D} = \begin{bmatrix} 1 & 0 & \xi_1 & 0 & \eta_1 & 0 \\ 0 & 1 & 0 & \xi_1 & 0 & \eta_1 \\ \vdots & \vdots & \vdots & \vdots & \vdots & \vdots \\ 1 & 0 & \xi_m & 0 & \eta_m & 0 \\ 0 & 1 & 0 & \xi_m & 0 & \eta_m \end{bmatrix}. \tag{2.44}$$

The unknown coefficients $\mathbf{a}$ can be estimated by solving, in the least squares sense,

$$\min_{\mathbf{a}}\{||\tilde{\mathbf{u}} - \mathbf{D}\mathbf{a}||\}, \tag{2.45}$$

where the objective function

$$||\tilde{\mathbf{u}} - \mathbf{D}\mathbf{a}|| = (\tilde{\mathbf{u}} - \mathbf{D}\mathbf{a})^\mathsf{T} (\tilde{\mathbf{u}} - \mathbf{D}\mathbf{a}), \tag{2.46}$$

represents the Euclidean distance between the vectors $\tilde{\mathbf{u}}$ and $\tilde{\mathbf{u}}^c$. The minimization problem stated in Eq.(2.45) is equivalent to the linear system

$$\mathbf{D}^\mathsf{T} \mathbf{D}\mathbf{a} = \mathbf{D}^\mathsf{T}\tilde{\mathbf{u}}, \tag{2.47}$$

whose solution, gives

$$\mathbf{a} = \left(\mathbf{D}^\mathsf{T}\mathbf{D}\right)^{-1} \mathbf{D}^\mathsf{T}\tilde{\mathbf{u}}. \tag{2.48}$$

Substituting Eq.(2.48) in Eq.(2.43) it is possible to obtain an approximation of the nodal values of the compatible displacements in terms of the nodal values of the virtual displacements

$$\tilde{\mathbf{u}}^c = \mathbf{\Pi}^s \tilde{\mathbf{u}}, \tag{2.49}$$

where $\mathbf{\Pi}^s \in \mathbb{R}^{2m \times 2m}$ is the matrix projector operator

$$\mathbf{\Pi}^s = \mathbf{D}\left(\mathbf{D}^\mathsf{T}\mathbf{D}\right)^{-1}\mathbf{D}^\mathsf{T}. \tag{2.50}$$

To reintroduce the amount of the deformation energy neglected by the consistency term, the energy associated to the displacement difference $\tilde{\mathbf{u}} - \tilde{\mathbf{u}}^c$ is taken in account by defining the following expression of the stabilization term

$$\begin{aligned}
s_E(u^h, v^h) &= [\tilde{v} - \tilde{u}^c]^\mathsf{T} \mu \, [\tilde{u} - \tilde{u}^c] \\
&= [\tilde{v} - \mathbf{\Pi}^s \tilde{v}]^\mathsf{T} \mu \, [\tilde{u} - \mathbf{\Pi}^s \tilde{u}] \\
&= [(\mathbf{I} - \mathbf{\Pi}^s) \, \tilde{v}]^\mathsf{T} \mu \, [(\mathbf{I} - \mathbf{\Pi}^s) \, \tilde{u}] \\
&= \tilde{v}^\mathsf{T} (\mathbf{I} - \mathbf{\Pi}^s)^\mathsf{T} \mu \, (\mathbf{I} - \mathbf{\Pi}^s) \, \tilde{u} \\
&= \tilde{v}^\mathsf{T} \mathbf{K}_E^s \, \tilde{u},
\end{aligned} \tag{2.51}$$

where $\mathbf{I} \in \mathbb{R}^{2m \times 2m}$ is the identity matrix, and

$$\mathbf{K}_E^s = (\mathbf{I} - \mathbf{\Pi}^s)^\mathsf{T} \mu \, (\mathbf{I} - \mathbf{\Pi}^s), \tag{2.52}$$

is the virtual element stabilization matrix $\mathbf{K}_E^s \in \mathbb{R}^{2m \times 2m}$. The constant parameter $\mu$ is used to ensure the correct scaling of the stability term with respect to the element size and material constants. For linear elasticity problems, following Ref.[34], this parameter can be expressed as

$$\mu = \tau \operatorname{tr}\left(\mathbf{K}_E^c\right), \tag{2.53}$$

where $\operatorname{tr}\left(\mathbf{K}_E^c\right)$ is the trace of the consistency matrix $\mathbf{K}_E^c$ and the factor $\tau > 0$ can be selected as $\tau = 1$ or $\tau = 0.5$ [13]. However, other choices can be found in the literature [77].

Eventually, the stiffness matrix $\mathbf{K}_E$ for a generic virtual element $E$ can be computed as the sum of the consistency and stability matrices

$$\mathbf{K}_E = \mathbf{K}_E^c + \mathbf{K}_E^s. \tag{2.54}$$

It is also worth noting that if the generic virtual element is a triangle, the approximated constant strain field computed trough the projector operator matrix is compatible with the boundary piece-wise linear approximation of the displacement field, the difference between the displacement vectors $\tilde{u}^c$ and $\tilde{u}$ is null and so is the virtual element stabilization matrix $\mathbf{K}_E^s$.

### 2.3.5 Loading vectors

For the lowest-order VEM, following Ref.[31] the local contribution $\mathscr{G}_E(\cdot)$ to the virtual work of the volume load $f$ appearing at the right-hand side of Eq.(2.17), if existing, can be approximated as

$$\mathscr{G}_E(v^h) \approx \mathscr{G}_E^h(v^h) = \int_E \bar{v}^h \cdot f^h \, dE, \tag{2.55}$$

where

$$\bar{v}^h = \frac{1}{m}\sum_{i=1}^m v^h(\tilde{x}_i) = \frac{1}{m}\sum_{i=1}^m \mathbf{N}(\tilde{x}_i)\tilde{v}, \tag{2.56}$$

denotes the average value of $v^h$ at the vertices of $E$ and

$$f^h = \Pi_0\left(f\right) := \frac{1}{|E|}\int_E f \, dE, \tag{2.57}$$

is the $L^2(E)$ projection onto constants of the load $f$.

The virtual work of distributed tractions $\bar{t}$, acting along a part of the element boundary denoted with $\partial E_t$ of the generic virtual element $E$ can be expressed as

$$\mathscr{T}_E(v^h) = \int_{\partial E_t} v^h \cdot \bar{t} \, ds = \sum_{i=1}^m \int_{e_i \in \partial E_t} v^h \cdot \bar{t} \, ds. \tag{2.58}$$

Analogously to standard FEM, the knowledge of the explicit expression of the restrictions of the shape functions $N_i$ to the element edges allows the computation of the virtual work of tractions and the convenient definition of nodal forces $f_t$ in terms of nodal values of tractions.

Eventually, it is worth noting that, since the shape functions $N_i$ are explicitly known on the element edges, non-homogeneous boundary conditions over the virtual elements can be enforced exactly as in standard FEM.

### 2.3.6 Assembly of global system

Denoting with $\tilde{f}_E$ the nodal forces, which in general include volume end edge contributions, and considering the definition of stiffness matrix given in Eq.(2.54), it is possible to write the elemental equilibrium equations of elasticity within the framework of the lowest-order VEM as

$$\mathbf{K}_E \, \tilde{u} = \tilde{f}_E. \tag{2.59}$$

Once the elemental stiffness matrices and load vectors are computed, they are assembled into global stiffness matrix **K** and load vector **F** by employing standard FE numbering and procedures, which motivates the appeal of VEM as a versatile method requiring minimum re-coding in existing software packages. Eventually, a system of equation for the overall discrete domain can be written as

$$\mathbf{K}\mathbf{U} = \mathbf{F}, \qquad (2.60)$$

where **U** is the global vector of degrees of freedom. After prescribing appropriate boundary conditions, Eq.(2.60) can be solved using common finite element solution methods.

### 2.3.7 Strains and Stresses

Once the global system in Eq.(2.60) is solved and the values of the displacement components at every node of the mesh are known, strains can be computed element-wise using the projector operator matrix $\mathbf{\Pi}$ and the vector of the local nodal values of the displacement components $\tilde{u}$

$$\varepsilon_\Pi = \mathbf{\Pi}\,\tilde{u}, \qquad (2.61)$$

where superscript $\Pi$ is used to remark that, since the displacement field is *virtual* in the interior of the element, the actual local strain field is approximated by its projection on the constant polynomial space.

Using Eq.(2.61), stresses can be computed element-wise using the constitutive law in Eq.(2.8)

$$\sigma = \mathbf{C}\,\varepsilon_\Pi. \qquad (2.62)$$

As previously highlighted, in the lowest-order VEM formulation, the deformation field inside the generic element, obtained through the projector's matrix expression, is a constant approximation of the actual strain field. Instead, the displacement field is assumed linear on the edges of the element while inside the element it is represented by functions whose Laplacian is zero. Therefore, for a virtual element with three sides, the lowest-order VEM formulation coincides with the standard FEM formulation of the linear triangular element, commonly known as CST (Constant Strain Triangle). In the case of polygonal virtual elements with more than three sides, the degrees of freedom that uniquely define the displacement field could give rise to a deformation field of higher order than the constant one. In this case, some loss of information might be expected when using Eq.(2.62) for calculating the stress field, especially in the case of polygonal elements with many sides. Although in literature there are studies of convergence of the VEM which show that this intrinsic loss of information of the method is negligible [132], it is worth noting that it is possible to extend to VEM [15] some stress-recovery techniques already widely used in FEM [4].

## 2.4 VESTA - A Virtual Element toolbox

The purpose of this Section is to introduce `VESTA`, the virtual element program that has been developed and used to perform all the analysis of the present thesis. `VESTA` is an acronym for Virtual Element for STructural Analysis, and it is a framework for the implementation of the Virtual Element Method for the solution of two-dimensional elasticity problems in plane strain or plane stress. `MATLAB` is the scripting language that has been chosen to write all the functions that are part of the current `VESTA` software library. The program is written using a proper mix of Object-oriented programming (OOP) and functional programming, using the former for core functions and organising and managing data structures and functional features where this makes coding more concise and flexible.

`VESTA` is essentially a finite element program and, in this respect, its structure is similar to other research-oriented finite element codes. The program features a "core and applications" approach where a set of core tools are available as building blocks in developing new applications that focus on the solution of particular problems of interest.
The program includes an integrated set of functions used to perform the following tasks:

- Generate, visualize and store the input data describing a finite element model;

- Solve a specific application trough appropriate solution algorithms;

- Return graphical and numerical output of analysis results and save them to file.

The program's main block includes a `MATLAB` class object that contains all the input of data describing the structural analysis model for the specific analysis. It stores all the relevant data on the analysis type, mesh structure, material properties, applied loads and boundary conditions. This function allows to save the analysis information to a text input file or to load a previously saved analysis model.

The current version of `VESTA` contains an element library with different types of elements. Each element type has a specific `MATLAB` class object that contains all the functions needed for computation of element matrices, element force vectors and output strains and stresses associated with each element. A separate function is used to project these quantities to nodes allowing graphical outputs of results. The main element type used by the program is the general polygonal VEM element with the first-order formulation. For accuracy comparison and testing purposes, the element library has been expanded to include first-order (linear) interpolation triangular standard finite elements and fully integrated, two-dimensional, linear and quadratic interpolation quadrilateral elements. Another relevant class object of this group is dedicated to the Boundary Element Method (BEM). The BEM class object contains all the functions to perform a stand-alone boundary element analysis with linear interpolation elements in the framework of linear elasticity. It also contains the functions needed to construct the equivalent stiffness matrix and the nodal force vectors for a BEM super-element. As shown later in this thesis, these functions are used to interface a BEM subdomain with another subdomain discretised with VEM elements.

A separate assembly function performs the task of assembling the global stiffness matrix and the global force vector. This function accesses a material model library used to compute the constitutive equations. The current implemented material models are linear elastic material and isotropic damage model material.

Mesh generation and manipulation are performed using both third-part and in-house developed codes. For the generation of a triangular mesh, two different third part codes are used. For relatively simple domains whose geometries can be specified by signed distance functions `VESTA` relies on `DistMesh` [145], a `MATLAB` code for generation of unstructured triangular and tetrahedral meshes that uses the Delaunay triangulation routine in `MATLAB` and optimizes the node locations by a force-based smoothing procedure. To generate triangular meshes on domains with more complex geometries, the software used is `MESH2D` [70], a `MATLAB`-based Delaunay mesh-generator designed to generate high-quality constrained Delaunay triangulations for general polygonal regions in the plane. `MESH2D` provides an effective implementations of "Delaunay-refinement" and "Frontal-Delaunay" triangulation techniques, in additional to "hill-climbing" type mesh-optimisation. It also provides sup-port for user-defined "mesh-spacing" functions and "multi-part" geometry definitions, allowing varying levels of mesh-resolution to be specified within complex domains.

To exploit the capabilities of the Virtual Element Method, a polygonal mesh generator is required. For simple analysis on geometric domain without internal partitions, `VESTA` relies on `Polymesher` [168], a simple and robust `MATLAB` code for polygonal mesh generation that relies on an implicit description of the domain geometry. An example of a polygonal mesh generated with `Polymesher` is shown in Fig.(2.2). To perform more complex polygonal mesh generation tasks, like the ones involved in the automatic discretisation of random generated composite microstructures, where the analysis domain consists of multiple subdomains, an in-house mesh generation code has been implemented. It is based on an initial conforming triangular discretisation of the domain and a subsequent generation of a bounded Voronoi tessellation generated using the centroids of the triangular mesh elements as seed points. More details and examples about this process will be given in the following Chapters.

## 2.5 VEM implementation aspects

As highlighted before, there are a few similarities between VEM and standard FEM that allow the use of virtual elements within existing software packages with minimum re-coding. The purpose of this Section is to summarize the main similarities and differences between the two methods concerning mesh generation and management, global system assembly procedure and related computational cost.

Regarding the management of the discretization input, it is evident that the main difference between the standard FEM and the VEM lies in the arbitrariness of the mesh elements' geometry. While in the standard FEM, in the two-dimensional case, the elements allowed are exclusively triangles or quadrilaterals, in the VEM generic polygonal elements are allowed. The arbitrary number of sides that the VEM admits for a generic element of the mesh requires appropriate input management. The latter consideration implies that the data structure used to store information on the mesh topology in a FEM code may not be directly applicable to a VEM-based code. A proposed way of approaching this issue [132] is to set a separate data structure for the VEM discretization that uses an associative array (map) where the key is the number of polygon nodes, and the value is the corresponding connectivity array (element freedom table). The approach followed in VESTA relies instead on considering all the mesh entries in the connectivity table as polygonal elements and associate with each element of the table an "element type" key to sort FEM elements from VEM elements. This approach is particularly useful when, as will be seen later in this thesis, the computation domain discretization includes BEM super-elements, i.e. polygonal elements discretised on the boundary with multiple BEM elements but considered, for the analysis, as single elements.

As already mentioned, the computational cost for generating the global system of equation for VEM's lowest-order formulation is comparable to FEM's for linear triangular elements. The main difference between the two formulations resides in the additional computation of the stabilization term of a VEM element stiffness matrix which requires the computation of a matrix inversion once per element.

Once an efficient scheme for polygonal mesh management has been set-up, VEM's assembling of the global system of equation proceeds in a similar fashion to standard FEM's one, actually, the same functions that would be used for a FEM program can be employed with minor modification for a VEM code when first- or second-order formulation are employed.

## 2.6 Linear patch test

This Section presents a numerical example used to assess the ability of the lowest-order VEM formulation to pass a polygonal element version [13] of the two-dimensional plane stress linear patch test [26]. The tests have been performed using the in-house developed VESTA program.

The analysis domain is a square whose side has length $L = 1$. The material properties are Young's modulus $E = 70000$ and Poisson's ratio $\nu = 0.3$. Geometry and boundary conditions are shown in Fig.(2.4). Two cases are considered: in case (a) a normal traction $q = 1000$ applied on the right edge of the square domain induces a constant normal stress; in case (b) a tangential traction $t = -400$ applied on the all the edge of the square domain induces a constant shear stress state. Both cases are analysed under plane stress using two different polygonal discretisation: mesh (1) shown in Fig.(2.5) and mesh (2) shown in Fig.(2.6). Node coordinates of both meshes are reported in Table (2.1).

The exact solution for case (a) is

$$u = \frac{q}{E}x, \quad v = -\frac{\nu q}{E}y, \quad \sigma_{xx} = q, \quad \sigma_{yy} = \tau_{xy} = 0. \tag{2.63}$$

The exact solution for case (b) is

$$u = 0, \quad v = \frac{t}{G}x, \quad \sigma_{xx}\sigma_{yy} = 0, \quad \tau_{xy} = t, \tag{2.64}$$

where $G$ is the shear modulus.

The error estimator for the displacement field is defined as

$$e_u = \left[\frac{\sum_i^N ||u_i^h - u_i||^2}{\sum_i^N ||u_i||^2}\right]^{\frac{1}{2}}, \tag{2.65}$$

where $i$ refers to the index of the $i$-th node and $N$ is the total number of mesh nodes.

The error estimator for the stress field is defined as

$$e_\sigma = \left[\frac{\sum_E \int_E ||\sigma(u^h) - \sigma(u)||^2 dE}{\sum_E \int_E ||\sigma(u)||^2 dE}\right]^{\frac{1}{2}}, \tag{2.66}$$

where $E$ denotes a generic virtual element of the mesh.

Results, reported in Tables (2.2)) and (2.3) shows that the lowest order VEM formulation is capable to recover the exact solution up to machine precision.

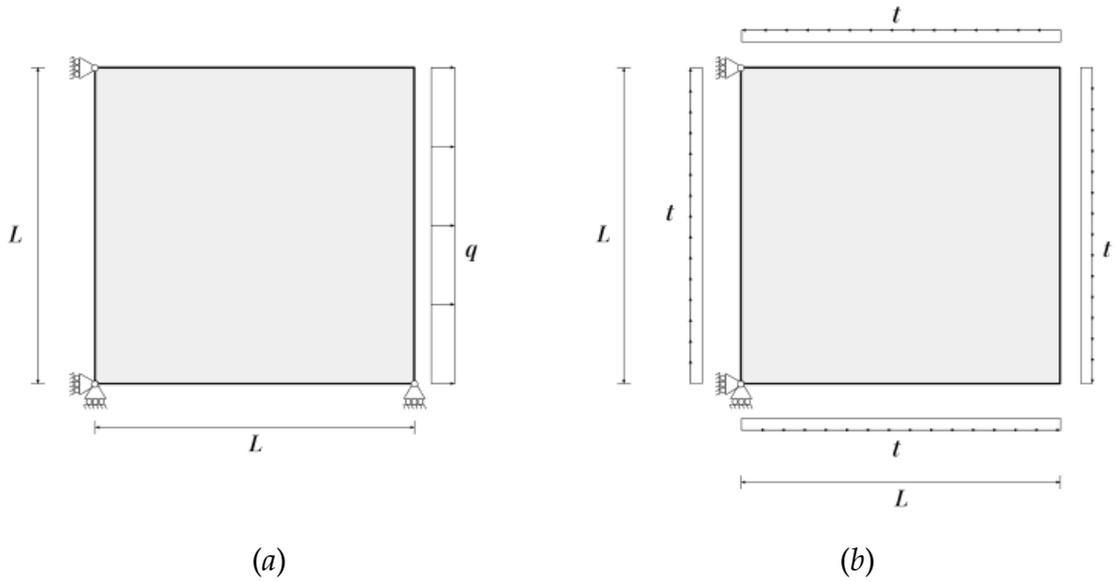

(*a*)  (*b*)

Figure 2.4: Linear patch test: geometry and boundary conditions for (*a*) normal and (*b*) shear constant stress states.

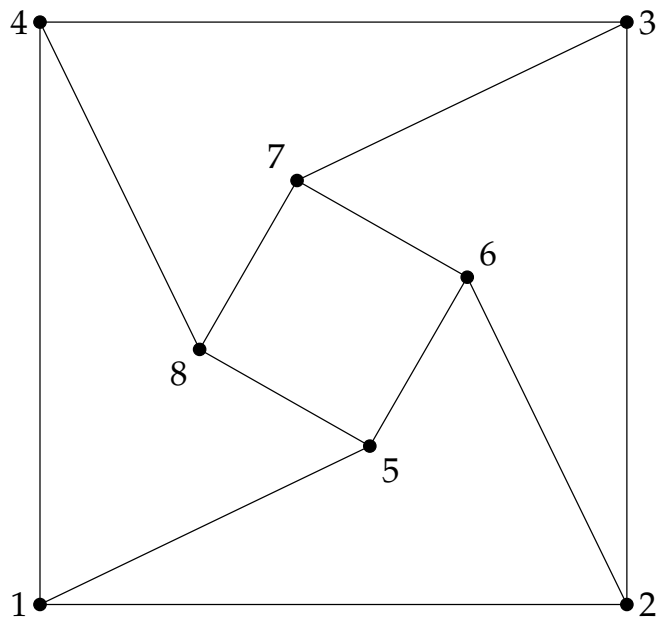

Figure 2.5: Linear patch test: mesh (1).

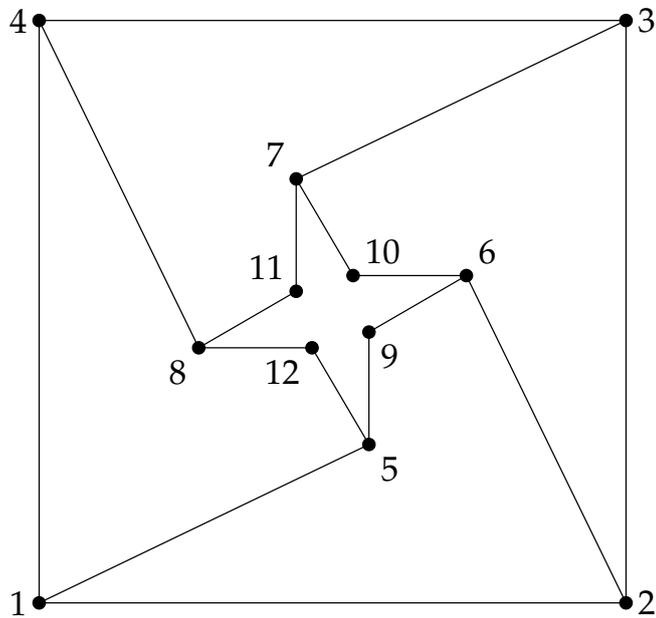

Figure 2.6: Linear patch test: mesh (2).

| Node | x | y |
|---|---|---|
| 1 | 0.000 | 0.000 |
| 2 | 1.000 | 0.000 |
| 3 | 1.000 | 1.000 |
| 4 | 0.000 | 1.000 |
| 5 | 0.562 | 0.272 |
| 6 | 0.728 | 0.562 |
| 7 | 0.438 | 0.728 |
| 8 | 0.272 | 0.438 |
| 9 | 0.562 | 0.465 |
| 10 | 0.535 | 0.562 |
| 11 | 0.438 | 0.535 |
| 12 | 0.465 | 0.438 |

Table 2.1: Linear patch test. Node coordinates for mesh (1), nodes 1 through 8 and mesh (2), nodes 1 through 12.

|       | Mesh (1) | Mesh (2) |
|-------|----------|----------|
| $e_u$ | 1.10e-15 | 2.44e-15 |
| $e_\sigma$ | 7.80e-16 | 1.68e-15 |

Table 2.2: Linear patch test. Case (a). Displacement error $e_u$ and stress error $e_\sigma$.

|       | Mesh (1) | Mesh (2) |
|-------|----------|----------|
| $e_u$ | 1.27e-15 | 5.39e-15 |
| $e_\sigma$ | 1.56e-15 | 4.54e-15 |

Table 2.3: Linear patch test. Case (b). Displacement error $e_u$ and stress error $e_\sigma$.

# Chapter 3

# Computational homogenization of composite and heterogeneous materials

## 3.1 Introduction

In Chapter (2), the VEM ability to deal with mesh elements of very general polygonal/polyhedral shape and to naturally address the presence of hanging nodes, providing accurate and consistent analysis results even with heavily distorted meshes, have been highlighted. Such flexibility makes the VEM an ideal candidate tool for computational homogenization studies, where the structure-property link is investigated homogenising the micro-fields over several statistical realisations of the material microstructure. In other words, being computational homogenization based on the analysis carried out over many micro representative volume elements, often generated and meshed automatically, the possibility to relieve the need of carefully assessing the quality of *each* mesh makes the VEM a suitable method for such analysis.

The present Chapter is intended to present the Virtual Element Method's application to the computational homogenization of polycrystalline and fibre-reinforced materials. The study has been conducted using the lowest-order VEM formulation presented in Chapter (2) for linear two-dimensional elastic problems. Emphasis is given to the method's flexibility in the analysis of randomly generated and meshed microstructures.

This Chapter is organised as follows. Section 3.2 details the computational aspects implemented to deal with generic polycrystalline and fibre-reinforced micro-morphologies and their meshing, describing the suitability of the virtual element method in dealing with specific features. Section 3.3 illustrates the application of the method to the computational homogenization of the considered materials and concludes the study.

## 3.2 Multi-domain implementation

In this Section, the multi-region implementation for computational material homogenization is described with reference to two classes of materials: polycrystalline and unidirectional fibre-reinforced composites, widely employed in engineering applications.

Some VEM applications to material homogenization of composites [12, 16, 148] and polycrystalline materials [124] have very recently appeared in the literature. Refs.[12, 16] consider unit cells with a single circular or elliptical inclusion, considered as the basic building block of composite materials with regular fibres distributions. Ref.[148] considers domains with a statistical distribution of fibres, but a single polygonal VEM element is used to model the individual fibres. Ref.[124] uses single polygonal or polyhedral VEM elements to model individual crystals in 2D and 3D, for homogenization purposes. In the present study, the focus is slightly different. Multi-domain microstructures obtained from random processes are considered, and no *a priori* assumption is made about the number of VEM elements used to model individual fibres or crystals. Emphasis is given to the flexibility offered by the features of VEM in meshing such general morphologies, which make it a convenient method for the analysis of complex random material microstructures.

The first step toward materials computational micro-mechanics is the adoption of an accurate representation of the material microstructure. This can be based either on the experimental reconstruction of real microstructures or on the computer generation of artificial models embodying the microstructural aggregate's relevant statistical features. Experimental techniques provide essential information, but they require suitable and generally expensive equipment and complicated and time-consuming post-processing. On the other hand, the use of reliable computer models offers the opportunity of simulating large numbers of microstructures, helping reduce the cost of the experimental effort [186].

In the context of the analysis of heterogeneous materials, the concepts of Representative Volume Element (RVE) and Statistical Volume Element (SVE) are notions of primary importance, see, e.g. Refs.[89, 87, 142, 81]. If a single microstructural *realization* is considered, it is crucial to determine the unit cell's minimum size needed to attain material representativity. The term *realization* is herein used to denote the specific morphology associated with a set of randomly scattered seed points, which can identify the centroids of polycrystals generated through Voronoi tessellations or the position of the fibres in fibre-reinforced composites; in this sense, the specific morphology has a role analogous to the value assumed by a random variable. For polycrystalline materials, the size of the RVE can be expressed in terms of the number of grains $N_g$ contained in the artificial microstructure. For unidirectional (UD) fibre-reinforced composites, the size of the RVE is measured by the parameter $\delta$, defined as the ratio between the length $L$ of the side of a square unit cell and the radius $r$ of the inclusion, typically a fibre, i.e.

$$\delta = \frac{L}{r}. \tag{3.1}$$

A definition of RVE can also be provided considering not only *volume* averages over individual realizations of different sizes but also *ensemble* averages over a set of realizations of the same size, provided that a sufficient number of samples is considered [102], which suggests the concept of SVE. With this approach, the estimation of the effective properties is obtained by computing the ensemble average of the apparent properties over a collection of realizations having the same size.

In the subsequent sections, the multi-region VEM strategy adopted for computational homogenization of heterogeneous materials is described, highlighting the VEM features that result particularly convenient for the meshing of irregular geometries, namely the VEM ability of naturally dealing with hanging nodes and non-convex or heavily distorted mesh elements. The method has been implemented for both polycrystalline and unidirectional (UD) fibre-reinforced composite materials, treated separately in the following sections to highlight the specific modelling requirements and the adopted solutions.

### 3.2.1 Polycrystalline materials

The modelling strategy employed for the analysis of polycrystalline materials at the micro-scale is described in this section, starting from the method adopted to construct the artificial microstructure.

**Generation of artificial polycrystalline micro-morphologies**

A reliable computer representation of the polycrystalline microstructure must retain the main topological, morphological and crystallographic features of the aggregate such as the number of vertices, edges and faces per grain, grain size distribution, grain shape and crystallographic orientation. Voronoi tessellations, which are analytically well defined and relatively simple to generate, have been successfully used to reproduce the main statistical features of real polycrystalline morphologies [23, 72, 150, 45], see Fig.(3.1).

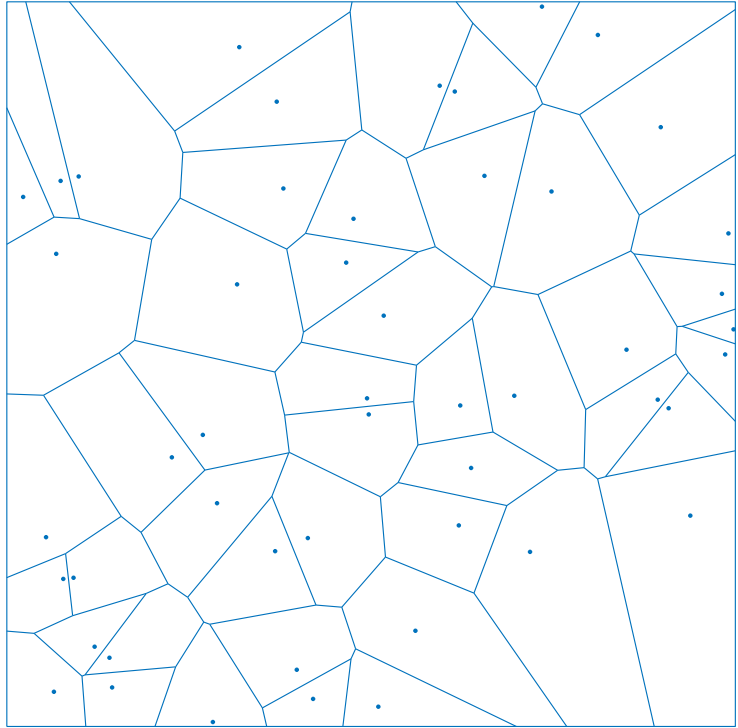

Figure 3.1: Example of a two-dimensional Voronoi tessellation on a square domain.

Given a bounded domain $\Omega \in \mathbb{R}^2$, its Voronoi tessellation is constructed starting from a set of $n$ seed points $S = \{x_i \in \Omega : i \in I_n\}$, with $I_n = \{1, 2, ..., n\}$. A Voronoi cell $\mathcal{G}_i$ having the seed $\mathbf{x}_i$ as its generator is defined as the set of points which are closer to $\mathbf{x}_i$ than to any other seed point, i.e.

$$\mathcal{G}_i = \{\mathbf{x} : \|\mathbf{x} - \mathbf{x}_i\| < \|\mathbf{x} - \mathbf{x}_j\| \ \forall j \neq i, \ j \in I_n\}. \tag{3.2}$$

Each seed is the generator of its Voronoi cell, and all cells form a Voronoi diagram, which divides the two-dimensional space into the union of convex, non-overlapping polygons with straight edges.

The tessellation's topology and morphology depend on the distribution of the seeds within the domain $\Omega$. It has been shown that a Voronoi tessellation, built on a set of randomly distributed seeds, referred to as Poisson-Voronoi tessellation, possesses statistical features that make it topologically close to real polycrystalline aggregates [106]. However, randomly distributed seed points tend to generate Voronoi tessellations with a high number of highly irregular or excessively distorted grains, particularly challenging from the point of view of mesh preparation for subsequent numerical analysis. Various techniques have been used to produce tessellations with non-pathological grain shapes or edges, e.g. enforcing a hardcore condition on the initial distribution of seed points or by employing more sophisticated regularization procedures, addressed at avoiding an excessively refined mesh induced by the presence of small edges in the mathematically exact built tessellation [83, 150].

In the present study, two-dimensional Voronoi tessellations are employed to generate artificial polycrystalline microstructures, where each Voronoi cell represents an individual grain. To demonstrate the ability of the Virtual Element Method to deal with mesh elements of very general polygonal shape, also generated over irregular geometries, no regularization scheme is adopted, and instead, a pure Poisson-Voronoi tessellation, with uniform random grain distribution and size, is used.

The tessellations have been generated using the Qhull [24] algorithm included in `MATLAB` to generate a uniform distribution inside the square domain representing the boundary of the unit cell. Since the edge length of the square domain is fixed, the only input required is the number of seed points equal to the number of grains. Fig.(3.2) shows different microstructural morphologies corresponding to different numbers of grains $N_g$.

**Micro-mechanical polycrystalline modelling**

A linear elastic orthotropic model is used to describe the mechanical behaviour of individual crystals. The orthotropic material hypothesis is not restrictive since most single-crystal metallic and ceramic materials present a general orthotropic behaviour. For an orthotropic material in a three-dimensional framework, the linear elastic constitutive laws introduced in Eq.(2.2) may be written

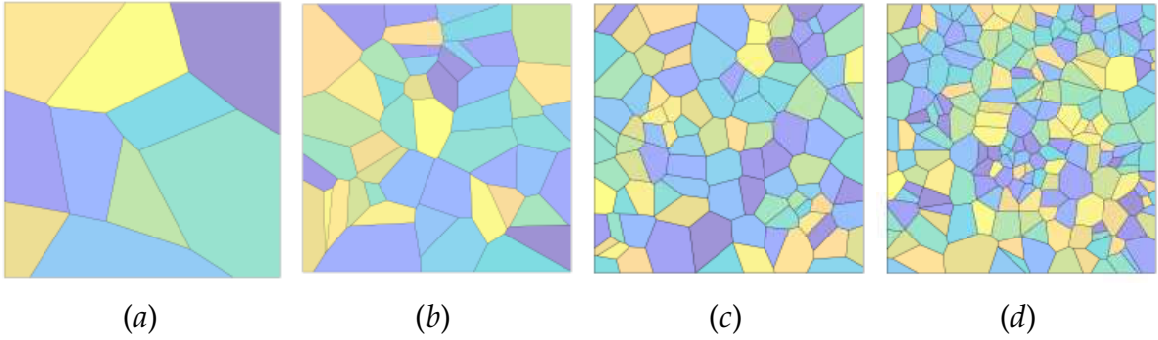

(a)                (b)                (c)                (d)

Figure 3.2: Polycrystalline morphologies with different numbers of grains: *a)* $N_g = 10$; *b)* $N_g = 50$; *c)* $N_g = 100$; *d)* $N_g = 200$.

as

$$\begin{bmatrix} \sigma_{11} \\ \sigma_{22} \\ \sigma_{33} \\ \sigma_{23} \\ \sigma_{13} \\ \sigma_{12} \end{bmatrix} = \begin{bmatrix} C_{11} & C_{12} & C_{13} & 0 & 0 & 0 \\ C_{12} & C_{22} & C_{23} & 0 & 0 & 0 \\ C_{13} & C_{23} & C_{33} & 0 & 0 & 0 \\ 0 & 0 & 0 & C_{44} & 0 & 0 \\ 0 & 0 & 0 & 0 & C_{55} & 0 \\ 0 & 0 & 0 & 0 & 0 & C_{66} \end{bmatrix} \begin{bmatrix} \varepsilon_{11} \\ \varepsilon_{22} \\ \varepsilon_{33} \\ \gamma_{23} \\ \gamma_{13} \\ \gamma_{12} \end{bmatrix}, \qquad (3.3)$$

where $\gamma_{ij} = 2\varepsilon_{ij}$ for $i = 1,...3$ and $i \neq j$.

Each grain of the microstructure is assumed to have random spatial orientation of the principal material directions $\{1, 2, 3\}$. Although two-dimensional problems are considered in the present study, the possibility of investigating the effect of the randomness of each grain's spatial orientation on the overall behaviour of the microstructure is preserved, as explained next. Following [72], each generated grain has, randomly, one of three principal material directions that coincides with the *z*-axis (normal to the analysis plane). Moreover, for each grain, the angle $\theta \in [0, 2\pi)$ between the global axes *x* and *y* and the axes of the two principal material directions lying in the plane $x - y$ is also randomly generated, Fig.(3.3).

The artificial polycrystalline morphology generated according to the procedure explained in Section 3.2.1 can be considered a multi-domain problem, in which different elastic properties and orientations are assigned to each grain. In the context of the FEM, several strategies have been used, in the literature, for the automatic generation of meshes for polycrystalline microstructures, and both structured and unstructured meshes have been used [150, 80, 74].

Structured meshes are generally unable to resolve the grain boundaries,

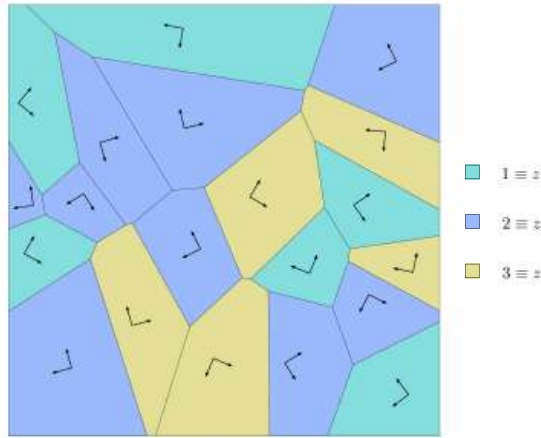

Figure 3.3: Example distribution of the principal material directions within each grain. The three different colours specify which principal direction coincides with the global $z$-axis. The orientations in the $x-y$ plane of the other two principal material directions is represented by two black vectors.

while unstructured meshes overcome this issue. However, given the morphological properties of random Voronoi tessellations, the generation of high quality conforming meshes requires a high degree of refinement that significantly increases the number of degrees of freedom.

In the present study, a multi-domain conforming meshing strategy is adopted, which takes advantage of the particular capability of the VEM of dealing with polygonal mesh elements with an arbitrary number of edges as well as with hanging nodes. Each grain of the microstructure has been independently *meshed* using a Centroidal Voronoi Tessellation (CVT; not to be confused with the tessellation used to generate the morphology), which allows subdividing the often very irregular grain geometry into quite regular polygonal elements, see Fig.(3.4). For this purpose, a modified version of `Polymesher` [168] is used; `Polymesher` is a mesh generator for polygonal elements written in `MATLAB`. The number of elements per grain is given as input, defined as the ratio between the grain area and the requested global mesh size.

Once all the grains have been *independently* meshed, in general, at the grain boundaries, there will be sets of collinear nodes belonging to different grains, which would induce a non-conformal mesh of the microstructure. However, since the VEM can deal with general polygonal elements, also presenting consecutive aligned edges, the creation of conformal meshes is conceptually straightforward, and it can be attained by just adding nodes on edges shared

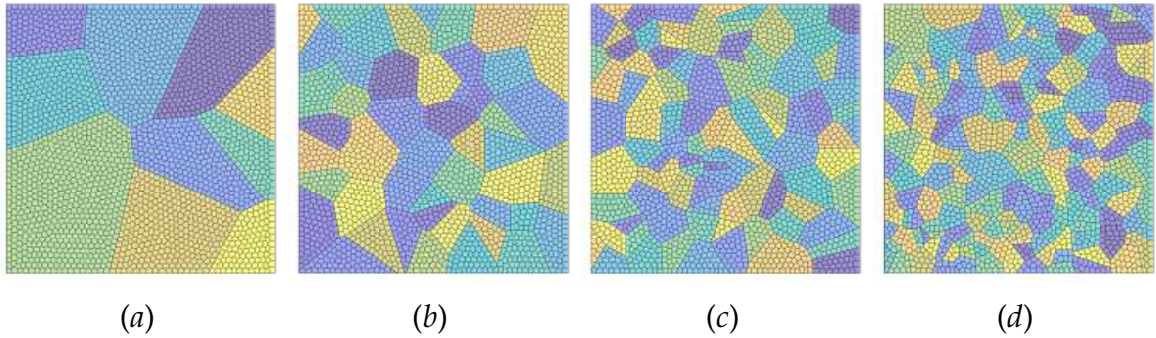

(*a*)  (*b*)  (*c*)  (*d*)

Figure 3.4: Polygonal meshes of different polycrystalline aggregates with increasing numbers of grains: *a*) $N_g = 10$; *b*) $N_g = 50$; *c*) $N_g = 100$; *d*) $N_g = 200$.

between different grains. Fig.(3.5) shows the creation of conforming meshes between adjacent grains: the presence of nodes initially hanging between contiguous grains is dealt with by transforming such nodes into vertices shared between the contiguous elements belonging to two adjacent grains. For the generic boundary polygonal element, such vertices are located between consecutive aligned edges, which are naturally dealt with by the VEM. In other words, the nodes that would be hanging in standard FEM implementations, are here treated as regular nodes, leveraging on the ability of the VEM of dealing with polygonal elements with an arbitrary number of edges and also with collinear consecutive edges.

Polycrystalline microstructures generated using the described strategies have been used to perform the computational material homogenization reported in Section (3.3).

### 3.2.2 Unidirectional fibre-reinforced composite materials

The modelling methodology adopted for the homogenization of composite fibre-reinforced materials is described in this Section. In general, reinforcing fibres may be randomly distributed within the matrix, which can induce irregular and complex meshes. The VEM's versatility in dealing with general polygonal elements, including non-convex or distorted elements, allows noticeable simplification of the pre-processing effort.

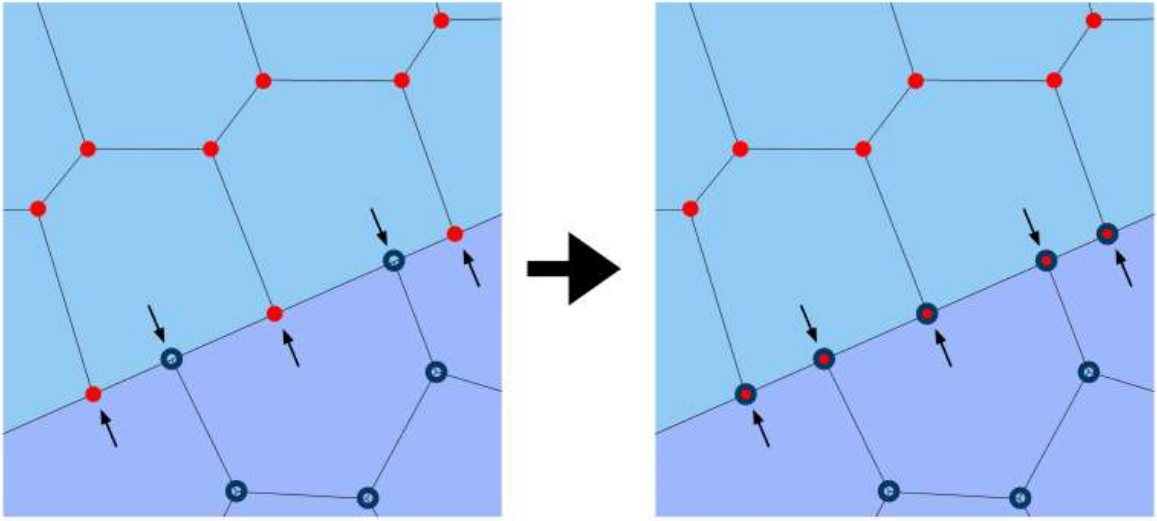

Figure 3.5: Creation of conforming meshes between adjacent grains: the nodes initially hanging between contiguous grains are transformed into vertices shared between the contiguous elements belonging to two adjacent grains.

**Generation of artificial composite micro-morphologies**

Several algorithms have been proposed in the literature to generate microstructures of UD fibre-reinforced composite materials, see, e.g. [128, 59, 144] and references therein. In the present study, artificial periodic microstructures of fibre-reinforced composites are generated as square unit cells with random circular disk-shaped inclusions representing the fibres' transversal sections. An individual morphology is generated starting from two input parameters: the target volume fraction $V_f$ and the size parameter $\delta$, see Eq.(3.1). The number of fibre inclusions $N_f$ in the unit cell is given by

$$N_f = \frac{V_f \delta^2}{\pi}. \tag{3.4}$$

A non-overlapping condition is enforced by setting a minimum allowed distance $d$ between the centres of the circular disk-shaped inclusions, with $d > 2r$, where $r$ is given by Eq.(3.1). In order to generate a valid periodic microstructure with random fibre distribution, the following iterative procedure is adopted:

1. A random uniform set of $N_f$ seed points is initially scattered within the squared bounding box representing the boundary of the unit cell;

2. To attain microstructural periodicity, the set of seed points is replicated within eight copies of the original box created around the original unit cell; each one of the surrounding boxes has the same size as that of the original one so that a total of $9N_f$ points are created overall (the process is similar to the one adopted, e.g. in Ref.[41]);

3. A Delaunay triangulation of the *extended domain* is generated starting from the $9N_f$ points;

4. For each edge of the triangulation, the distance between the end vertices is computed; if such length is $\leq 2r$, the vertices are moved apart along the direction identified by the edge itself of distance proportional to the original edge length;

5. The new coordinates of the original $N_f$ points are extracted. If in step 3, any point has been translated, the set of $N_f$ points with the new coordinates is sent to step 2 for a new iteration; otherwise, the process is terminated.

Once a set of points respecting the non-overlapping condition is obtained, a disk-shaped inclusion can be associated to each seed; the desired periodic microstructure is then extracted by trimming the original bounding box, with circular inclusions, out of the extended domain. Some realizations obtained for different values of $\delta$ and $V_f = 0.29$ are shown in Fig.(3.6).

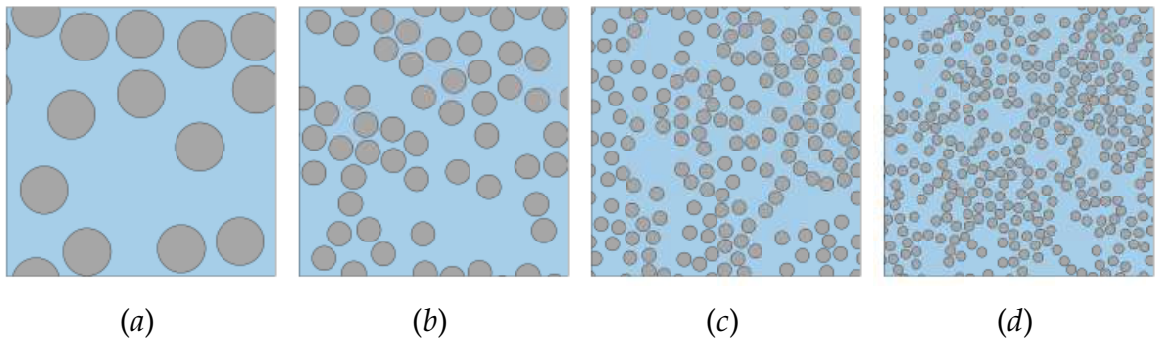

(a)     (b)     (c)     (d)

Figure 3.6: Different realizations of fibre-reinforced composite unit cells for $V_f = 0.29$ and different values of the parameter $\delta = L/r$: a) $\delta = 10$; b) $\delta = 20$; c) $\delta = 35$; d) $\delta = 50$.

**Micro-mechanical composites modelling**

Unidirectional fibre-reinforced composites can be macroscopically considered transversely isotropic materials, whose properties emerge from the features and interplay of their constituents, i.e. from the properties of fibres, matrix, fibre-matrix interface and the ratio $V_f$ between the volume of fibres and the total volume of the composite.

For representing the composite microstructure, a multi-domain meshing strategy is adopted that is slightly different from the one used for the polycrystalline microstructure. Still, the ability of VEM to handle elements of very general shape is exploited. The adopted meshing strategy is based on the three following steps:

1. A conforming triangular mesh of the considered artificial micro-morphology is generated using the software `DistMesh` [145];

2. A polygonal mesh is built from the bounded Voronoi diagram generated using the centroids of the triangular mesh elements as seed points;

3. The polygonal element of the mesh obtained which intersect the fibre inclusions boundaries are trimmed so to conform to such boundaries.

The above process allows the generation of a regular polygonal discretization over the whole computational domain except for the areas close to the fibre boundaries, where the ability of VEM to handle elements of arbitrary shapes, including non-convex shapes, is exploited. Fig.(3.7) shows an example of polygonal mesh generated for a composite unit cell sample, and the detail in the inset on the right shows how irregular polygonal elements may appear in proximity of the inclusions boundaries; the capability of the VEM to address irregular, distorted or non-convex elements allows to retain meshes that would require regularization or further treatment otherwise.

Indeed, the use of VEM may also simplify the implementation of straightforward regularization schemes. An example is provided in Fig.(3.8): the meshing of a fibre-reinforced composite through the operations summarised above may induce the presence of polygonal elements of size comparatively too small with respect to the average mesh size, represented as the blue elements in Fig.(3.8*a*); in this case, it may be useful to *absorb* such small entities within contiguous elements, the red ones in Fig.(3.8*a*-3.8*b*). While such "absorption" operation would require nodes/edges shifting in standard FEM, it can be performed using VEM by just retaining the external polygonal edges of the absorbing/absorbed element couples, as shown in Fig.(3.8*b*). This simple strategy, discussed here as an example of the VEM's flexibility in dealing with meshing, noticeably reduces the number of small elements in general composite unit cells, also reducing the possibility of artefacts in the local fields. However, the subsequent computational analysis cost is generally little affected unless very large numbers of such small elements are present.

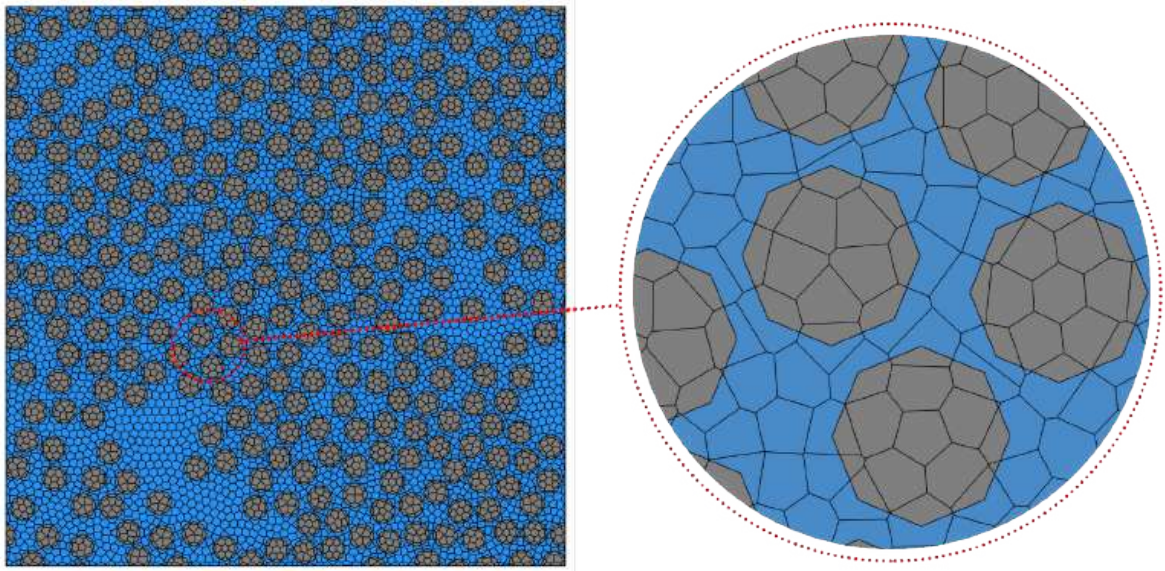

Figure 3.7: Generation of a polygonal mesh for a composite unit cell morphology with $V_f = 0.44$ and $\delta = 40$ (*left*).

## 3.3 Computational homogenization tests

This section describes the numerical tests performed to validate the developed homogenization procedure and the Virtual Element Method's reliability with respect to such application. The numerical tests' purpose is to estimate the effective transverse elastic properties of polycrystalline and unidirectional fibre-reinforced composite materials. The obtained numerical results are compared with available analytical bounds.

For each microstructural sample, assuming plane strain conditions, the apparent transverse elastic properties are calculated from the solution of three different boundary value problems, differing only in the prescribed set of boundary conditions. Kinematic uniform boundary conditions, i.e. linear displacements boundary conditions, are enforced at all external nodes of the con-sidered microstructure. Such enforced boundary displacements correspond to a macro-strain $\bar{\Gamma}$. More specifically, if a reference system $x - y$ with the axes aligned with the external edges of the unit cell is adopted, the three different sets of displacement boundary conditions correspond to: *a*) a uniaxial direct macro-strain along the $x$ direction; *b*) a uniaxial direct macro-strain along the $y$ direction; *c*) a pure shear macro-strain acting to modify the angle between the axes $xy$. The enforced displacement micro-BCs are related to the macroscopic strain by the relation

$$\bar{u}_i = \bar{\Gamma}_{ij} x_j \quad \forall \mathbf{x} \in \partial\Omega. \tag{3.5}$$

The relation between macro-stress and macro-strain is given by:

$$\Sigma_{ij} = \hat{\mathbb{C}}_{ijkl} \bar{\Gamma}_{kl}, \tag{3.6}$$

where $\hat{\mathbb{C}}$ is the apparent macroscopic fourth-order elastic tensor, while $\Sigma_{ij}$ are the components of the macroscopic stress tensor, which can be computed upon the solution of the micro boundary value problem by the volume average of the local micro stress tensor over the domain of the RVE, i.e.

$$\Sigma_{ij} = \frac{1}{|\Omega|} \int_\Omega \sigma_{ij}(\mathbf{x}) \mathrm{d}\Omega. \tag{3.7}$$

In Voigt notation, the apparent macroscopic elastic tensor $\hat{\mathbb{C}}$ is expressed through the apparent stiffness matrix $\hat{C}$ whose components can be determined column-wise from the solution of the three linearly independent boundary value problems mentioned above. Once an estimate of the apparent stiffness matrix is available, the apparent elastic modula can be readily estimated.

### 3.3.1 Computational homogenization of FCC polycrystals

The determination of the macroscopic properties of materials presenting microscopic cubic symmetry has been previously addressed in the literature, see, e.g. Ref.[46]. Numerical simulations are performed in order to estimate the effective transverse elastic properties, namely the macroscopic isotropic Young's modulus $\hat{E}$ and shear modulus $\hat{G}$, for three different polycrystalline materials presenting cubic symmetry at crystal level: copper, gold and nickel. In the case of materials with cubic symmetry, such as FCC (Face Centred Cubic) metals, the specification about the grains' orientation, as mentioned in Section 3.2.1, is unnecessary, and the three principal axes are equivalent.

Grains with cubic symmetry present only three distinct elastic constants $C_{11}$, $C_{12}$ and $C_{44}$ and the reduced stiffness matrix for plain strain $C$ reads

$$C = \begin{bmatrix} C_{11} & C_{12} & 0 \\ C_{12} & C_{11} & 0 \\ 0 & 0 & C_{44} \end{bmatrix}. \tag{3.8}$$

The elastic constants for the three selected materials are summarized in Table (3.1), as taken from Ref.[46].

For each material, aggregates with $N_g = 10, 20, 50, 100, 200$ grains have been tested. For given material and number of grains, 50 different realisations have been generated and analysed. Each realisation differs from the others in terms of both geometry and grains orientation. Table (3.2) reports the minimum, the average and the maximum number of degrees of freedoms, related to the number of grains in the analysed polycrystalline microstructures.

The homogenization is performed following the procedure employed in Ref.[46]. For a macroscopically isotropic aggregate, the range of the Young and shear effective moduli is bounded by a lower (Reuss) bound and an upper (Voigt) bound. Such limits are also referred to in the literature as first-order bounds. The Reuss [152] lower bound is obtained by assuming that all the grains undergo uniform stress, while the Voigt [175] upper bound is obtained assuming that all the grains undergo uniform strain. Since a two-dimensional model is being considered, the bounds are computed by averaging the single-crystal plain strain reduced stiffness matrix over all possible orientations of the random angle $\theta$ formed between the material direction 1 and the lower horizontal edge of the square unit cell, as shown, e.g. in Ref.[135].

The obtained numerical results, in terms of the effective Young's modulus $\hat{E}$ and shear modulus $\hat{G}$, are shown in Fig.(3.9). The Reuss and Voigt bounds are also shown for comparison purpose. The effective properties are estimated as the ensemble average over realizations containing the same number of grains. It is noticed how, as the number of grains per realization increases, the scatter of the apparent properties reduces. When realizations with $N_g = 200$ are considered, the apparent moduli always fall within the first-order bounds.

|        | $C_{11}$ | $C_{12}$ | $C_{44}$ |
|--------|------|------|------|
| Copper | 168  | 121  | 75   |
| Gold   | 185  | 158  | 40   |
| Nickel | 251  | 150  | 124  |

Table 3.1: Single crystal elastic constants used for the analysed materials from Ref.[46]; the values are given in [GPa].

|       |         | 10    | 20    | 50    | 100   | 200  |
|-------|---------|-------|-------|-------|-------|------|
|       | $N_g$   |       |       |       |       |      |
|       | Min     | 9982  | 9976  | 9966  | 9938  | 9886 |
| $n_{dof}$ | Average | 9995  | 9994  | 9988  | 9975  | 9942 |
|       | Max     | 10006 | 10012 | 10020 | 10022 | 9986 |

Table 3.2: Minimum, average and maximum number of DOFs for the analyzed polycrystalline aggregates.

### 3.3.2 Computational homogenization of fibre-reinforced composites

Two of the unidirectional fibre-reinforced composite materials considered in Ref.[159] are selected for the numerical tests on composite unit cells. The first composite, here labelled COMP1, is made of AS4 carbon fibres embedded in $3501 - 6$ epoxy matrix. The second composite, here labelled COMP2, is made of *Silenka* E-glass 1200 tex fibres embedded in MY750/HY917/DY063 epoxy matrix. The fibre volume fractions considered in the performed tests are $V_f = 0.22$, $V_f = 0.29$, $V_f = 0.36$ and $V_f = 0.44$.

The axis (1) is parallel to the fibres, and it is normal to the (2-3) plane, in which the 2D unit cell lies. The mechanical properties of the constituents, themselves isotropic in the (2-3) plane, are given in Table (3.3) in terms of transverse Young's modulus $E_{22}$ and transverse shear modulus $G_{23}$.

Table 3.3: Mechanical properties for the matrix and fibres of COMP1 and COMP2, as taken from Ref.[159].

| Mechanical Properties           | $E_{22}$ [GPa] | $G_{23}$ [GPa] |
|---------------------------------|----------------|----------------|
| AS4 carbon fibres               | 15             | 7              |
| 3501-6 epoxy matrix             | 4.2            | 1.567          |
| Silenka E-Glass 1200 tex fibres | 74             | 30.8           |
| MY750/HY917/DY063 epoxy matrix  | 3.35           | 1.24           |

A unidirectional fibre-reinforced composite lamina is macroscopically transversely isotropic so that only two elastic modula are needed to completely characterize the transverse behaviour in the plane of isotropy (2-3). In this study, the numerical tests results are given in terms of the plain strain bulk modulus $K_{23}$ and the shear modulus $G_{23}$.

The minimal RVE size for unidirectional fibre-reinforced composites similar to those considered here has been investigated in Ref.[172], where it was found that, when the purpose of the analysis is the estimation of the effective properties, a minimum size parameter of $\delta \geq 30$ is required. The effective properties' convergence is assessed in the range $10 \leq \delta \leq 50$.

Figs.(3.10-3.11) show, for both composite materials and for each considered value of the volume fraction $V_f$, the average and the scatter range of the computed elastic properties as a function of the unit cell size, as expressed by $\delta$. The average is computed over ensembles of 50 realizations for each value of $\delta$. It can be observed that, for both considered materials and for $\delta \geq 30$, there is no appreciable variation in the values of either the average elastic modula or the scatter, which confirms convergence of the effective properties (please note the tight scale used in the graphs).

Figs.(3.12-3.13) show the numerical predictions about the computed transverse mechanical properties $K_{23}$ and $G_{23}$ versus the fibre-volume fraction $V_f$ at $\delta = 50$. The Voigt and Reuss bounds and the Hashin-Hill bounds [89, 86] for the effective elastic modula are also shown for comparison purpose. The obtained numerical estimates are in agreement with the theoretical predictions.

Table (3.4) shows the minimum, average and maximum number of degrees of freedoms, in the analysed composite microstructures, for $V_f = 0.22$ and at different values of the size parameter $\delta$.

## 3.4 Discussion

In this Chapter, a lowest-order Virtual Element framework for computational materials homogenisation has been developed, and it has been applied to both polycrystalline materials and unidirectional fibre-reinforced composites.

Table 3.4: Minimum, average and maximum number of DOFs for the analysed composite microstructures for $V_f = 0.22$.

|  | $N_g$ | 10 | 15 | 20 | 25 | 30 | 35 | 40 | 45 | 50 |
|---|---|---|---|---|---|---|---|---|---|---|
|  | Min | 1836 | 3700 | 6604 | 10328 | 14670 | 19750 | 25828 | 32438 | 39842 |
| $n_{dof}$ | Average | 1932 | 3772 | 6711 | 10349 | 14838 | 19906 | 26018 | 32651 | 40117 |
|  | Max | 1984 | 3856 | 6810 | 10470 | 15022 | 20098 | 26174 | 32868 | 40354 |

General polycrystalline Voronoi microstructures have been analysed, addressing the occurrence of hanging nodes at the interface between independently meshed contiguous grains through the ability of VEM of dealing with elements with aligned edges. The ability of VEM of addressing non-convex polygonal element, on the other hand, has been employed in the analysis of general composite fibre-reinforced morphologies obtained from the random scattering of fibres with circular sections.

This study has shown how the VEM's capability to deal with very general polygonal mesh elements, including non-convex and highly distorted elements, can be profitably exploited to relax the requirements on the mesh quality that may hinder the automatic analysis of micro-morphologies presenting complex or highly statistically varying features, commonly met in computational materials micro-mechanics and homogenisation, where materials microstructures are often generated resorting to stochastic algorithms.

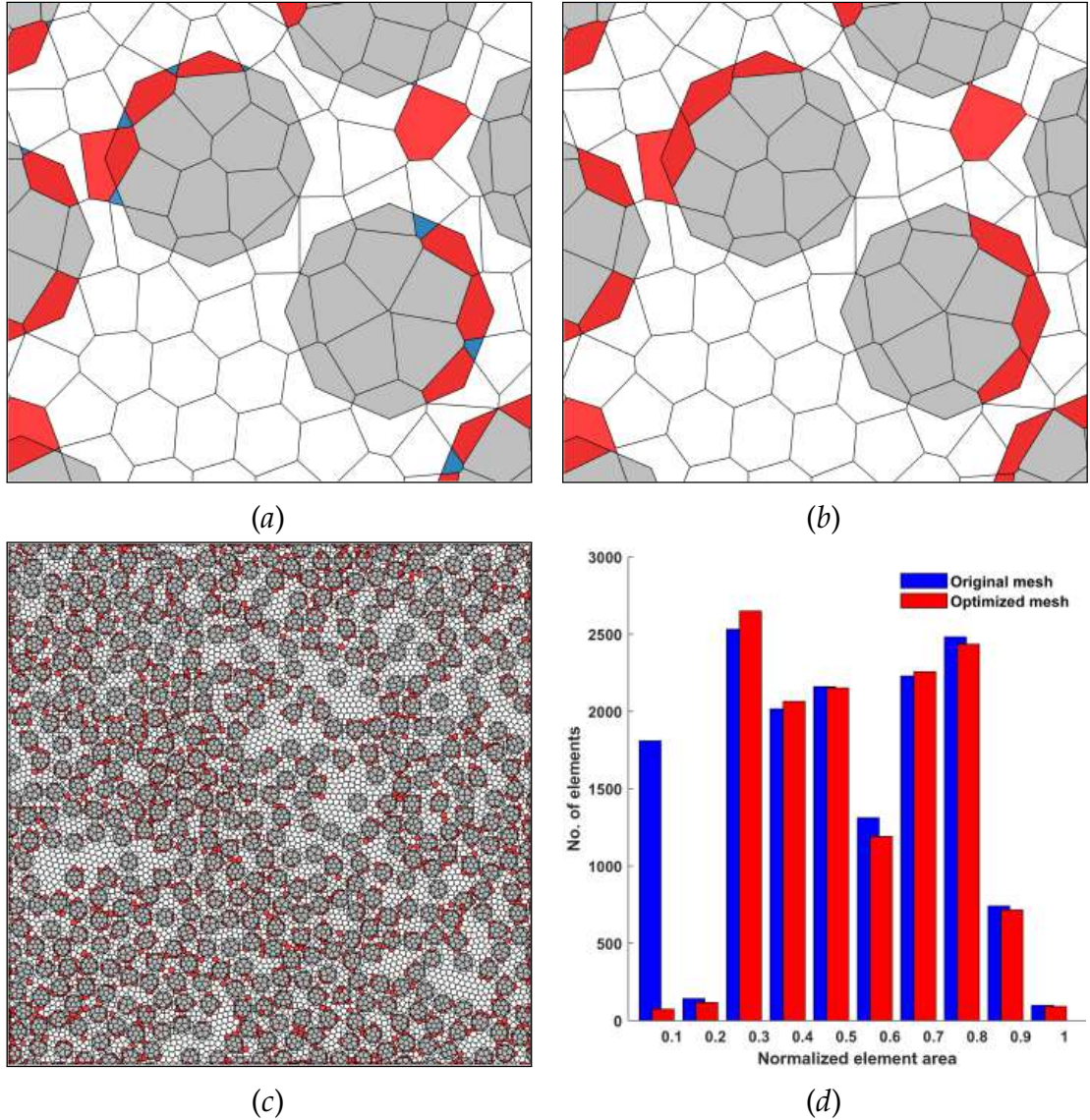

(*a*)   (*b*)

(*c*)   (*d*)

Figure 3.8: A simple VEM-based regularization scheme: *a*) elements considerably smaller than the average mesh size may be present in the mesh of the composite fibre-reinforced unit cell (blue in the online version of the paper); *b*) the small elements can be *absorbed* within contiguous elements of larger size (red in the online version of the manuscript); with VEM such operation is performed by simply creating larger polygonal elements bounded by the external edges of the absorbing/absorbed element couple. In unit cells with large numbers of fibres, e.g. the one shown in (*c*) with 477 fibres, $\delta = 50$ and $V_f = 0.44$, the regularization scheme noticeably reduces the presence of small elements, as shown by the histogram in (*d*).

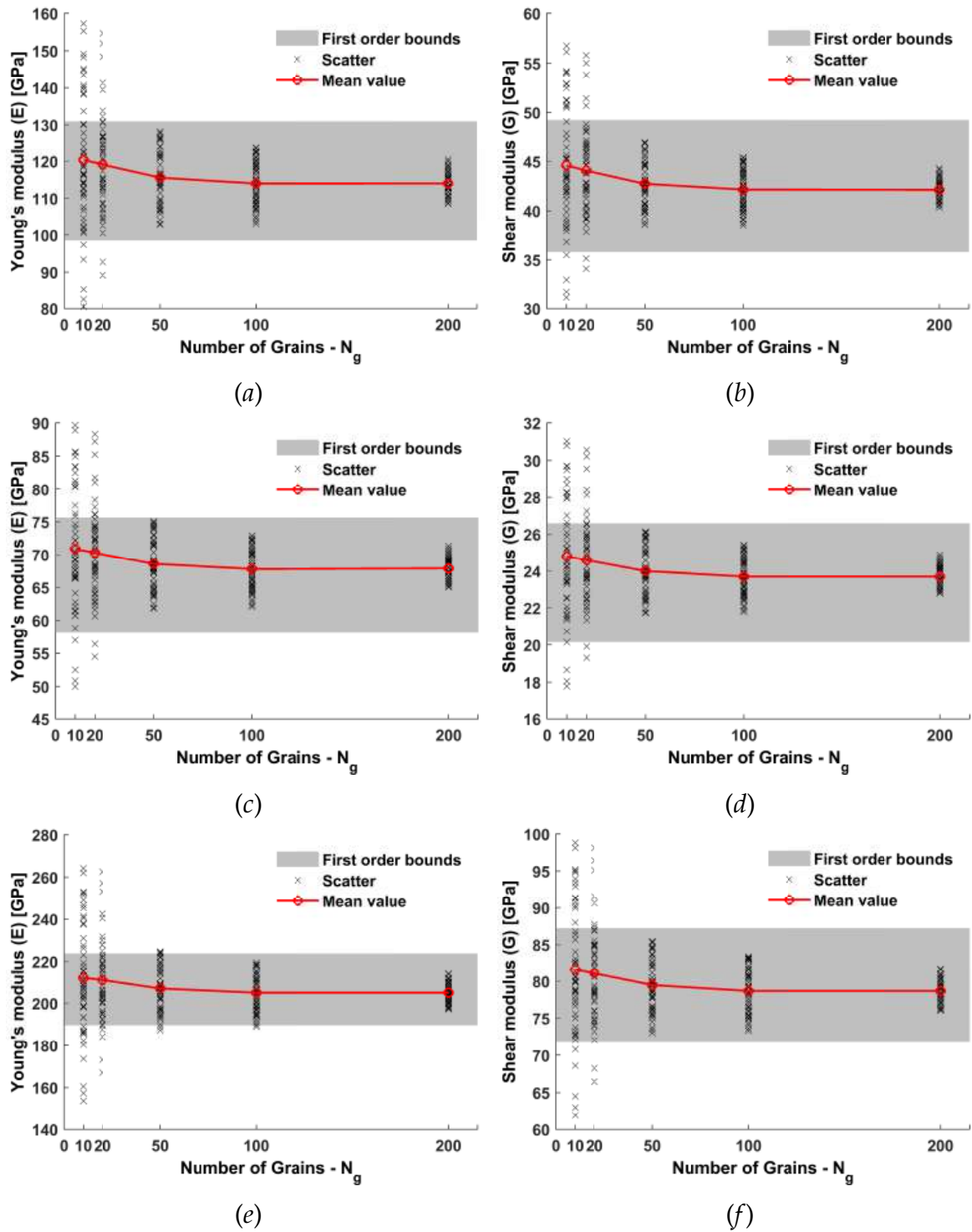

Figure 3.9: Computed effective Young's modulus $E$ and shear modulus $G$ for polycrystalline aggregates of copper (*a-b*), gold (*c-d*), nickel (*e-f*).

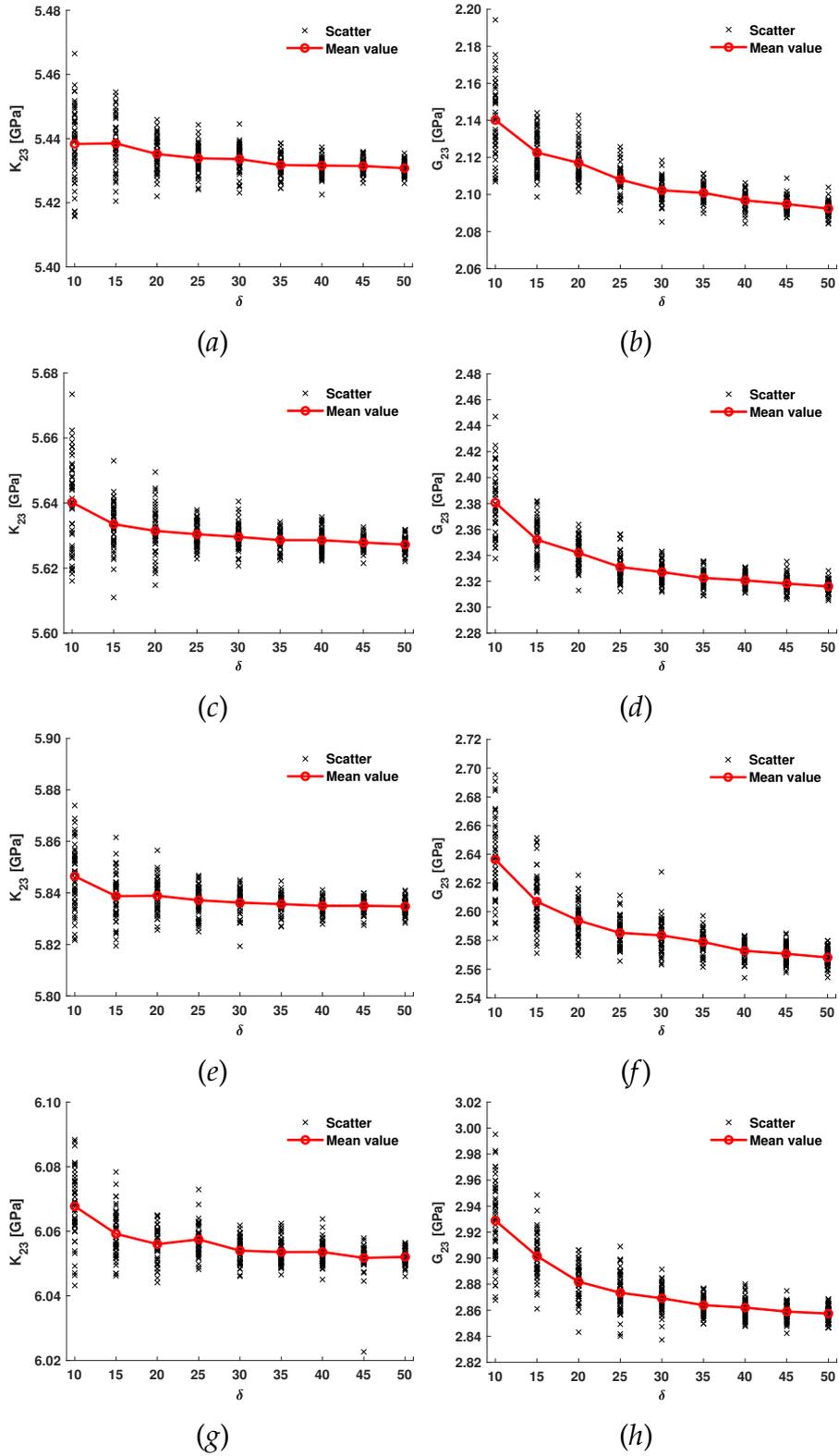

Figure 3.10: Computed effective transverse elastic properties as a function of $\delta$ for different values of $V_f$ for COMP1: $V_f = 0.22$ (*a-b*); $V_f = 0.29$ (*c-d*); $V_f = 0.36$ (*e-f*); $V_f = 0.44$ (*g-h*).

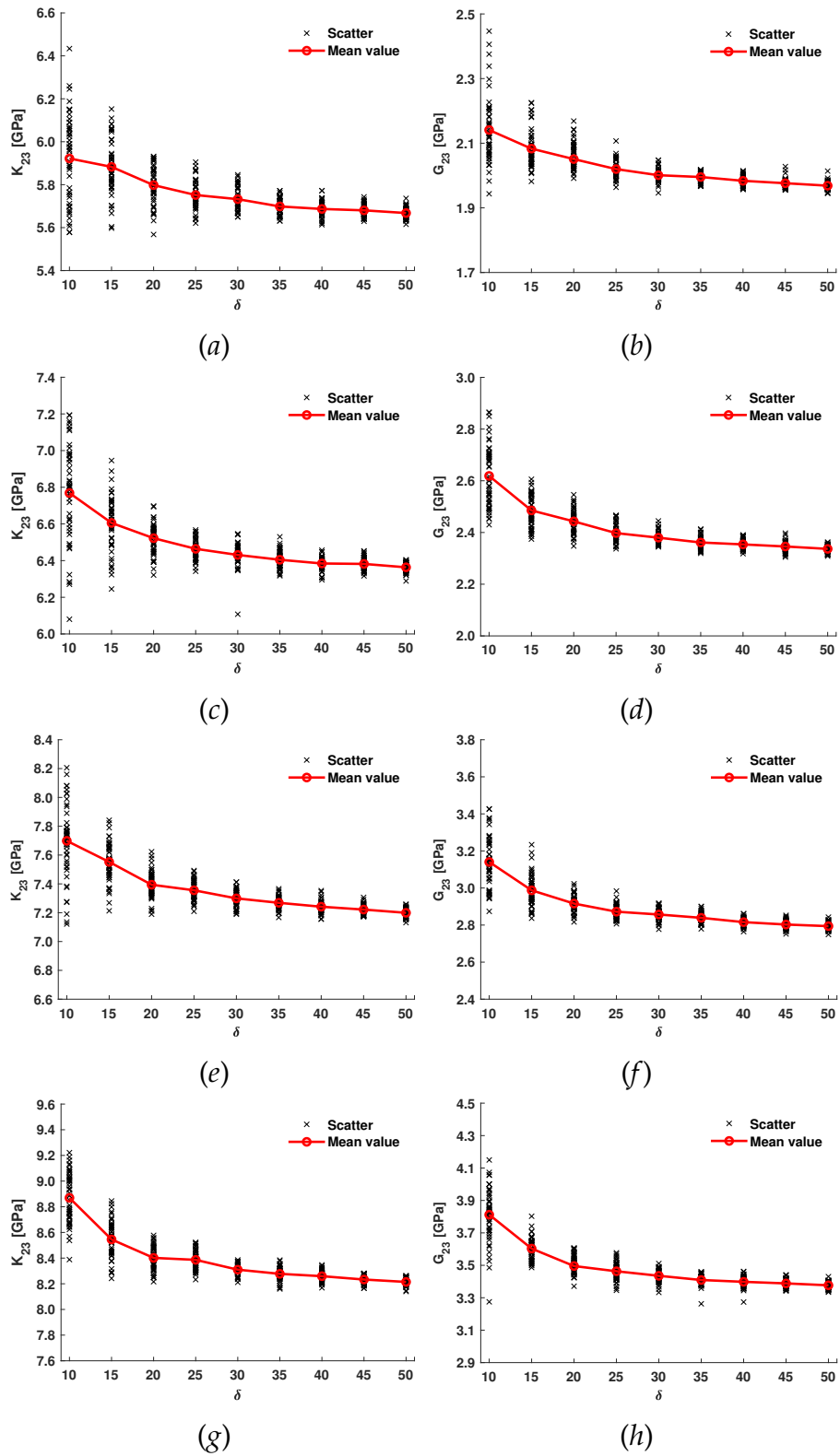

Figure 3.11: Computed effective transverse elastic properties as a function of $\delta$ for different values of $V_f$ for COMP2: $V_f = 0.22$ (*a-b*); $V_f = 0.29$ (*c-d*); $V_f = 0.36$ (*e-f*); $V_f = 0.44$ (*g-h*).

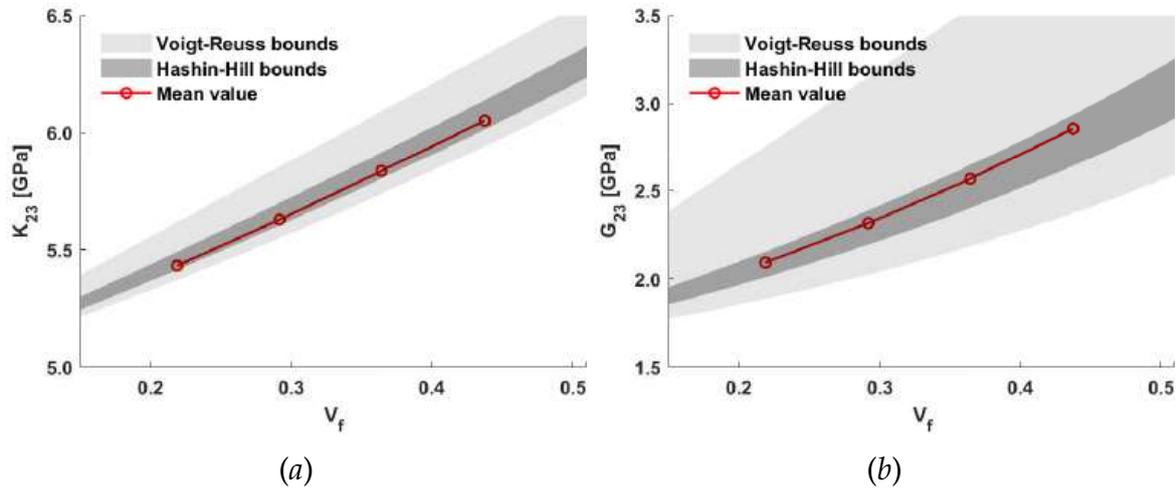

Figure 3.12: Computed transverse elastic properties and Hashin-Hill bounds as a function of the volume fraction $V_f$ for COMP1 and $\delta = 50$.

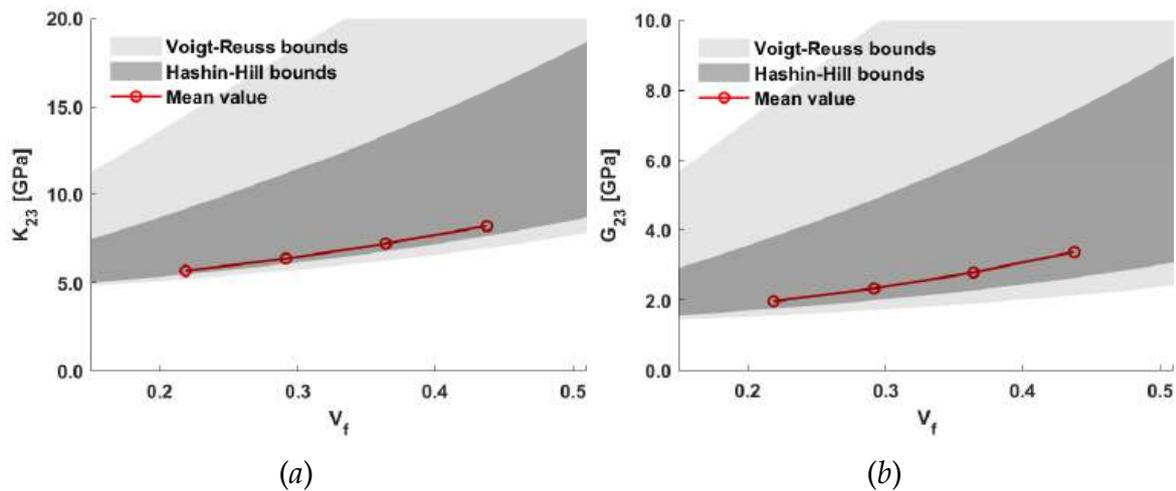

Figure 3.13: Computed transverse elastic properties and Hashin-Hill bounds as a function of the volume fraction $V_f$ for COMP2 and $\delta = 50$.

# Chapter 4

# A hybrid virtual-boundary element formulation

In this Chapter, a hybrid formulation based on the conjoined use of the recently developed Virtual Element Method (VEM) and the Boundary Element Method (BEM) is proposed for the effective computational analysis of multi-region domains. VEM allows the straightforward employment of elements of general polygonal shape, maintaining a high level of accuracy. For its inherent features, it allows the use of meshes of general topology, including non-convex elements. BEM is an effective technique for the numerical solution of sets of boundary integral equations, employed as the original model of the represented physical problem. For several classes of problems, BEM offers some advantages over more popular techniques, namely the reduction of the dimensionality of the problem, with associated computational savings. The inherent advantages of the VEM and the BEM are simultaneously employed to facilitate the study of heterogeneous material microstructures.

## 4.1 Introduction

VEM has been presented in Chapter 2 as a rapidly emerging generalisation of the Finite Element Method (FEM). In Chapter 3 it has been shown how VEM can provide accurate results when applied to mesh whose elements can have very general shape, such in the case of polygonal elements with an arbitrary number of edges non-convex and highly distorted elements.

The Boundary Element Method has been developed over the last four decades as a powerful numerical tool for the analysis and solution of many physical and engineering problems [21, 55, 180] and today, it represents a viable alternative to other numerical approaches in many fields of engineering analysis [7].

BEM is based on the use and the numerical solution of integral equations, i.e. equations where the unknown functions appear under the integral sign. The use of integral equations for the analysis of potential field problem appeared in the pioneering classical work of Fredholm [73]. The Boundary Integral Equation (BIE) method for the analysis of elastic problems was introduced by Betti [53] and Somigliana [161]. Starting from these pioneering works, many authors have contributed to developing the method [125, 153, 65]. Among such contributions, the work of Lachat and Watson [108] was particularly influential for the development of a general numerical treatment of boundary-value equation problems. They introduced an isoparametric formulation, analogous to the one used for FEM analysis, and described the numerical integration procedure in detail, thus demonstrating the possibility of addressing complex three-dimensional elastic problems.

The first step for the boundary element analysis of any problem is describing the problem itself through a boundary integral equation. There are many ways to obtain such representation, but a particularly attractive approach for solids mechanics problems uses classical reciprocity theorems, such as Betti's theorem for elasticity, and special solutions of an auxiliary problem called *fundamental solutions*. The use of fundamental solutions, which represent the solution to the problem governing equations for some special cases, is the main difference between BEM and other numerical methods. Such solutions are used as weighting functions and allow to reduce the discretisation requirements, incorporating in the formulation some knowledge about the governing equation solution itself. However, they must be known in advance for the method to be effectively viable. This point often has constituted a hindrance for the application of BEM to some fields of application. However, today the fundamental solutions are known for many engineering problems.

The main advantage of boundary element techniques is reducing the degrees of freedom needed to model a given physical system. Such reduction is allowed by the underlying boundary integral formulation, which requires only the discretisation of the boundary of the analysed domain for its numerical solution. Consequently, the analysis of a two-dimensional domain requires the discretion of its one-dimensional boundary, while for three-dimensional bodies, only their boundary surfaces have to be discretised. More-over, due to the boundary only nature of the computational grid, stress concentrations areas can be modelled more effectively by increasing the mesh density locally around the area of interest without affecting the mesh's quality elsewhere. As a result, BEM can capture high-stress gradients with very good accuracy and a limited pre-processing effort. Another aspect of noticeable interest is the capability to represent interior point displacements and stresses continuously, without making the analysis heavier. The interior points quantities are computed as a post-processing task without increasing the solving system's size.

Besides the computational cost savings, the reduction of the model dimensionality generally induces a pre-processing simplification, thanks to the need of discretising curves instead of surfaces, in the two-dimensional case, or surfaces instead of volumes, in the three-dimensional case. Such a feature may result particularly appealing when materials morphologies with high statistical variability have to be automatically generated, meshed and analysed [46]. BEM produces a linear system of equations whose coefficient matrix is neither symmetric nor definite. Moreover, it results fully populated. Such features make the system solution generally more demanding in memory storage and computational time with respect to FEM systems with the same number of degrees of freedom. This aspect becomes particularly relevant for *large scale* systems, as discussed in-depth in the next Chapter. Despite the last consideration, the above mentioned general advantages make BEM particularly attractive in some fields of investigations, such as Computational Mechanics, where the internal domain meshing can present some difficulty. BEM has been successfully employed to the solution of several classes of problems in fluids [180] and solids [7] mechanics and, more recently, in multi-scale materials modelling [156, 83, 42, 48, 79, 47, 50].

In the present Chapter, a hybrid computational technique, based on the simultaneous use of the Virtual Element Method and the Boundary Element Method is proposed with the idea that the conjoined use of VEM and BEM might provide some benefits in the modelling of heterogeneous materials with complex microstructures [3, 61]. Unit cells with stiff inclusions embedded within a more compliant matrix, representative e.g. of the transverse section fibre-reinforced composites are considered. BEM is employed to model the material inclusions, while VEM is used to represent the matrix. This choice could be further motivated by the assumption that, under progressive loading, the stiffer inclusions would remain in the linear behaviour range, while the matrix might undergo complex non-linear phenomena, e.g. hardening, damaging or fracturing processes, which could be modelled with the framework of VEM, taking advantage of the generality inherited by FEM and its peculiar ability to deal with elements of very general shapes.

This Chapter is organised as follows. Section 4.2 addresses the generation and meshing strategy for illustrative artificial morphologies. The BEM formulation for two-dimensional linear elastic problems is recalled in Section 4.3.4. In Section 4.4, the formulation of the hybrid virtual-boundary element method is described introducing the VEM-BEM coupling procedure. Section 4.5 discusses the application of the hybrid procedure to a case study represented by a matrix with complex-shaped inclusions, assessing the accuracy in terms of displacements and stresses. Section 4.6 the advantages of the hybrid VEM-BEM approach are exploited by applying the proposed method to the computational homogenization of unidirectional fibre-reinforced composites. Eventually, Section 4.7 recalls the key features of the proposed numerical technique.

## 4.2 Reference morphology

In this Section, the procedures adopted for generating and meshing the artificial representation of the considered material microstructures are described. As it will be shown in Section 4.4, the present formulation is based on a multi–region approach, in which different phases are modelled using either a *virtual* or a *boundary* element approach, depending on several considerations, including the phase physics, as discussed in Section 4.7. An example is provided by the unit cell representative of a fibre–reinforced polymer composite, for which a certain number of inclusions, modelled, e.g. with the boundary element method, may represent the transverse section of the fibres, while the surrounding domain, modelled with the virtual element method, may represent the polymer matrix.

In general, the considered two-dimensional unit cell may contain $N_V$ regions modelled with the virtual element method and $N_B$ domains modelled with the boundary element method, so that the overall domain $\Omega$ is given by

$$\Omega = \left( \bigcup_{k=1}^{N_V} \Omega_k^V \right) \cup \left( \bigcup_{k=1}^{N_B} \Omega_k^B \right) = \Omega^V \cup \Omega^B, \tag{4.1}$$

where the superscripts *V* and *B* refer to virtual and boundary element regions respectively. The overall domain is bounded by the contour $\Gamma = \partial \Omega$, while the *k–th* subdomain $\Omega_k^B$ is bounded by the contour $\Gamma_k^B = \partial\Omega_k^B$ and the *k–th* subdomain $\Omega_k^V$ is bounded by the contour $\Gamma_k^V = \partial\Omega_k^V$.

The virtual element regions can be meshed with generic polygonal elements, which ensures certain meshing flexibility, as discussed, e.g. in Ref.[116]. On the other hand, the boundary element regions only require consistent meshes of their contours $S_k$ and do not need any internal mesh, at least when they do not experience any non-linear material process (e.g. plasticity and/or damage). The meshing procedure must then interface the polygonal virtual element mesh with the one-dimensional boundary element mesh. Due to the inherent features of the virtual element method, which allows the natural treatment of generic polygonal elements and hanging nodes, the meshing can be implemented without resorting to complex pre–processing algorithms.

The simple example geometry shown in Fig.(4.1), consisting of a square unit cell with an inclusion of arbitrary shape, is considered to describe the implemented procedures. In this case, $\Omega = \Omega^V \cup \Omega^B$ and $\Gamma^B = \partial\Omega^B$ is the interface between the two regions. Once the morphology of the unit cell is geometrically reconstructed, the meshing procedure is based on the generation of a Voronoi tessellation [19] of the overall domain and the subsequent clipping of the internal inclusions to be modelled with BEM.

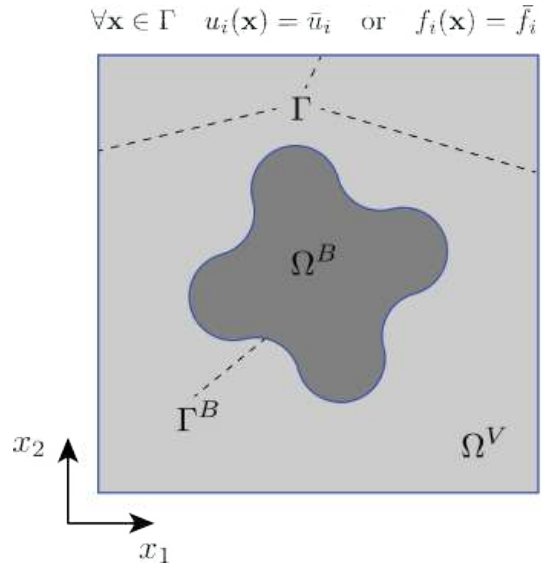

Figure 4.1: A microstructure consisting of an arbitrary shaped inclusion within a surrounding matrix as an example two-dimensional heterogeneous unit cell. Boundary conditions are enforced on the boundary $\Gamma = \partial\Omega$ of the microstructure; $\Gamma^B$ is the interface between $\Omega^V$ and $\Omega^B$.

The workflow of the overall procedure can be summarised as follows:

*a*) The micro–morphology is created as a two-dimensional geometric entity;

*b*) A *conformal* triangular mesh of the overall domain is generated;

*c*) A non-conformal polygonal mesh is generated as the Voronoi dual of the triangular discretisation;

*d*) The Voronoi cells falling within the inclusion domain are removed;

*e*) The Voronoi cells intersecting the contour $\Gamma^B$ are clipped using the nodes and edges of the conformal triangular mesh, thus providing the sought *conformal* polygonal mesh.

The adjective *conformal* used above refers to the circumstance that the vertices of the initial triangular mesh and those of the target polygonal one lie on the interface $S$ between the two regions, thus identifying the interface mesh nodes, where suitable continuity conditions will be enforced to retrieve the integrity of the domain.

The above procedure, schematically represented in Fig.(4.2), has been implemented in `MATLAB`. The geometry is represented as a collection of points and curves identifying each subdomain, which forms the input for generating the first conformal triangular mesh of the overall domain. This task has been performed using an unstructured mesh-generator for two-dimensional geometries [70]. The target polygonal mesh is generated from the triangular mesh output using an in–house developed code that performs the following sequence of operations: *a*) retrieves the triangular mesh data structure; *b*) constructs a two-dimensional Voronoi diagram from the given triangulation; *c*) clips the polygonal mesh elements intersecting the interface $\Gamma^B$, providing the target conformal polygonal mesh of the domain.

## 4.3 Boundary integral formulation

In this Section, the boundary element formulation for two-dimensional linear elasticity problems is reviewed, starting from classical reciprocity principles.

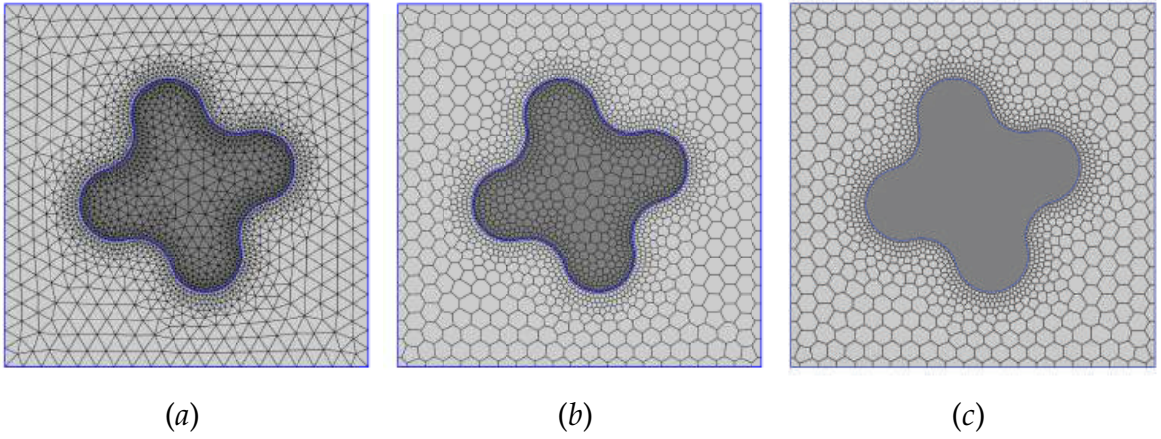

(a) (b) (c)

Figure 4.2: Generation of the polygonal mesh of the artificial multi–region morphologies: *a)* initial conformal triangular mesh of the overall domain; *b)* two-dimensional Voronoi diagram associated with the previous triangulation; *c)* target conformal polygonal mesh obtained by removing the Voronoi cells within the inclusion domain and clipping those intersecting the sub–domains interface.

### 4.3.1 Somigliana identity

The Betti reciprocity work theorem states that if two linear elastic self equilibrated systems $(u_i, t_i, b_i)$ and $(u_i^*, t_i^*, b_i^*)$ exist in a domain $\Omega^B$ having $\Gamma^B$ as its boundary, then the work done by the forces of the first system on the displacements of the second is equal to the work done by the forces of the second system on the displacements of the first

$$\int_{\Omega^B} b_j^* u_j \, d\Omega + \int_{\Gamma^B} t_j^* u_j \, d\Gamma = \int_{\Omega^B} u_j^* b_j \, d\Omega + \int_{\Gamma^B} u_j^* t_j \, d\Gamma. \tag{4.2}$$

Let the second system represent the solution to the case of a unit point load acting on a point $x_0$ of an infinite elastic domain

$$b_j^* = \delta_{ij} \delta(x - x_0), \tag{4.3}$$

where $\delta_{ij}$ is the Kronecker delta

$$\delta_{ij} = \begin{cases} 1 & \text{when } i = j \\ 0 & \text{when } i \neq j \end{cases} \tag{4.4}$$

representing the *j*-th component of the *i*-th unit vector of the two-dimensional standard basis, and $\delta(x - x_0)$ is the two-dimensional Dirac's delta. Under such assumption and also assuming that the body forces $b_i$ can be neglected, Eq.(4.2) can be rewritten in the form

$$u_i(x_0) + \int_{\Gamma^B} H^*_{ij}(x_0,x) u_j(x)\, d\Gamma(x) = \int_{\Gamma^B} G^*_{ij}(x_0,x) t_j(x)\, d\Gamma(x). \qquad (4.5)$$

The notation $G^*_{ij}$ and $H^*_{ij}$ has been introduced to denote the components of displacements and tractions corresponding to a unit point load. More specifically $G^*_{ij}(x_0,x)$ and $H^*_{ij}(x_0,x)$ represent respectively the displacement and traction component along the direction $j$ at the point $x$ when a unit point load is applied at the point $x_0$ along the direction $i$. Eq.(4.5) is the well known Somigliana identity, which expresses the displacement at an interior point $x_0 \in \Omega^B$ in terms of the displacements and tractions at the boundary $\Gamma^B$.

### 4.3.2   Kelvin fundamental solutions

The knowledge of the kernels $G^*_{ij}$ and $H^*_{ij}$ plays a fundamental role in the formulation of any boundary element model. These functions are the *Kelvin fundamental solutions* and express displacements and tractions at the *field point* $x$ due to the application of a unit point load at the *source point* $x_0$.
Under plane strain assumptions, the components of the Kelvin fundamental displacements are given by

$$G^*_{ij}(x_0,x) = C_1 \left( C_2 \delta_{ij} \ln r - r_{,i} r_{,j} \right), \qquad (4.6)$$

while the components of the Kelvin fundamental tractions have the form

$$H^*_{ij}(x_0,x) = \frac{C_3}{r} \left[ n_k r_{,k} \left( C_4 \delta_{ij} + 2 r_{,i} r_{,j} \right) - C_4 \left( r_{,i} n_j - r_{,j} n_i \right) \right]. \qquad (4.7)$$

In the previous expressions $r = ||x - x_0||$ is the Euclidean distance between points $x_0$ and $x$, the index notation $f_{,i} = \partial f / \partial x_i$ is adopted to denote differentiation, $n_i$ are components of the outward unit normal vector to the boundary $\Gamma^B$ at the generic smooth point $x$. The coefficients $C_1$, $C_2$, $C_3$ and $C_4$ are given

by

$$C_1 = -\frac{1+\nu}{4\pi(1-\nu)E}, \qquad (4.8)$$

$$C_2 = 3 - 4\nu, \qquad (4.9)$$

$$C_3 = -\frac{1}{4\pi(1-\nu)}, \qquad (4.10)$$

$$C_4 = 1 - 2\nu, \qquad (4.11)$$

with $E$ and $\nu$ denoting respectively the Young's modulus and the Poisson's ratio of the isotropic material of the domain $\Omega^B$.

### 4.3.3 Displacement Boundary Integral Equation

Eq.(4.5) allows to express the displacements at an internal point $x_0$ once displacements and tractions at the boundary $\Gamma^B$ are known. Its use alone, however, does not allow to solve the elastic problem. The solution of the elastic problem can be found by writing Eq.(4.5) for points belonging to the boundary itself. However, since the kernels of the boundary integral representation present a singularity when $r = \|x - x_0\| \to 0$, the *collocation* on the boundary requires some attention. The boundary collocation can be accomplished through a limiting process involving an augmented domain $\Gamma^{B\prime} = \Gamma^B - \Gamma^B_\varepsilon + \Gamma^{B\prime}_\varepsilon$, (see Fig.4.3). Eq.(4.5) can then be written

$$u_i(x_0) + \lim_{\varepsilon \to 0} \int_{\Gamma^{B\prime}} H^*_{ij}(x_0, x) u_j(x)\, d\Gamma(x) = \lim_{\varepsilon \to 0} \int_{\Gamma^{B\prime}} G^*_{ij}(x_0, x) t_j(x)\, d\Gamma(x). \quad (4.12)$$

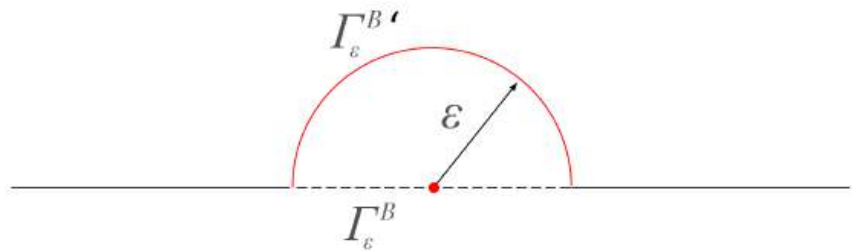

Figure 4.3: Limiting process for collocation on the boundary.

It is easy to realize that when $\varepsilon \to 0$ then $\Gamma^{B\prime} \to \Gamma^B$. In calculating the two limiting expressions, attention should be paid to the order of the kernels' singularity. The limit involving the fundamental displacements can be written

$$\lim_{\varepsilon \to 0} \int_{\Gamma_\varepsilon^{B\prime}} G_{ij}^*(x_0, x) t_j(x) \, d\Gamma(x) + \lim_{\varepsilon \to 0} \int_{\Gamma^B - \Gamma_\varepsilon^B} G_{ij}^*(x_0, x) t_j(x) \, d\Gamma(x). \qquad (4.13)$$

Considering that $G_{ij}^*$ is of order $O(\ln r)$, it can be shown that the first integral goes to zero, while the second one can be evaluated as an improper integral. On the other end, for the traction fundamental solutions, considering that $H_{ij}^*$ contains a strong singularity of order $O(r^{-1})$, the integral over $\Gamma_\varepsilon'$ gives rise to a jump term, which can be computed by expanding the displacements about the source point $x_0$ in a Taylor series. It follows

$$\lim_{\varepsilon \to 0} \int_{\Gamma_\varepsilon^{B\prime}} H_{ij}^*(x_0, x) u_j(x) \, d\Gamma(x) = \alpha_{ij}(x_0) u_j(x_0). \qquad (4.14)$$

The other contribution is treated in a Cauchy principal value sense

$$\lim_{\varepsilon \to 0} \int_{\Gamma^B - \Gamma_\varepsilon^B} H_{ij}^*(x_0, x) u_j(x) \, d\Gamma(x) = \fint_{\Gamma^B} H_{ij}^*(x_0, x) u_j(x) \, d\Gamma, \qquad (4.15)$$

where the symbol $\fint$ stands for Cauchy principal value.
Eq.(4.12) can then be written

$$[\delta_{ij} + \alpha_{ij}(x_0)] u_j(x_0) + \fint_{\Gamma^B} H_{ij}^*(x_0, x) u_j(x) \, d\Gamma(x) = \int_{\Gamma^B} G_{ij}^*(x_0, x) t_j(x) \, d\Gamma(x) \qquad (4.16)$$

It can be shown that for smooth boundary points $x_0$, where the unit outward vector $\mathbf{n}(x_0)$ is uniquely defined, the jump term $\alpha_{ij} = -\frac{1}{2}\delta_{ij}$. It follows then, for smooth boundary points

$$c_{ij}(x_0) u_j(x_0) + \fint_{\Gamma^B} H_{ij}^*(x_0, x) u_j(x) \, d\Gamma = \int_{\Gamma^B} G_{ij}^*(x_0, x) t_j(x) \, d\Gamma \qquad (4.17)$$

with $c_{ij}(x_0) = 1/2\delta_{ij}$. Eq.(4.17) represents the Displacement Boundary Integral Equation (DBIE) and it is the starting point for the construction of a boundary element model for two.dimensional elasticity.

### 4.3.4 The Boundary Element Method

The numerical solution of Eq.(4.17) is based on the discretisation of $\Gamma^B$ and the subsequent approximation of the boundary displacement and traction components in terms of shape functions and nodal values. More specifically, $\Gamma^B$ is subdivided into $m$ straight segments $s_k$, and two nodes are associated with each segment's ends. In plane problems, each node carries two components of displacements and two components of tractions. Assuming $\Gamma^B$ as smooth, it follows that a tangent can be associated to any $x \in \Gamma^B$, so that the existence of a unique value of traction at the node is ensured; corner points are not considered in the present formulation, although these could be treated resorting to known boundary element techniques [7].

Displacement and traction components are here assumed to be globally continuous over $\Gamma^B$ and to vary linearly over each boundary element $s_k$ according to

$$u(\zeta) = \mathbf{N}(\zeta)\,\tilde{u}^k, \tag{4.18}$$

$$t(\zeta) = \mathbf{N}(\zeta)\,\tilde{t}^k, \tag{4.19}$$

where the vectors $u(\zeta)$ and $t(\zeta)$ collect the components of displacements and points belonging to the segment $s_k$, the matrix $\mathbf{N}(\zeta) \in \mathbb{R}^{2\times 4}$ collects the one-dimensional linear shape functions for the boundary segment $s_k$, expressed as function of the natural coordinate $\zeta$ and $\tilde{u}^k, \tilde{t}^k \in \mathbb{R}^{4\times 1}$ collect the nodal components of displacements and tractions associated with the two ends of the boundary element $s_k$.

It is worth noting that the shape functions $N(\zeta)$, used for the boundary element modelling of the inclusions, could be seen as restrictions over the element edges of the shape functions $N(\xi,\eta)$ appearing in Eq.(2.25), used in the approximation of the virtual elements fields. Indeed, in the lowest order VEM, the restriction of the shape functions over the edges of a polygonal virtual element is linear, which ensures consistency at the interface between matrix (VEM) and inclusions (BEM).

Writing Eq.(4.17) for the generic boundary node $p$ and $i = 1,2$ in matrix form gives

$$\mathbf{c}\,\tilde{u}_p + \sum_{q=1}^{m}\left[\int_{s_q}\mathbf{H}_{pq}(\zeta)\,\mathbf{N}(\zeta)J(\zeta)\,d\zeta\right]\tilde{u}^q = \sum_{q=1}^{m}\left[\int_{s_q}\mathbf{G}_{pq}(\zeta)\,\mathbf{N}(\zeta)J(\zeta)\,d\zeta\right]\tilde{t}^q \tag{4.20}$$

where $\mathbf{c} \in \mathbb{R}^{2\times 2}$ depends on the geometry of the boundary at the considered collocation point $p$, smooth in this case, $\tilde{\mathbf{u}}_p \in \mathbb{R}^{2\times 1}$ collects the components of displacements at the node $p$, $\mathbf{H}_{pq}(\zeta), \mathbf{G}_{pq}(\zeta)$ collect the components of the fundamental solution, when the integral equations are collocated at the node $p$ and integrated over the element $q$, $\tilde{\mathbf{u}}^q, \tilde{\mathbf{t}}^q \in \mathbb{R}^{4\times 1}$ collect the nodal displacements and tractions associated with the ends of the generic boundary element $s_q$, according to Eq.(4.18), and $J(\zeta)$ is the Jacobian of the transformation between segment local and natural coordinates. After numerical integration [1], Eq.(4.20) may be rewritten in compact form as

$$\mathbf{H}_p \mathbf{U}^B = \mathbf{G}_p \mathbf{T}^B \qquad (4.21)$$

where $\mathbf{H}_p, \mathbf{G}_p \in \mathbb{R}^{2\times 2m}$ denote the rectangular matrices obtained by collocating at the node $p$ and integrating over the whole boundary $\Gamma^B$, while $\mathbf{U}^B, \mathbf{T}^B \in \mathbb{R}^{2m\times 1}$ collect the components of displacements and tractions for all the nodes identified on $\Gamma^B$, with the superscript B introduced to highlight that such quantities are associated with the BEM domain. Writing Eq.(4.21) $\forall p \in [1, ..., m]$ produces the set of linear algebraic equations

$$\mathbf{H}\,\mathbf{U}^B = \mathbf{G}\,\mathbf{T}^B, \qquad (4.22)$$

where $\mathbf{H}, \mathbf{G} \in \mathbb{R}^{2m\times 2m}$ collect matrix blocks of the form appearing in Eq.(4.21). It is worth noting that, when the BEM domain identifies an inclusion in the analysed domain, both $\mathbf{U}^B$ and $\mathbf{T}^B$ are unknown quantities that must be determined by interfacing Eq.(4.22) with the equations produced by the model employed for the matrix domain.

### 4.3.5 Displacements and stresses within BEM domains

Once the displacements and tractions at the boundary of the inclusions modelled with BEM are known, the value of displacements and stresses at points *within* the inclusions may be computed in post-processing.

Interior points displacements may be computed employing the boundary integral representation

$$u_j(\mathbf{x}_0) + \int_{\Gamma^B} H_{ij}(\mathbf{x}_0, \mathbf{x})\, u_j(\mathbf{x})\, d\Gamma(\mathbf{x}) = \int_{\Gamma^B} G_{ij}(\mathbf{x}_0, \mathbf{x})\, t_j(\mathbf{x})\, d\Gamma(\mathbf{x}), \qquad (4.23)$$

---

[1] The numerical integration over the boundary is not a trivial process. The reader interested to the details of the numerical integration of kernels is referred to Ref.[7] and the references therein.

which differs from Eq.(4.17) for the absence of the coefficients $c_{ij}(\bm{x}_0)$, arising from the limiting boundary collocation process.

Internal stresses, on the other hand, can be computed from the boundary integral representation

$$\sigma_{ij}(\bm{x}_0) + \int_{\Gamma^B} S_{ijk}(\bm{x}_0,\bm{x})\, u_k(\bm{x})\, d\Gamma(\bm{x}) = \int_{\Gamma^B} D_{ijk}(\bm{x}_0,\bm{x})\, t_k(\bm{x})\, d\Gamma(\bm{x}), \quad (4.24)$$

obtained by differentiating Eq.(4.23), to obtain the integral representation of strains at the considered interior point, and then using the constitutive equations, see, e.g. Refs.[20, 7].

Eqs.(4.23-4.24) express displacements and stresses at internal points as a function of known displacements and tractions at points along the boundary of the inclusion. The numerical integration of such equations is generally straightforward, except for internal points whose distance from the boundary is less than the employed boundary elements' size. In such cases, the integrals appearing in Eqs.(4.23-4.24) become *nearly singular*, as the distance $r(\bm{x}_0,\bm{x})$ between the collocation and integration points appears at the denominator of the kernels $H_{ij}$, $G_{ij}$, $S_{ijk}$, $D_{ijk}$. In such cases, specific integration schemes may be employed to enhance the accuracy of the integration, see e.g. Refs.[99, 181]. In the present work, a simple technique has been implemented: *i)* internal points are selected so that their distance from the boundary is no less than half the boundary element length; *ii)* the Gauss quadrature order over the elements closer to the selected point is increased with respect to the order of integration employed for the far elements. The method's accuracy has been assessed in simple benchmark tests, and the absence of artefacts has been verified in the analysed test cases. However, the technique is not general, and the use of specific schemes for nearly singular integrals should be considered in general implementations [7, 99, 181].

For further details about the use of Eqs.(4.23-4.24) and their numerical treatments, the interested readers are referred to Refs.[20, 7].

## 4.4 VEM-BEM coupling

The coupling between boundary and finite elements has been achieved in the literature using various approaches [185, 56, 90, 39, 64]. The approach herein adopted to couple the virtual and the boundary element equations consist in treating the BEM subdomains as macro-finite elements and in transforming the *traction*-displacement equations associated with them into *force*- displacement equations that will eventually be assembled with the VEM equilibrium equations, already expressed in terms of nodal forces and displacements.

In Chapter 1 the equilibrium equation for the overall virtual element domain have been obtained. Here they are recalled for convenience

$$\mathbf{K}^V \mathbf{U}^V = \mathbf{F}^V, \tag{4.25}$$

where the superscript $V$ is employed to identify terms stemming from the virtual element model and differentiate them from those associated with the boundary element model of the inclusions.

The vectors $\mathbf{U}^V$ and $\mathbf{F}^V$ appearing in Eq.(4.25) collect the nodal components of displacements and forces of all the VEM nodes in the considered domain. Since only some of such nodes belong to the interface $\Gamma^B$ between boundary and virtual elements, it is possible to partition the vectors as

$$\mathbf{U}^V = \begin{bmatrix} \mathbf{U}^V_\Gamma \\ \mathbf{U}^V_\Omega \end{bmatrix}, \qquad \mathbf{F}^V = \begin{bmatrix} \mathbf{F}^V_\Gamma \\ \mathbf{F}^V_\Omega \end{bmatrix}, \tag{4.26}$$

where $\mathbf{U}^V_\Gamma$ and $\mathbf{F}^V_\Gamma$ identify components related to nodes belonging to $\Gamma^B$. In lieu of the decomposition in Eq.(4.26), the equilibrium equation (4.25) for the VEM subdomain can be rewritten as

$$\begin{bmatrix} \mathbf{K}^V_{\Gamma\Gamma} & \mathbf{K}^V_{\Gamma\Omega} \\ \mathbf{K}^V_{\Omega\Gamma} & \mathbf{K}^V_{\Omega\Omega} \end{bmatrix} \begin{bmatrix} \mathbf{U}^V_\Gamma \\ \mathbf{U}^V_\Omega \end{bmatrix} = \begin{bmatrix} \mathbf{F}^V_\Gamma \\ \mathbf{F}^V_\Omega \end{bmatrix}, \tag{4.27}$$

Along the interface $\Gamma^B$ between the two subdomains, the nodal displacements and forces must satisfy the compatibility conditions for displacements

$$\mathbf{U}^B = \mathbf{U}^V_\Gamma, \tag{4.28}$$

and equilibrium conditions

$$\mathbf{F}^B + \mathbf{F}^V_\Gamma = \mathbf{0}, \tag{4.29}$$

which have been written considering that no external nodal forces act on the nodes belonging to $\Gamma^B$. The displacement continuity equations can be readily written, as the displacement components appearing in the VEM system (4.27) and in the BEM system (4.22) carry the same physical meaning.

On the contrary, while nodal *forces* appear in Eq.(4.27), related to the VEM subdomain, nodal components of *tractions* appear in Eq.(4.22), related to the BEM subdomain, so that it is necessary to retrieve a consistent expression of the nodal forces associated to the BEM tractions, before writing the equilibrium equations appearing in Eq.(4.29).

For a generic boundary element node, this can be accomplished by resorting to appropriate energy considerations. In the scheme adopted in this work, since two-node piecewise linear continuous boundary elements are used, a generic node always lies at the conjunction between two contiguous *boundary* elements. It is here recalled that, in the considered two-dimensional background, boundary elements are one-dimensional segments, which are interfaced with the *edges* of the polygonal virtual elements. If the generic node $i$ lies between the boundary elements $s_k$ and $s_{k+1}$, then, for a virtual displacement $\delta \tilde{u}(x_i) \equiv \delta \tilde{u}_i$ of the node $i$, the unknown nodal force $\tilde{F}_i^B$ will perform some work that has to be equivalent to the work performed by the tractions acting on the two contiguous boundary elements. Thus, the following equivalence holds

$$\delta u_i^\mathsf{T} \tilde{F}_i^B = \sum_{j=k}^{k+1} \int_{s_j} \delta u^\mathsf{T}(\zeta)\, t(\zeta)\, J(\zeta)\, d\zeta, \qquad (4.30)$$

which, recalling the interpolation expressed in Eq.(4.18), may be written as

$$\delta u_i^\mathsf{T} \tilde{F}_i^B = \sum_{j=k}^{k+1} \delta \tilde{u}^{j\mathsf{T}} \left[ \int_{s_j} \mathbf{N}(\zeta)^\mathsf{T} \mathbf{N}(\zeta) J(\zeta)\, d\zeta \right] \tilde{t}^j = \sum_{j=k}^{k+1} \delta \tilde{u}^{j\mathsf{T}} \mathbf{M}^j \tilde{t}^j, \qquad (4.31)$$

where the matrices $\mathbf{M}^j \in \mathbb{R}^{4\times 4}$ stem from the integration over the considered elements of the shape functions matrices, while the vectors $\delta \tilde{u}^j, \tilde{t}^j \in \mathbb{R}^{4\times 1}$ collect the components of displacements at the two end nodes belonging to the element $j$, so that

$$\delta \tilde{u}^k = \begin{bmatrix} \delta \tilde{u}_{i-1} \\ \delta \tilde{u}_i \end{bmatrix} = \begin{bmatrix} \mathbf{0} \\ \delta \tilde{u}_i \end{bmatrix}, \qquad \delta \tilde{u}^{k+1} = \begin{bmatrix} \delta \tilde{u}_i \\ \delta \tilde{u}_{i+1} \end{bmatrix} = \begin{bmatrix} \delta \tilde{u}_i \\ \mathbf{0} \end{bmatrix}. \qquad (4.32)$$

Taking into account Eqs.(4.32), Eq.(4.31) may be rewritten

$$\delta \tilde{u}_i^\mathsf{T} \tilde{F}_i^B = \delta \tilde{u}_i^\mathsf{T} \sum_{j=k}^{k+1} \mathbf{M}_i^j \tilde{t}^j \quad \Rightarrow \quad \tilde{F}_i^B = \sum_{j=k}^{k+1} \mathbf{M}_i^j \tilde{t}^j, \qquad (4.33)$$

where $\mathbf{M}_i^j \in \mathbb{R}^{2\times 4}$ is the sub-matrix extracted from $\mathbf{M}^j$ selecting the appropriate rows corresponding to the displacements associated with the node $i$. It is important to realise that Eq.(4.33) allows expressing $\tilde{F}_i^B$ in terms of the traction components associated with the two elements containing the node $i$; for two-node linear boundary elements such expression could be written as

$$\tilde{\mathbf{F}}_i^B = \sum_{k=i-1}^{i+1} \mathbf{M}_k \, \tilde{\mathbf{t}}_k, \qquad (4.34)$$

where $\tilde{\mathbf{t}}_k$ collects the components of tractions associated with the node $k$ and $\mathbf{M}_i \in \mathbb{R}^{2\times 2}$. Once Eq.(4.34) is written for all the boundary element nodes belonging to $\Gamma^B$, the nodal forces $\mathbf{F}^B$ appearing in Eq.(4.29) can be expressed in terms of the boundary tractions $\mathbf{T}^B$ appearing in Eq.(4.22) as

$$\mathbf{F}^B = \mathbf{M}\,\mathbf{T}^B, \qquad (4.35)$$

where $\mathbf{F}^B, \mathbf{T}^B \in \mathbb{R}^{2m\times 1}$ and $\mathbf{M} \in \mathbb{R}^{2m\times 2m}$, with $m$ expressing the total number of boundary nodes/elements. Exploiting Eq.(4.35), Eq.(4.22) can be written in a form to be used in conjunction with the VE equations. In particular, remembering that $\mathbf{T}^B = \mathbf{G}^{-1}\,\mathbf{H}\,\mathbf{U}^B$, it is possible to write

$$\mathbf{F}^B = \mathbf{M}\mathbf{T}^B = \left(\mathbf{M}\mathbf{G}^{-1}\,\mathbf{H}\right)\mathbf{U}^B = \mathbf{K}^B\,\mathbf{U}^B. \qquad (4.36)$$

The above BEM equations can now be combined with the VEM equations in Eq.(4.27), leading to the following system of equations valid for the whole hybrid VEM-BEM domain

$$\begin{bmatrix} \mathbf{K}^V_{\Gamma\Gamma} & \mathbf{K}^V_{\Gamma\Omega} \\ \mathbf{K}^V_{\Omega\Gamma} & \mathbf{K}^V_{\Omega\Omega} \end{bmatrix} \begin{bmatrix} \mathbf{U}^V_\Gamma \\ \mathbf{U}^V_\Omega \end{bmatrix} = \begin{bmatrix} -\mathbf{K}^B\,\mathbf{U}^V_\Gamma \\ \mathbf{F}^V_\Omega \end{bmatrix}. \qquad (4.37)$$

After prescribing suitable external boundary conditions, Eq.(4.37) can be solved to obtain the problem solution.

## 4.5 Analysis of a microstructure with multiple inclusions

In this Section, the proposed methodology's accuracy and robustness are tested by solving, under the plane strain assumption, the elastic problem of a unit cell with some inclusions of involved shape and assessing the developed method's reliability in reconstructing the local elastic fields.

All the numerical experiments have been performed using an in-house developed code written in MATLAB; this software addresses all the stages of the computations, starting from morphology generation and meshing, handles FEM, VEM and BEM elements in the processing stage as well as all the interface and post-processing subroutines.

　　The geometry of the microstructure, shown in Fig.(4.4), is a two-dimensional square box with four inclusions of involved shape. The external edges of the square box are aligned with the global Cartesian reference system $x - y$.

　　The purpose of this numerical test is to compare the displacement and stress fields obtained with the developed technique with a benchmark finite element solution, obtained employing an unstructured mesh of linear triangular elements. The analysis of the microstructure is performed with three different sets of homogeneous displacement boundary conditions corresponding to prescribed macro-strains $\bar{\epsilon}_{ij}$: two uniaxial macro-strains acting along the $x$ and $y$ directions ($BC_x$, $BC_y$) and a pure shear macro-strain acting to modify the angle between the axes $x - y$ ($BC_{xy}$). The values of the displacement components, enforced over all the nodes belonging to the external boundary $\Gamma$ of the computational domain, is given by

$$u_i = \bar{\epsilon}_{ij} x_j \quad \forall x \in \Gamma. \tag{4.38}$$

Additionally, a parametric analysis is also performed by varying the contrast of material properties between matrix and inclusions. Both phases are assumed to be linear elastic and isotropic in the plane of the analysis, and their relevant mechanical properties are given in Table 4.1 in terms of Poisson's ratio $\nu$ and of the ratio $\frac{E_f}{E_m}$, between the Young's modulus of the inclusions $E_f$ and the matrix $E_m$.

Table 4.1: Mechanical properties for the matrix and the inclusions.

| Material Code | $E_f/E_m$ | $\nu$ |
|---|---|---|
| M10 | 10 | 0.3 |
| M100 | 100 | 0.3 |
| M1000 | 1000 | 0.3 |

### 4.5.1 Benchmark finite element solutions

Before assessing the proposed hybrid VE-BE scheme's convergence, some benchmark finite element solutions are selected by performing a *h*-convergence analysis on triangular meshes. The elastic problem is solved for each set of boundary conditions and for each material. When passing from a coarser to a finer mesh with a smaller average element size, and then a higher number of associated degrees of freedom $N_{dof}$, the *distance* between the two related FE

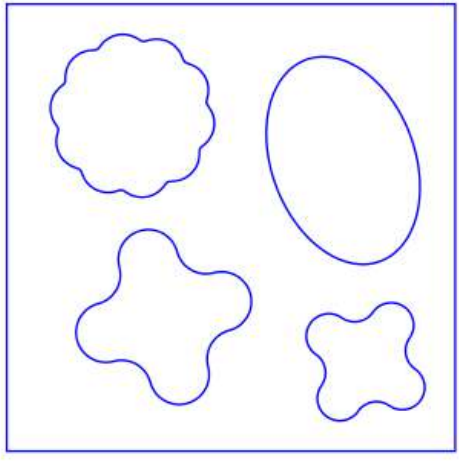

Figure 4.4: Morphology of the considered unit cell.

solutions is assessed employing a relative error measure for the displacement field $e_u$ defined as

$$e_u = \left[ \frac{\sum_{p=1}^{N_p} ||\boldsymbol{u}_a(\boldsymbol{x}_p) - \boldsymbol{u}_b(\boldsymbol{x}_p)||^2}{\sum_{p=1}^{N_p} ||\boldsymbol{u}_b(\boldsymbol{x}_p)||^2} \right]^{\frac{1}{2}}, \qquad (4.39)$$

computed with reference to a fixed set of $N_p$ sampling points $p$. In Eq.(4.39), $\boldsymbol{u}_a(\boldsymbol{x}_p)$ and $\boldsymbol{u}_b(\boldsymbol{x}_p)$ are the point-wise interpolated displacement vectors, computed at points having coordinates $\boldsymbol{x}_p$, for two different meshes $a$ and $b$, where $h_a > h_b$ and $N_{dof,a} < N_{dof,b}$. Fig.(4.5) shows an example of triangular FE mesh and highlights the $N_p$ fixed grid points selected for the computation of the measure given in Eq.(4.39). The evaluation points are selected to suitably sample the considered morphology and remain fixed as different meshes are considered; they will also be used to assess the accuracy of the hybrid virtual-boundary element scheme with respect to the benchmark solution. An analogous relative error measure can be introduced for the stress field as

$$e_\sigma = \left[ \frac{\sum_{p=1}^{N_p} ||\boldsymbol{\sigma}_a(\boldsymbol{x}_p) - \boldsymbol{\sigma}_b(\boldsymbol{x}_p)||^2}{\sum_{p=1}^{N_p} ||\boldsymbol{\sigma}_b(\boldsymbol{x}_p)||^2} \right]^{\frac{1}{2}}, \qquad (4.40)$$

Table 4.2 reports information about the size and corresponding number of degrees of freedom for the considered FE meshes.

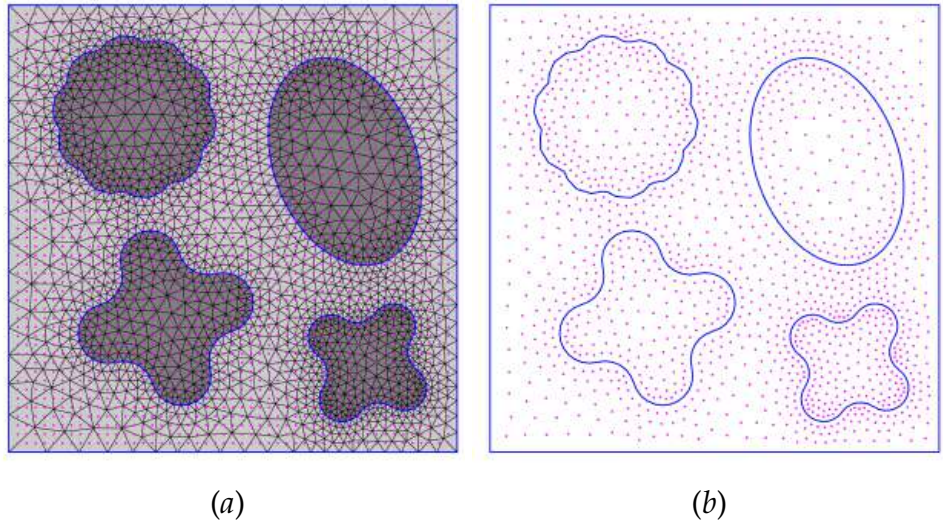

(a) (b)

Figure 4.5: *a*) Example of a FE triangular mesh for the considered unit cell; *b*) Sampling points selected for the convergence analysis; their position remains fixed as the FE mesh is refined and the same sampling points will be employed to assess the performance of the VE-BE scheme.

Table 4.2: Features of the finite element meshes considered in the convergence analysis.

|  | $F_1$ | $F_2$ | $F_3$ | $F_4$ | $F_5$ | $F_6$ | $F_7$ | $F_8$ |
|---|---|---|---|---|---|---|---|---|
| $N_{dof}$ | 3574 | 10712 | 23612 | 36520 | 93104 | 144034 | 256142 | 575836 |
| $N_{el}$ | 3483 | 10442 | 23210 | 36018 | 92302 | 143032 | 254804 | 573834 |

Table 4.3 reports the convergence data obtained by using the considered FE meshes. The generic column $e_{ufi}$ reports, for the boundary condition and materials combination identified by the considered rows, the relative error defined in Eq.(4.39) obtained by considering the coarser mesh $F_i$ and the finer mesh $F_{i+1}$. Such results are graphically shown in Fig.(4.6). It is possible to note that, for all the considered Young's modulus ratios and for all the sets of boundary conditions, convergence may be considered achieved with the mesh $F_5$. The results obtained for this mesh are taken as a benchmark for any further comparison.

The artificial linear sample shown in Fig.(4.4) has also been analysed by employing a pure boundary element approach for both the matrix and the inclusions. The adopted boundary element mesh is built starting from the nodes lying over the external boundary and the matrix/inclusion interfaces in the benchmark finite element mesh and consists of discontinuous linear elements for a total of 2788 nodes. The relative displacement error $e_{ub}$ with respect to the benchmark FEM solutions (mesh $F_5$) for the three sets of boundary conditions and for each material considered is reported in Table 4.4.

Table 4.3: Convergence analysis for the considered FE solutions: $e_{ufi}$ represents the relative error between the displacement field computed with the mesh $i+1$ (finer) and that computed with the mesh $i$ (coarser) at the selected sampling points.

|        |       | $e_{uf1}$ | $e_{uf2}$ | $e_{uf3}$ | $e_{uf4}$ | $e_{uf5}$ | $e_{uf6}$ | $e_{uf7}$ |
|--------|-------|-----------|-----------|-----------|-----------|-----------|-----------|-----------|
|        | M10   | 9.94e−4   | 4.01e−4   | 1.63e−4   | 8.75e−5   | 4.03e−5   | 3.27e−5   | 2.17e−5   |
| $BC_x$ | M100  | 1.34e−3   | 5.67e−4   | 2.24e−4   | 2.06e−4   | 5.58e−5   | 4.63e−5   | 3.10e−5   |
|        | M1000 | 1.38e−3   | 5.89e−4   | 2.32e−4   | 2.13e−4   | 5.79e−5   | 4.80e−5   | 3.21e−5   |
|        | M10   | 1.15e−3   | 4.82e−4   | 1.82e−4   | 1.64e−4   | 4.77e−5   | 3.96e−5   | 2.63e−5   |
| $BC_y$ | M100  | 1.61e−3   | 7.17e−4   | 2.61e−4   | 2.44e−4   | 6.99e−5   | 5.74e−5   | 3.84e−5   |
|        | M1000 | 1.68e−3   | 7.47e−4   | 2.71e−4   | 2.53e−4   | 7.29e−5   | 5.95e−5   | 3.97e−5   |
|        | M10   | 6.95e−4   | 3.07e−4   | 1.24e−4   | 1.07e−4   | 2.94e−5   | 2.45e−5   | 1.83e−5   |
| $BC_{xy}$ | M100 | 9.05e−4 | 4.32e−4   | 1.72e−4   | 1.50e−4   | 4.03e−5   | 3.40e−5   | 2.55e−5   |
|        | M1000 | 9.32e−4   | 4.46e−4   | 1.79e−4   | 1.54e−4   | 4.17e−5   | 3.51e−5   | 2.63e−5   |

Table 4.4: Relative displacement error $e_{ub}$ of the BE solutions with respect to the benchmark FEM solutions.

|       | $BC_x$  | $BC_y$  | $BC_{xy}$ |
|-------|---------|---------|-----------|
| M10   | 9.30e−5 | 1.23e−4 | 7.12e−5   |
| M100  | 1.34e−4 | 1.81e−4 | 1.02e−4   |
| M1000 | 1.39e−4 | 1.87e−4 | 1.05e−4   |

### 4.5.2 Virtual element solutions

In this Section, the morphology shown in Fig.(4.4) is analysed by employing a pure virtual element approach. Fig.(4.7a) shows an example polygonal mesh of the considered morphology, built by using the meshing strategy described in Section 4.2. Table 4.5 summarises the features of the five polygonal mesh refinements used to assess the convergence of the virtual element scheme with respect to the benchmark finite element solution. In particular $e_{uvi}$ represents the relative error for displacements, with respect to the reference FE solution, of the virtual element solution obtained by the $i$-th virtual element mesh $V_i$, computed using Eq.(4.39). Analogously, Table 4.7 reports a global measure of relative error for the stress vector computed by the virtual element method, with respect to the stresses provided by the reference FE solution.

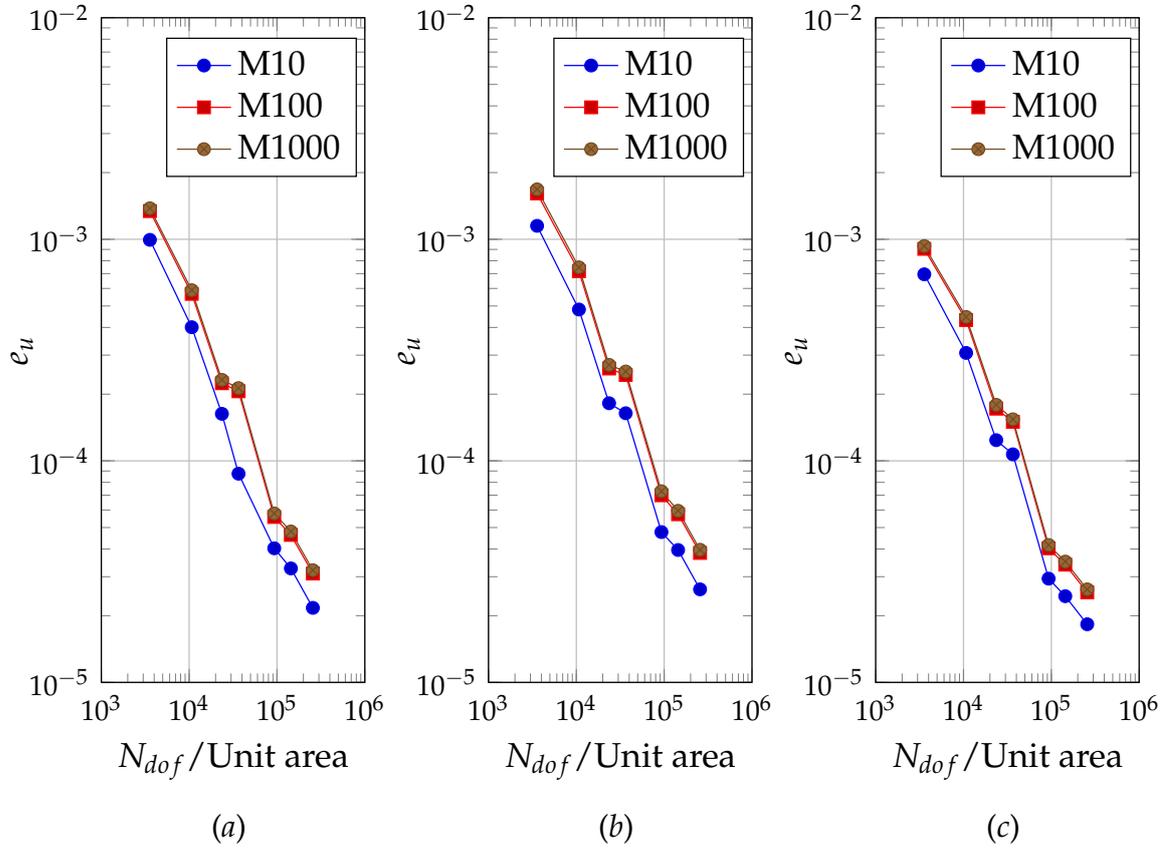

Figure 4.6: Convergence of the FE solutions: displacements relative error $e_u$ for (a) $BC_x$, (b) $BC_y$, (c) $BC_{xy}$.

Table 4.5: Features of the polygonal mesh refinements used for the virtual element analysis of the considered morphology.

|          | $V_1$ | $V_2$ | $V_3$ | $V_4$ | $V_5$ |
|----------|-------|-------|-------|-------|-------|
| $N_{dof}$ | 16516 | 30616 | 60388 | 98140 | 132844 |
| $N_{el}$  | 4128  | 7653  | 15096 | 24534 | 33210 |

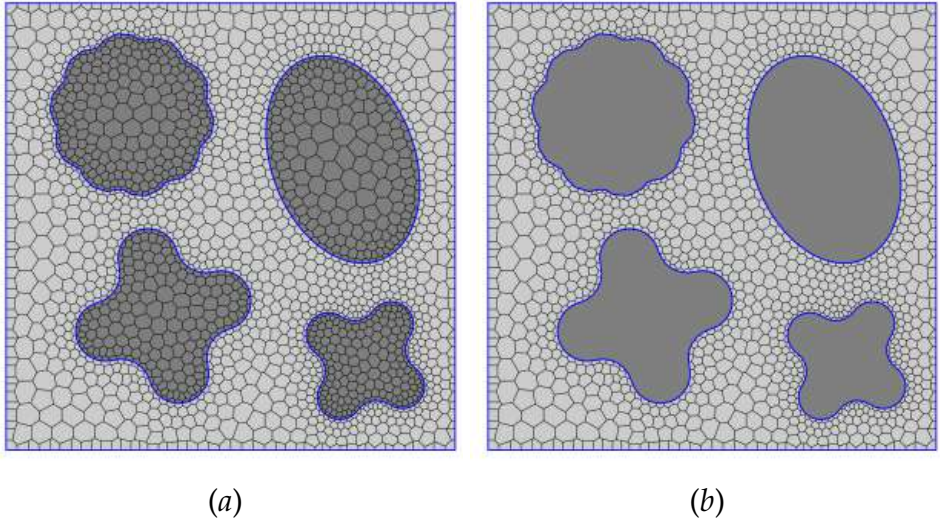

(a) (b)

Figure 4.7: *a*) Example polygonal mesh for the virtual element analysis; *b*) Example mesh for the hybrid virtual element - boundary element analysis.

Table 4.6: Relative displacement error $e_u$ with respect to the benchmark FEM solutions of the VE solutions obtained by using the considered progressive polygonal mesh refinements.

|  |  | $e_{uv1}$ | $e_{uv2}$ | $e_{uv3}$ | $e_{uv4}$ | $e_{uv5}$ |
|---|---|---|---|---|---|---|
|  | M10 | 6.30e−4 | 3.70e−4 | 1.63e−4 | 8.58e−5 | 6.57e−5 |
| $BC_x$ | M100 | 9.19e−4 | 5.41e−4 | 2.37e−4 | 1.28e−4 | 9.45e−5 |
|  | M1000 | 9.58e−4 | 5.63e−4 | 2.48e−4 | 1.35e−4 | 9.98e−5 |
|  | M10 | 8.89e−4 | 4.92e−4 | 2.06e−4 | 1.11e−4 | 8.21e−5 |
| $BC_y$ | M100 | 1.37e−3 | 7.55e−4 | 3.21e−4 | 1.78e−4 | 1.25e−4 |
|  | M1000 | 1.43e−3 | 7.91e−4 | 3.39e−4 | 1.90e−4 | 1.34e−4 |
|  | M10 | 5.14e−4 | 2.67e−4 | 1.22e−4 | 6.43e−5 | 4.50e−5 |
| $BC_{xy}$ | M100 | 7.02e−4 | 3.72e−4 | 1.76e−4 | 9.17e−5 | 6.33e−5 |
|  | M1000 | 7.46e−4 | 3.86e−4 | 1.85e−4 | 9.60e−5 | 6.65e−5 |

### 4.5.3 Hybrid virtual-boundary element solutions

In this Section, the developed hybrid virtual-boundary element scheme is employed to analyse the reference morphology in Fig.(4.4). Fig.(4.7b) shows an example discretisation of the considered morphology employed for the hybrid analysis, built by using the strategy described in Section 4.2. Table 4.8 reports some features of the mesh employed for the hybrid analysis, high-lighting the number of both 2D polygonal virtual elements and 1D linear continuous boundary elements employed in the analyses. Table 4.9 show the error, computed using Eq.(4.39), of the displacements field reconstructed using the hybrid strategy with respect to the FE benchmark solution, while Table 4.10 reports data about the measure of the relative error in the stress field computed using Eq.(4.40).

Table 4.7: Relative stress error $e_\sigma$ with respect to the benchmark FEM solutions of the VE solutions obtained by using the considered progressive polygonal mesh refinements.

|  |  | $e_{\sigma v1}$ | $e_{\sigma v2}$ | $e_{\sigma v3}$ | $e_{\sigma v4}$ | $e_{\sigma v5}$ |
|---|---|---|---|---|---|---|
| | M10 | 2.74e−2 | 2.65e−2 | 2.09e−2 | 1.98e−2 | 1.94e−2 |
| $BC_x$ | M100 | 4.15e−2 | 4.13e−2 | 3.18e−2 | 3.16e−2 | 3.05e−2 |
| | M1000 | 4.40e−2 | 4.41e−2 | 3.38e−2 | 3.39e−2 | 3.27e−2 |
| | M10 | 2.96e−2 | 2.98e−2 | 2.34e−2 | 2.30e−2 | 2.19e−2 |
| $BC_y$ | M100 | 4.61e−2 | 4.82e−2 | 3.69e−2 | 3.79e−2 | 3.61e−2 |
| | M1000 | 4.90e−2 | 5.16e−2 | 3.93e−2 | 4.07e−2 | 3.88e−2 |
| | M10 | 6.71e−2 | 5.89e−2 | 5.12e−2 | 4.21e−2 | 3.92e−2 |
| $BC_{xy}$ | M100 | 9.28e−2 | 7.77e−2 | 7.29e−2 | 5.69e−2 | 5.41e−2 |
| | M1000 | 9.68e−2 | 8.05e−2 | 7.64e−2 | 5.92e−2 | 5.63e−2 |

Table 4.8: Features of the hybrid virtual-boundary element mesh refinements employed in the comparative analysis. The total number of degrees of freedom $N_{dof}$, the number of virtual elements $N_{VEs}$ and the number of boundary elements $N_{BEs}$ are reported.

| | $H_1$ | $H_2$ | $H_3$ | $H_4$ | $H_5$ | $H_6$ |
|---|---|---|---|---|---|---|
| $N_{dof}$ | 11060 | 20496 | 39960 | 64656 | 87424 | 124248 |
| $N_{VEs}$ | 2614 | 4903 | 9703 | 15841 | 21459 | 30561 |
| $N_{BEs}$ | 584 | 864 | 1128 | 1272 | 1568 | 1984 |

Figs.(4.8-4.9) compare graphically the convergence of the hybrid virtual-boundary element solution with that of the pure virtual element solution. For both techniques, the accuracy of the solution is measured with respect to the assumed benchmark solution for all the investigated combinations of boundary conditions and materials. Specifically, the plots report the relative error measures of displacement and stress versus the number of degrees of free-

Table 4.9: Relative displacement error $e_u$ with respect to the benchmark FEM solutions of the hybrid VE-BE solutions obtained by using the considered progressive mesh refinements.

|  |  | $e_{uh1}$ | $e_{uh2}$ | $e_{uh3}$ | $e_{uh4}$ | $e_{uh5}$ | $e_{uh6}$ |
|---|---|---|---|---|---|---|---|
| $BC_x$ | M10 | 5.93e−4 | 3.33e−4 | 1.51e−4 | 8.39e−5 | 6.97e−5 | 6.35e−5 |
|  | M100 | 8.99e−4 | 5.23e−4 | 2.31e−4 | 1.23e−4 | 9.19e−5 | 7.29e−5 |
|  | M1000 | 9.55e−4 | 5.61e−4 | 2.47e−4 | 1.34e−4 | 9.92e−5 | 7.50e−5 |
| $BC_y$ | M10 | 8.08e−4 | 4.29e−4 | 1.91e−4 | 1.09e−4 | 9.01e−5 | 8.43e−5 |
|  | M100 | 1.32e−3 | 7.21e−4 | 3.09e−4 | 1.69e−4 | 1.21e−4 | 9.79e−5 |
|  | M1000 | 1.43e−3 | 7.86e−4 | 3.37e−4 | 1.88e−4 | 1.33e−4 | 1.01e−4 |
| $BC_{xy}$ | M10 | 4.61e−4 | 2.40e−4 | 1.08e−4 | 6.13e−5 | 4.99e−5 | 4.76e−5 |
|  | M100 | 6.97e−4 | 3.61e−4 | 1.69e−4 | 8.73e−5 | 6.20e−5 | 5.38e−5 |
|  | M1000 | 7.44e−4 | 3.85e−4 | 1.84e−4 | 9.55e−5 | 6.63e−5 | 5.52e−5 |

Table 4.10: Relative stress error $e_\sigma$ with respect to the benchmark FEM solutions of the hybrid VE-BE solutions obtained by using the considered progressive mesh refinements.

|  |  | $e_{\sigma h1}$ | $e_{\sigma h2}$ | $e_{\sigma h3}$ | $e_{\sigma h4}$ | $e_{\sigma h5}$ | $e_{\sigma h6}$ |
|---|---|---|---|---|---|---|---|
| $BC_x$ | M10 | 2.06e−2 | 1.87e−2 | 1.69e−2 | 1.58e−2 | 1.59e−2 | 1.54e−2 |
|  | M100 | 2.92e−2 | 2.69e−2 | 2.47e−2 | 2.36e−2 | 2.37e−2 | 2.33e−2 |
|  | M1000 | 3.09e−2 | 2.85e−2 | 2.62e−2 | 2.52e−2 | 2.52e−2 | 2.48e−2 |
| $BC_y$ | M10 | 2.29e−2 | 2.09e−2 | 1.89e−2 | 1.78e−2 | 1.73e−2 | 1.76e−2 |
|  | M100 | 3.37e−2 | 3.13e−2 | 2.88e−2 | 2.77e−2 | 2.73e−2 | 2.74e−2 |
|  | M1000 | 3.58e−2 | 3.33e−2 | 3.07e−2 | 2.96e−2 | 2.92e−2 | 2.93e−2 |
| $BC_{xy}$ | M10 | 4.59e−2 | 4.15e−2 | 3.47e−2 | 3.33e−2 | 3.20e−2 | 3.21e−2 |
|  | M100 | 5.68e−2 | 5.19e−2 | 4.57e−2 | 4.42e−2 | 4.31e−2 | 4.32e−2 |
|  | M1000 | 5.86e−2 | 5.36e−2 | 4.75e−2 | 4.60e−2 | 4.49e−2 | 4.50e−2 |

dom per unit area employed in the analysis.
It is observed that, in this sense, the convergence of the hybrid solution is quicker with respect to the convergence of the pure virtual element scheme, both for the displacement and the stress fields. However, while for the displacement field the two techniques show closer convergence rates, it emerges that, for the stress field, the hybrid VE-BE technique approaches convergence noticeably more rapidly than the pure virtual element scheme, at least when measured with respect to the number of degrees of freedom per unit area. The reason for such behaviour is twofold: *i*) in the hybrid technique, the nodes within the inclusions are removed due to the employment of the boundary integral formulation, which contributes to the reduction in the number of degrees of freedom per unit area, this explaining the convergence patterns observed for the displacement field; *ii*) in the hybrid scheme, the stresses within the inclusion are computed by employing, in post-processing, the boundary integral representation given in Eq.(4.24), which generally ensures higher accuracy, with respect to standard FE methods, in the reconstruction of the internal stresses. The interplay between the reduction in the number of degrees of freedom associated with nodes within the inclusions and better rendering of the stresses due to the employment of the boundary integral representation of stresses explains the convergence patterns shown in Fig.(4.9).

Eventually Fig.(4.10) shows the plot of stress components $\sigma_{xx}$, $\sigma_{yy}$ and $\sigma_{xy}$, corresponding to an enforced uniaxial strain $\bar{\epsilon}_{xx} = 0.05$, computed with the finite element benchmark scheme, the virtual element implementation and the hybrid strategy, highlighting satisfying agreement among the three schemes.

## 4.6 Computational homogenization of fibre-reinforced composites via the hybrid VEM-BEM approach

This Section describes the application of the proposed hybrid VEM-BEM method to the computational homogenization of unidirectional fibre-reinforced composites. In computational homogenization, the macroscopic material properties are computed by simulating the micro-scale response of properly selected material domains referred to as *unit cells*, and then averaging, over such domains, the fields of interest, to identify macroscopic links between such averaged quantities. Unit cells become *representative volume elements* (RVEs) when they can be considered representative of the material's mechanical behaviour at the macro-scale. Interested readers are referred to Ref.[137] for an in-depth treatment of materials homogenization.

This test case is presented in this study for the following reason. One of the strategies employed for the computational homogenization of heterogeneous materials is based on the generation of a certain number of artificial digital samples of the considered material, with random features, and on the computation of the ensemble averages, over the set of considered specimens, of suitable volume-averaged quantities, either stresses of strains, see, e.g. Refs.[102, 74, 46]. In this procedure, it is essential to ensure a suitable mesh quality for all the generated random microstructures, which may result in a particularly challenging task. It is believed that the inherent features of VEM may benefit the meshing procedures in this kind of problems, i.e. when a certain number of morphologies with statistical features need to be considered.

The unit cells for the present test lie in the plane $x_2 - x_3$, normal to the fibres' axes, which are parallel to the axis $x_1$. They are generated by randomly scattering a given number of arbitrarily shaped inclusions in a rectangular domain through an algorithm avoiding pathological superposition of the inclusions. The inclusions considered here present the transversal section shown in Fig.(4.11), are all of the same size and have and random orientation $\theta$ with respect to the $x_2$ axis. The average number of inclusions is determined by the parameter $\delta = L/r$, where $L$ is the unit cell's side length and $r$ is the radius of the circle that circumscribes the fibre inclusion. Fig.(4.12) shows two example geometries for a random microstructure and the subsequent VE-BE discretisations for $\delta = 20$ and $\delta = 45$.

The material constants of the composite constituents, isotropic in the $x_2 - x_3$ plane, are given in Table 4.11, in terms of transverse Young modulus $E_{22}$ and transverse shear modulus $G_{23}$. Assuming a Poisson random distribution of fibres within the unit cell and considering the constituents' in-plane isotropy, the composite will be isotropic in the plane $(x_2 - x_3)$ at the macroscopic level. Its transverse behaviour is then completely defined by two elastic modula. In this study, the plane strain bulk modulus $\bar{K}_{23}$ and the transverse shear modulus $\bar{G}_{23}$ are considered.

Table 4.11: Material properties for epoxy matrix and carbon fibres in transverse direction, as taken from Ref.[160].

| Mechanical Properties | $E_{22}$ [GPa] | $G_{23}$ [GPa] |
|---|---|---|
| AS4 carbon fibres | 15 | 7 |
| 3501-6 epoxy matrix | 4.2 | 1.567 |

The problem of determining the appropriate size of the unit cell, or the appropriate number of inclusions within it, so to identify an RVE has been extensively investigated in the literature [170, 102, 166, 173?, 120, 104]. In general, given a random microstructural sample subjected to a suitable set of boundary conditions, see, e.g. [102], the link between homogenized stresses and strains is provided by *apparent* properties, which may not be representative of the macro-material if the microstructural sample, or unit cell, is too small. As the unit cell size or the number of inclusions within it increase, the unit cell becomes more representative of the macro-material, and the apparent properties approach the *effective* properties. Besides considering the behaviour of the averaged properties versus the size or number of inclusions of the unit cell, the homogenization procedure can be enriched by considering *ensemble* averages of the volume homogenized properties over a set of unit cells with the same size and number of inclusions, but different spatial distribution of the inclusions themselves. The procedure generally produces an estimate of the effective properties with unit cells smaller with respect to the case in which only individual microstructures are considered, see, e.g. Ref.[102] for a detailed discussion. In the present study, this homogenization procedure is used. The interested readers are referred to Refs.[74, 46] for further examples about the application of the methodology.

For the considered composite, sets of unit cells at varying values of the parameter $\delta$ are considered, while the fibre volume fraction is kept constant at $V_f = 0.25$. For each value of $\delta$, $N_s = 50$ different random sample micro-morphologies have been generated and analysed using the proposed hybrid approach. Each unit cell $\mathcal{U}_m$, comprising several randomly located and orientated inclusions as in Fig.(4.12), has been discretised using arbitrary polygonal virtual elements for the matrix and a single boundary element domain for each inclusion. Each $\mathcal{U}_m$ is the subjected to three linearly independent sets of displacement boundary conditions, corresponding to three sets of enforced macro-strains expressed in Voigt notation as $\bar{\epsilon} = \{\bar{\epsilon}_{22}, \bar{\epsilon}_{33}, 2\bar{\epsilon}_{23}\}$. More specifically, the unit cells are loaded through displacements given by Eq.(4.38), where the following three sets of macro-strains

$$\bar{\epsilon}^a = \{1,0,0\}, \qquad \bar{\epsilon}^b = \{0,1,0\}, \qquad \bar{\epsilon}^c = \{0,0,1\}, \qquad (4.41)$$

are considered. Once a prescribed boundary condition is enforced, the microstructural problem is solved employing the proposed hybrid scheme, thus providing the micro displacement, strain and stress fields within the microstructure. The averaged stresses $\bar{\sigma} = \{\bar{\sigma}_{22}, \bar{\sigma}_{33}, \bar{\sigma}_{23}\}$ are then computed as volume averages of the local micro-stress tensor over the domain of the unit cells, as

$$\bar{\sigma}_{ij} = \frac{1}{\Omega} \int_\Omega \sigma_{ij}(x)\, d\Omega = \frac{1}{\Omega} \left( \int_{\Omega^V} \sigma_{ij}(x)\, d\Omega + \int_{\Omega^B} \sigma_{ij}(x)\, d\Omega \right), \qquad (4.42)$$

where the domain integral is subdivided into contributions coming separately from the VE and BE regions. The integral over the BE regions can be further expressed as a sum of integrals over each BE modelled inclusion $\Omega^B_k$, and it

may be demonstrated that

$$\int_{\Omega^B} \sigma_{ij}(\boldsymbol{x})\,d\Omega = \sum_i \int_{\Omega_k^B} \sigma_{ij}(\boldsymbol{x})\,d\Omega = \sum_i \int_{\Gamma_k} t_i\,n_j\,d\Gamma, \quad (4.43)$$

which implies that the integration of stresses over the BE inclusions only require the computation of integrals along the boundary $\Gamma_k^B$ of the inclusion of the traction components $t_i$, which are readily available from the BE solution, thus avoiding the more expensive use of Eq.(4.24). The use of Eq.(4.43) into Eq.(4.42) allows remarkable computational savings in computational homogenization problems and constitutes a benefit of the presented technique.

For a given unit cell $\mathcal{U}_m$, the computation of the averaged stresses corresponding to the three considered sets of boundary conditions given in Eq.(4.41) allows populating the columns of the apparent elastic matrix $\bar{\mathbf{C}}_m$, which links averaged stresses and strains according to

$$\bar{\sigma} = \bar{\mathbf{C}}_m\,\bar{\epsilon}. \quad (4.44)$$

For each value of the parameter $\delta$, once the components of $\bar{\mathbf{C}}_m$ are computed for all the $N_s = 50$ generated random unit cells, a macroscopic *apparent* constitutive matrix $\langle\bar{\mathbf{C}}\rangle$ is computed from the ensemble average of the components of $\bar{\mathbf{C}}_m$ over the $N_s$ samples, i.e.

$$\langle\bar{\mathbf{C}}\rangle = \frac{1}{N_s}\sum_{m=1}^{N_s} \bar{\mathbf{C}}_m. \quad (4.45)$$

The apparent transverse elastic properties $\bar{K}_{23}$ and $\bar{G}_{23}$ associated to the considered value of $\delta$ are eventually obtained from the ensemble averaged matrix $\langle\bar{\mathbf{C}}\rangle$.

Fig.(4.13) shows the values of $\bar{K}_{23}$ and $\bar{G}_{23}$ versus $\delta$ in plain strains, reporting both the values corresponding to individual samples $\mathcal{U}_m$ and the ensemble averaged values. In general, the scatter of the individual values decreases as $\delta$ increases as the unit cells approach the RVE by including a higher number of fibres. In the literature, several theoretical models have been introduced to provide rigorous bounds for heterogeneous materials' effective macroscopic properties. In the present study, the computed effective material properties are compared with the Hashin-Hill bounds [89, 86], identified in Fig.(4.13) by the boundaries of the grey region: it is observed that the values computed through the developed technique fall within such bounds, confirming its usefulness in computational homogenization applications.

## 4.7   Key features of the hybrid VEM-BEM formulation

In this Chapter, a hybrid computational technique has been developed to analyse multi-region two-dimensional elastic problems for computational micromechanics applications. The method suggests the simultaneous use of the recently emerged virtual element method and the highly accurate boundary element method. Each of the two techniques offers some definite advantages.

The use of VEM in addressing complex mesh morphologies and problems inducing high mesh distortion has been demonstrated in the literature, as reviewed in Section 4.1, and in Chapter 3 when dealing with materials micromechanics problems. In the present framework, the advantages offered by VEM are twofold: *i*) in general, the method offers a powerful tool for meshing morphologically complex domains, as those often encountered in statistical homogenization procedures, see, e.g. Fig.(4.12), in which the regularity of the regions related to the different phases cannot be *a-priori* assumed; *ii*) thanks to the possibility of extending to VEM the features of FEM, in particular its generality in dealing with non-linear constitutive behaviours, in the proposed framework, the method can be employed for meshing phases likely to exhibit, in the loading process, non-linear behaviours such as plasticity, viscosity or damage [61]. This is the case of composite materials subjected to loadings that can initiate visco-plastic flows and/or damage in the matrix.

On the other hand, the Boundary Element Method has proven effective in the accurate reconstruction of the elastic fields through a discretisation procedure involving only the boundary of the analysed domains, thanks to the underlying integral formulation, alternative to methods based on the weak formulation of the considered boundary value problems. In particular, BEM is known for providing accurate solutions at reduced computational costs [7]. The method can be used for analysing non-linear problems, although its employment in linear problems is more widespread and straightforward. An important caveat about the use of BEM is related to the fact that the method induces non-symmetric and non-definite fully populated solving matrices, see, e.g. Ref.[20]. As long as the number of elements used for modelling each inclusion is limited, this does not require additional consideration, and the potential of BEM in reducing the computational burden is preserved. However, should an inclusion need several hundred boundary elements, the presence of fully populated blocks in the solving systems could reduce the computation's effectiveness and increase the computational costs. These aspects could be mitigated and effectively addressed by using fast iterative solvers in conjunction with special matrix representations, e.g. fast multipoles [112], or hierarchical matrices [29, 43, 51, 44]. In the proposed framework, BEM is proposed for modelling microstructural phases not expected to develop non-linear constitutive behaviour. This use is beneficial for two reasons, as already mentioned in the previous Section: *i*) it allows to reduce the number of degrees of freedom needed for modelling the inclusions, thus reducing the computational burden of the analysis; *ii*) it generally provides a more accurate representation of stresses within the inclusions, thus inducing a faster convergence in the stress fields, as shown in Figs.(4.8-4.9).

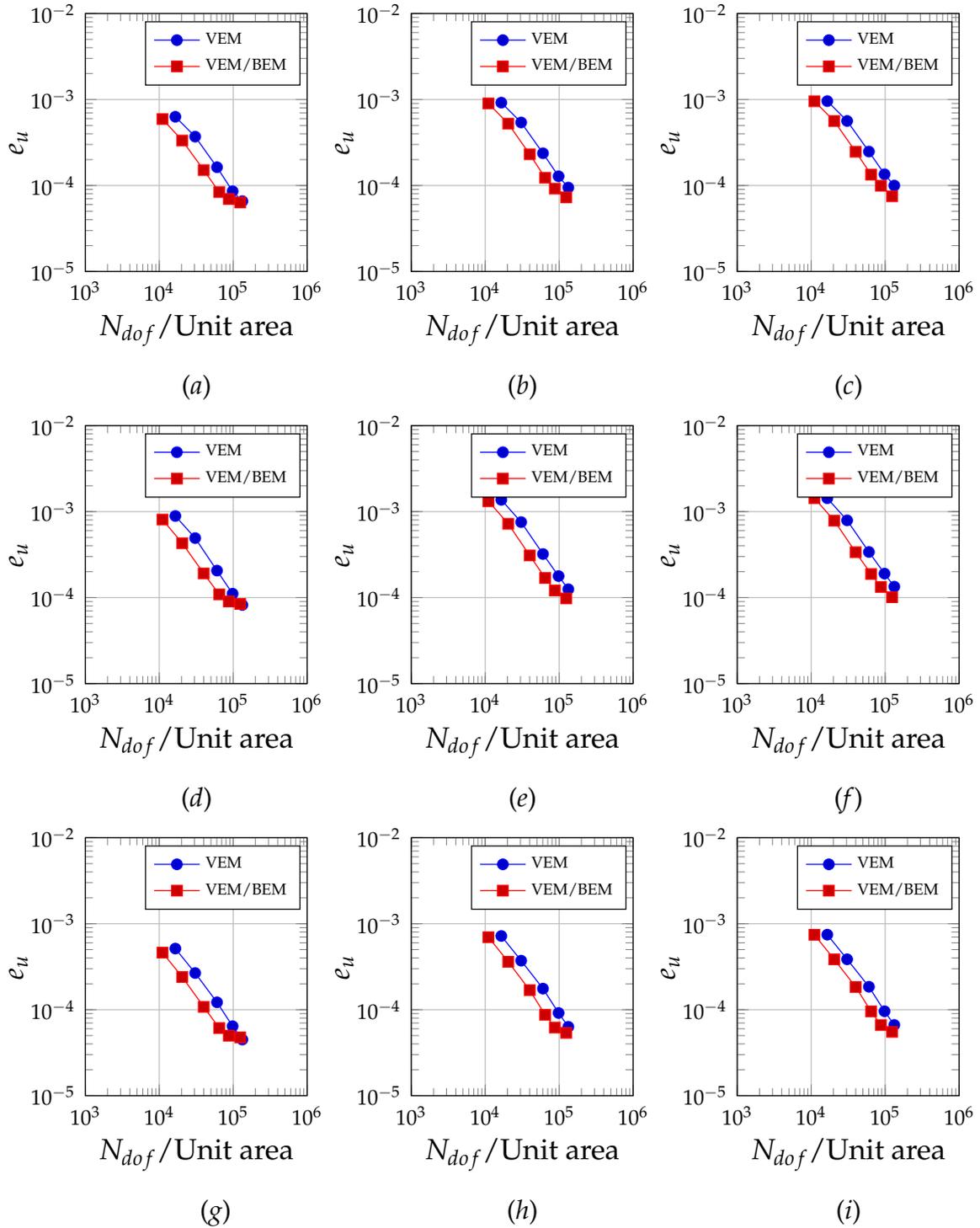

Figure 4.8: Comparison between the convergence of the VE solutions and that of the hybrid VE-BE solution in terms of displacements. The rows of the plots grid correspond to the different considered boundary conditions, namely $BC_x$ (a,b,c), $BC_y$ (d,e,f), $BC_{xy}$ (g,h,i). The columns correspond to the different materials, i.e. M10 (a,d,g), M100 (b,e,h), M1000 (c,f,i).

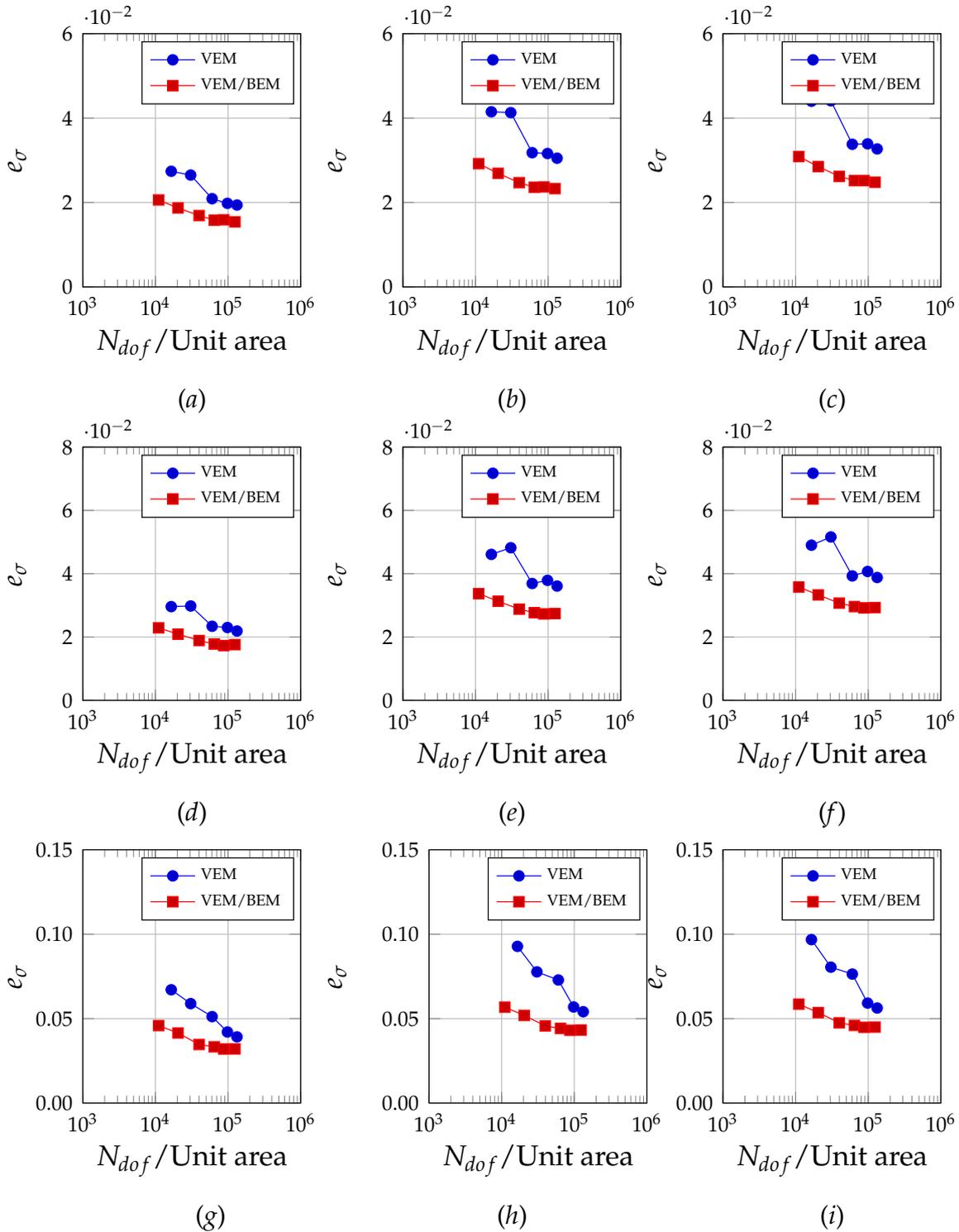

Figure 4.9: Comparison between the convergence of the VE solutions and that of the hybrid VE-BE solution in terms of stresses. The rows of the plots grid correspond to the different considered boundary conditions, namely $BC_x$ (a,b,c), $BC_y$ (d,e,f), $BC_{xy}$ (g,h,i). The columns correspond to the different materials, i.e. M10 (a,d,g), M100 (b,e,h), M1000 (c,f,i).

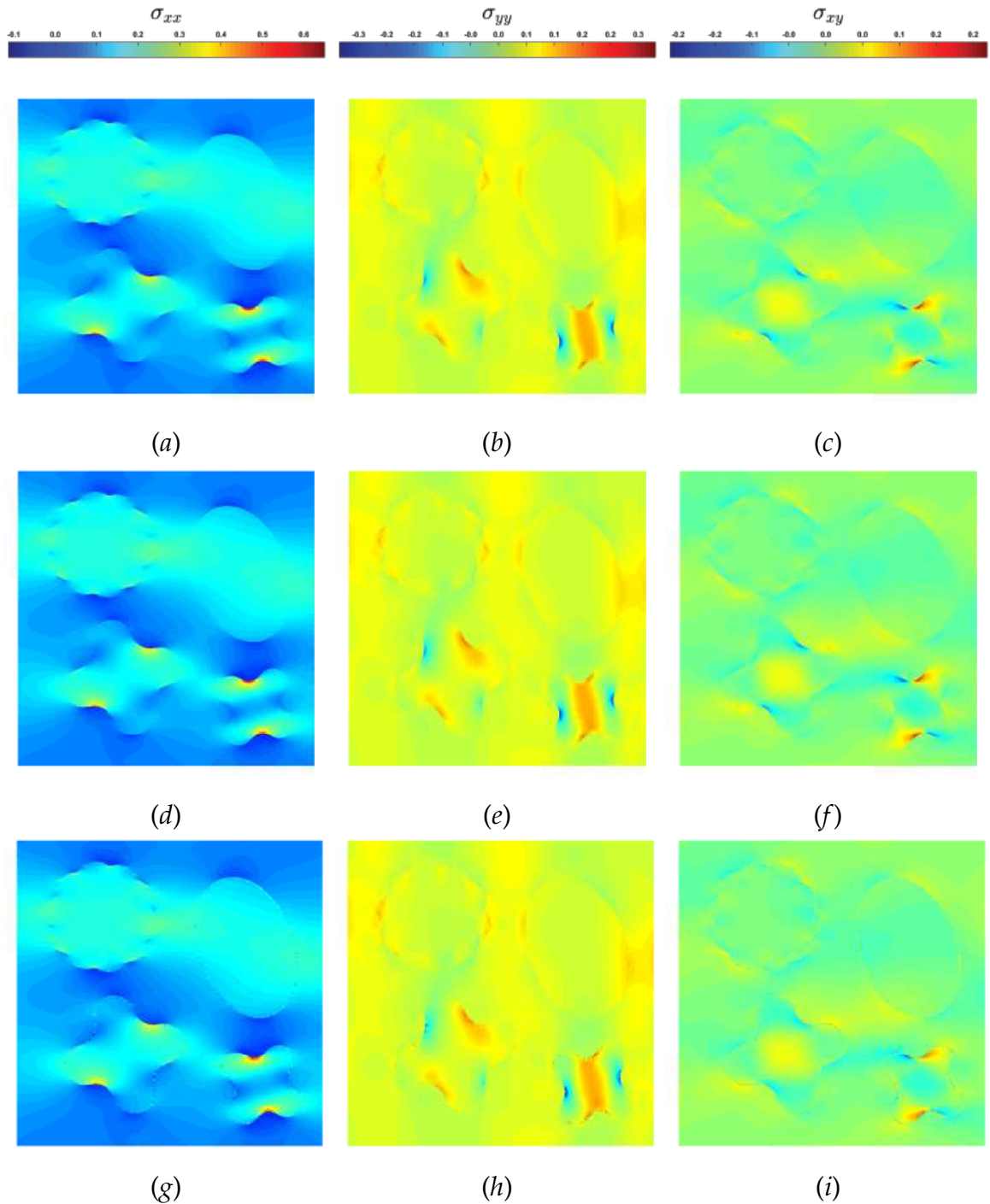

Figure 4.10: From left to right, plot of stress components $\sigma_{xx}$, $\sigma_{yy}$ and $\sigma_{xy}$ [GPa] corresponding to an enforced uniaxial strain $\bar{\epsilon}_{xx} = 0.05$ computed by using (a,b,c) FEM, (d,e,f) VEM and (g,h,i) the hybrid VEM-BEM scheme. The material considered is *M1000*. The comparison highlights remarkable agreement among the three different solutions.

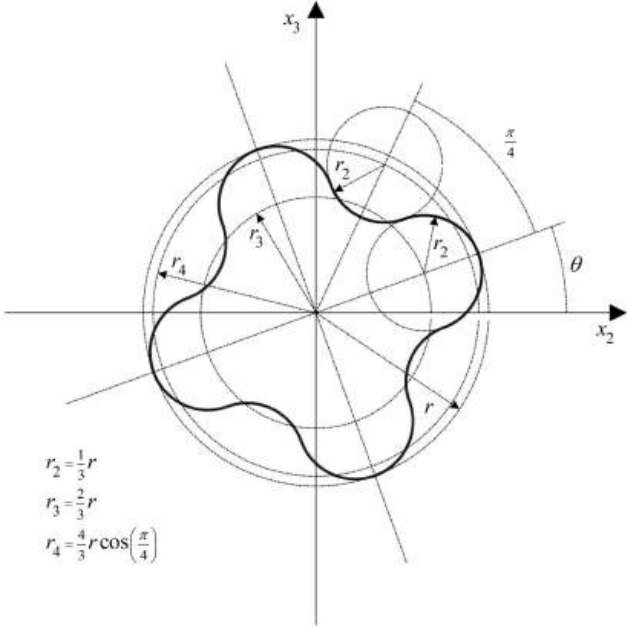

Figure 4.11: Geometry of the transversal section of the inclusions randomly placed within the analysed unit cells.

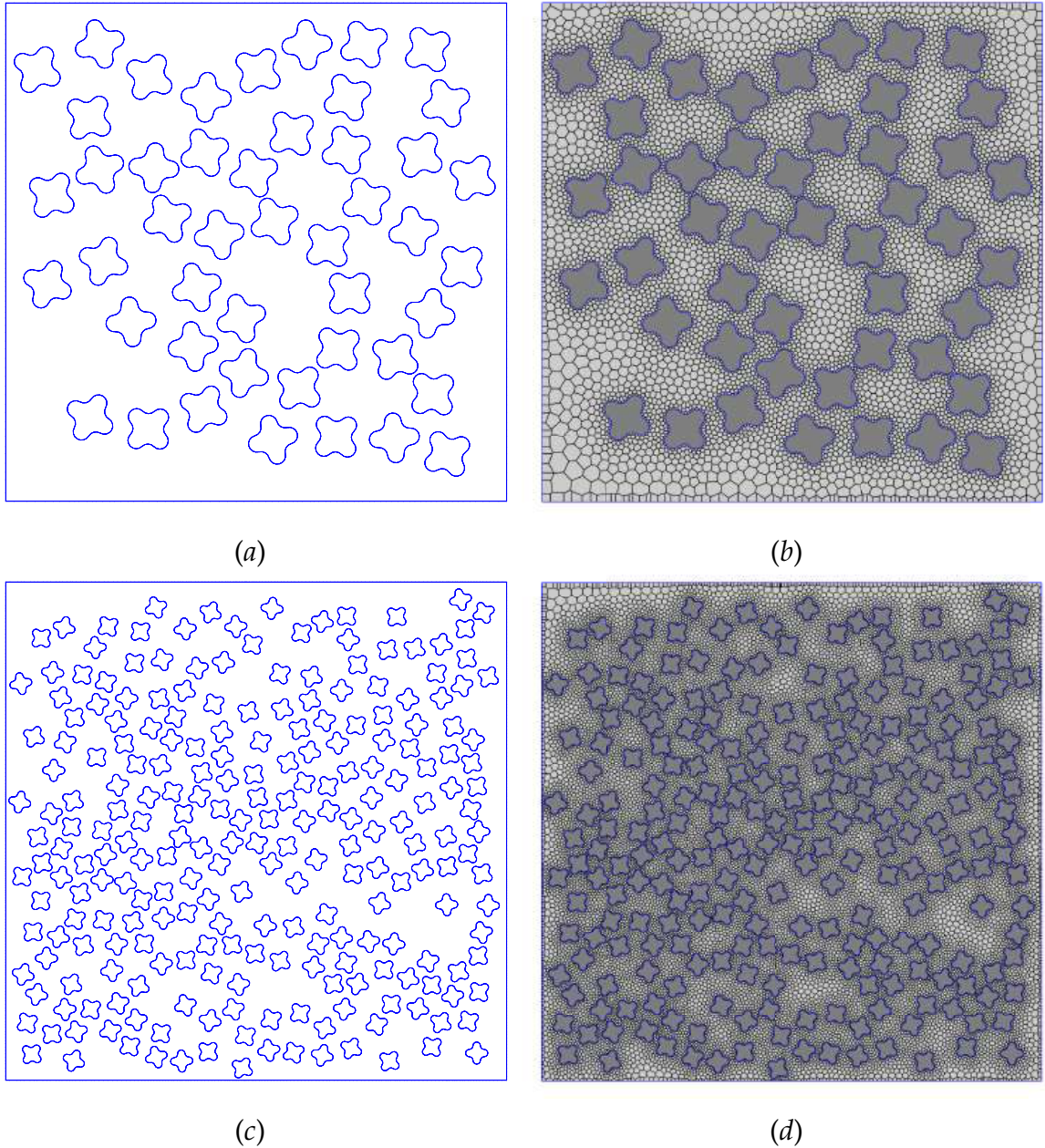

Figure 4.12: Examples of unit cells employed in the computational homogenization tests: random geometries obtained by setting $V_f = 0.25$ and *a*) $\delta = 20$ *c*) $\delta = 45$; *b,d*) meshes employed for the VE-BE analyses.

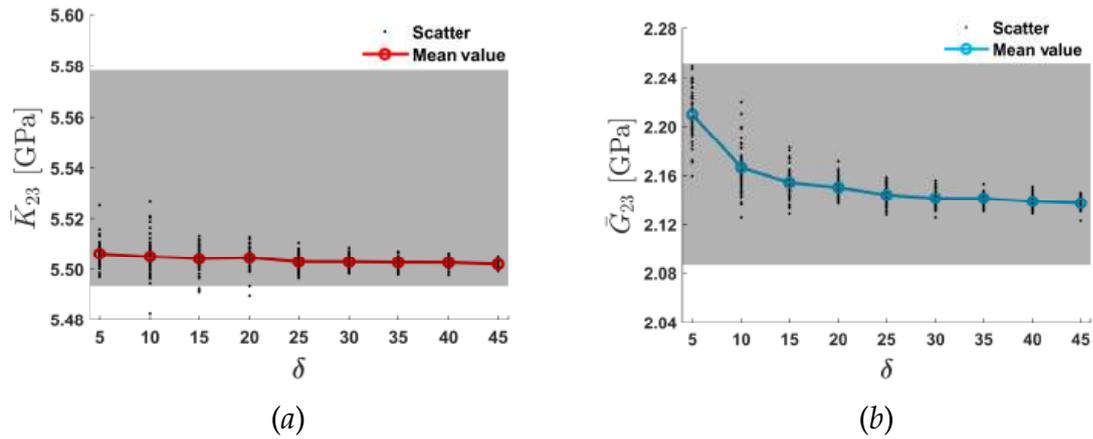

Figure 4.13: Apparent transverse elastic properties $\bar{K}_{23}$ and $\bar{G}_{23}$ as a function of $\delta$ for $V_f = 0.25$ as computed using the hybrid virtual-boundary element technique. The Hashin-Hill bounds for the considered composite are identified by the grey area.

## Chapter 5

# Damage and fracture damage in heterogeneous materials via the hybrid VEM-BEM approach

Common advanced structures are expected to undergo quasi-static and dynamic operating loading during their service life. Such load conditions are likely to induce a loss of the structure's mechanical performance that might ultimately end with its failure. Thus, the structure's effectiveness widely depends on the capability to predict damage initiation and propagation up to the point where the structure is no longer able to sustain its design loads and adequately perform its intended function. Such knowledge is essential during both the initial material selection and the detailed engineering design stage of a structure.

Most failure phenomena in engineering material are due to the propagation and coalescence of microscopic defects. These microstructure modifications lead to irreversible material degradation, characterized by a loss of stiffness observed at the macroscopic scale. In the last decades, the scientific community has put extensive and remarkable efforts into developing numerical techniques to simulate the damage and failure process at the microscopic scale with increasing computational efficiency and accuracy. The nowadays broad use of heterogeneous materials like fibre-reinforced composite in the most advanced structures broaden the design spectrum and add complexity to the simulation process due to some composite materials' peculiar features, such as the inherent complexity of both their microscopic structure and their failure process.

This Chapter aims to present further applications of the hybrid VEM-BEM formulation introduced in Chapter (4) by exploiting its peculiar features for modelling damage phenomena in heterogeneous materials [115, 118]. This Chapter is organised as follows: in Section 5.1 a crack propagation study in the matrix phase of fibre-reinforced composite materials is performed within the framework of Linear-Elastic Fracture Mechanics (LEFM). In Section 5.2 the proposed hybrid formulation is extended to a non-linear framework by adopting a constitutive law based on an isotropic damage model for the VEM subdomain. Several applications are discussed, including the analysis of matrix degradation in a fibre-reinforced composite unit cell under progressive loading by implementing a damage model combined with a non-local integral regularisation technique for the matrix phase modelled with VEM.

## 5.1 Crack propagation in FRC

Fibre-reinforced composites (FRC) materials may experience different types of damage mechanisms that strongly influence a structural component's overall behaviour. Many of these damage mechanisms often occur at the microscopic level, with the initiation and propagation of cracks. Analysis of such damage mechanisms is often performed via computational methods, and many numerical techniques such as FEM, BEM [103, 18, 155, 8, 94], XFEM [184, 95], among others, have been employed to model crack propagation processes .

The need for an accurate description of the evolution of micro-defect may lead to simulations requiring heavy computational effort. In this Section, crack propagation analysis in the matrix phase of fibre-reinforced composite materials is performed, taking advantage of the hybrid VEM-BEM formulation features developed in Chapter (4).

### 5.1.1 Modelling approach

With finite element techniques, crack propagation is modelled explicitly by consecutively adapting the domain's spatial discretization. The simplest method to simulate crack propagation is the nodal release technique, where a crack is extended by a certain increment along the element edge up to the next node. However, such a method restricts the crack propagation along the element edges, introducing a dependence of the computed crack path on the mesh topology. More powerful but complex methods may combine element modification techniques, such element splitting, with advanced re-meshing strategies, in which the overall mesh is progressively updated starting from a finer mesh in the proximity of the crack tip region. Such a continuous mesh refinement/coarsening process may generally affect large regions of the analysis domain, due to the need of preserving a conform transition between the re-meshed propagation region and the surrounding areas, and it is thus a computationally intensive operation [107].

On the other hand, VEM can handle polygonal elements of arbitrary shapes. This distinctive feature allows a substantial simplification of the re-meshing procedure following the propagation of a crack. Indeed, using VEM elements allows avoiding mesh dependency of the crack propagation direction, as any computed crack path can be represented by modifying the topology of the virtual element over which the crack propagation is occurring, including the crack edges as new element edges, without the need of further re-meshing.

The algorithm that governs crack path generation and tracking for the proposed methodology may be broken down into two steps: (i) linear static analysis and crack path computation; (ii) cracking and mesh modification.

Initiation of the fracturing process is assumed to start from a node of a boundary element and the same procedure can be employed either when the computed crack length increment is large enough to split an element completely or when it splits this element only partially. In both the aforementioned cases, the resulting element topology is valid from the VEM's standpoint and does not need any further mesh modification. Moreover, to improve numerical accuracy near the crack tip, a local mesh refinement can be introduced by subdividing one or more of the local elements in any number of elements of arbitrary shape, without necessarily affecting large portions of the analysed domain. This can be achieved by exploiting another peculiar VEM feature, namely its capability to naturally handle hanging nodes since a VEM element can be a polygon with an arbitrary number of aligned vertices. A schematic of the VEM's modelling benefits in crack propagation modelling is shown in Fig.(5.1).

Summarising, using VEM for crack propagation modelling allows to: *a)* avoid any crack path mesh-dependency, thanks to the possibility of capturing the crack propagation direction by modifying the topology of few elements in the proximity of the crack tip; *b)* improve the accuracy of the fields reconstruction in the proximity of the propagation region, thanks to the possibility of performing *local* mesh-refinements at low computational costs, without the need to update and optimize the mesh in large regions of the computational domain.

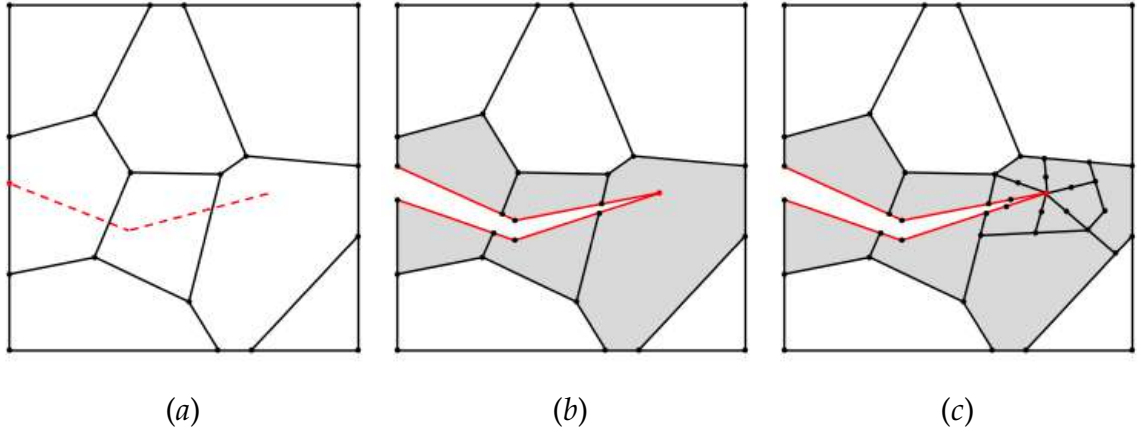

(a)           (b)           (c)

Figure 5.1: Schematic of crack propagation modelling with VEM elements: (a) computed crack path; (b) elements splitting; (c) local mesh refining.

### 5.1.2 Numerical example

In this Section, the hybrid technique based on the simultaneous use of BEM and VEM is applied to the analysis of crack propagation in fibre reinforced composite materials with the analysis of the growth of kinked cracks originating from the partial fibre-matrix debonding in a unit cell of a fibre-reinforced composite material subjected to a transverse load [122].

BEM is used to model the fibres, which are not expected to develop non-linear behaviours, while the VEM is employed to model the matrix, which may experience crack initiation and propagation. The inherent flexibility of the VEM with respect to the admissible elements shapes is fully exploited to avoid mesh-dependency in the crack propagation modelling, through a mesh topology modification restricted only to the elements containing the crack tip node. Such hybrid usage should in principle provide a reduction of the cost of the analysis and flexibility in the study of the matrix cracking. In this work we present some preliminary results and establish a workflow for future in-depth investigations.

The problem domain is the multi-region two-dimensional domain $\Omega \subset \mathbb{R}^2$ with external boundary $\Gamma = \partial \Omega$, shown in Fig.(5.2). It is assumed that no body forces act within $\Omega$, but either displacements or tractions can be enforced on the boundary $\Gamma$. $\Omega$ is the union of two sub-domains, namely $\Omega^{BEM}$ and $\Omega^{\text{VEM}}$, which represent, respectively, the transverse section of a fibre and the the surrounding polymer matrix in a polymer fibre-reinforced composite. The two sub-domains share the interface $S$. $\Omega^{\text{VEM}}$ is partitioned into a number of polygons of general shape, while the boundary $S = \partial\Omega^{\text{BEM}}$ is divided into a number of straight segments, which form the edges of the polygonal elements in $\Omega^{\text{VEM}}$ lying in proximity of the interface between the two sub-domains.

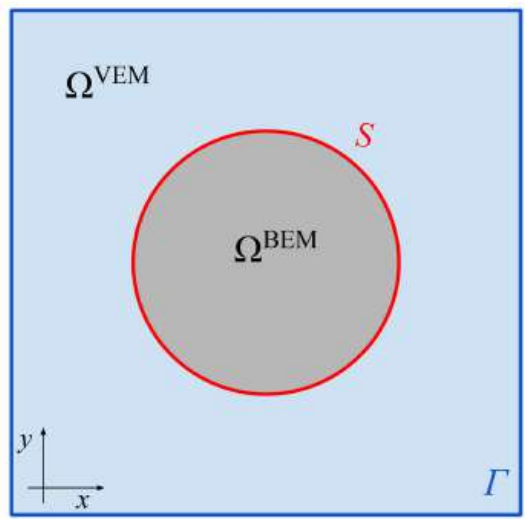

Figure 5.2: Geometry of the analysis domain: the BEM is employed to model the inclusion while the VEM models the surrounding matrix.

The test case, shown in Fig.(5.3), represents the transverse section of a composite material unit cell consisting of a single fibre and the surrounding matrix, subjected to a tensile load $\bar{\sigma}$ acting parallel to the $x$ axis. It is supposed that a crack has first grown along the interface, partially debonding the fibre from the matrix [143]. The extension of the debonded zone is identified by the angle $\theta_d = 65°$. Outside the debonded zone, the inclusion is perfectly bonded to the matrix. Two kinked cracks start from both ends of the debonded zone. The growth of the kinked cracks takes place only in the matrix material. The external boundary of the matrix domain is a square whose side has length $L = 1\,\text{mm}$. The inclusion is represented by a circle of diameter $D = 0.15\,\text{mm}$. The centre of the circle coincides with the centre of the square. The initial crack length is $a = D/10$. The initial kinked crack runs parallel to the $y$-axis. The model is symmetric with respect to the $x$-axis.

A carbon fibre/epoxy matrix composite is considered. Both the fibre and the matrix materials are treated as linear elastic under plane strain assumptions. The materials of matrix and fibre, respectively, are considered as isotropic and transversely isotropic in the analysis plane. Transverse elastic properties are: Young's modulus $E_F = 13.5\,\text{GPa}$ and Poisson's ratio $\nu_F = 0.25$ for the fibre and $E_M = 2.79\,\text{GPa}$, $\nu_M = 0.33$ for the matrix.

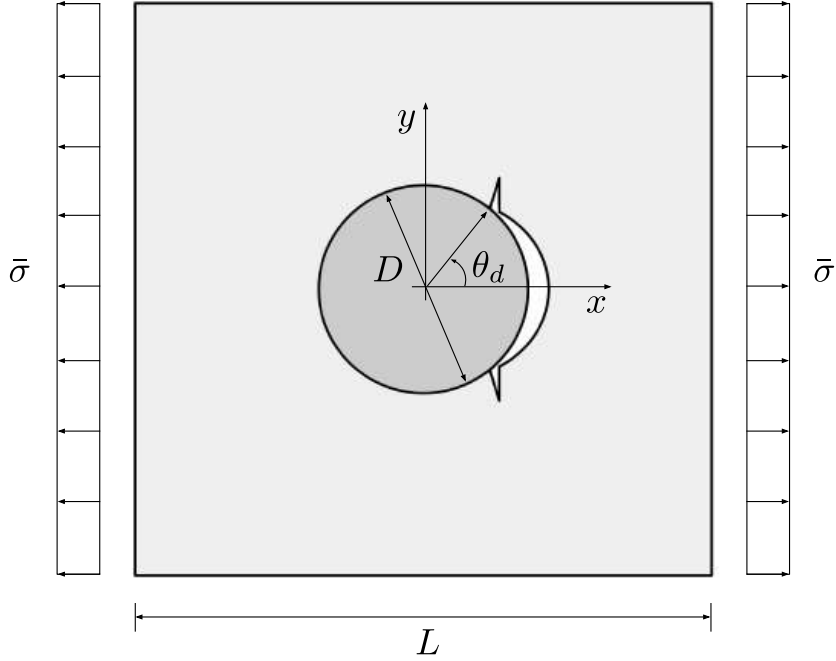

Figure 5.3: Schematic of the unit cell containing a single fibre partially debonded from the matrix.

Overall, 96 linear boundary elements are employed to model the fibre and 12034 polygonal virtual elements to model the matrix, giving a total of 48306 degrees of freedoms.

In this example, Linear-Elastic Fracture Mechanics (LEFM) is applied, i.e. geometrical and material non-linearities are excluded. For isotropic linear-elastic material behaviour the stress field near the crack tip of the mixed Mode I-II can be expressed, in polar coordinates, as [10]

$$\begin{aligned}\sigma_{11} =\;& \frac{K_I}{\sqrt{2\pi r}} \cos\left(\frac{\theta}{2}\right)\left[1 - \sin\left(\frac{\theta}{2}\right)\sin\left(\frac{3\theta}{2}\right)\right] + \\ & -\frac{K_{II}}{\sqrt{2\pi r}} \sin\left(\frac{\theta}{2}\right)\left[2 + \cos\left(\frac{\theta}{2}\right)\cos\left(\frac{3\theta}{2}\right)\right],\end{aligned} \qquad (5.1)$$

$$\sigma_{22} = \frac{K_I}{\sqrt{2\pi r}} \cos\left(\frac{\theta}{2}\right) \left[1 + \sin\left(\frac{\theta}{2}\right) \sin\left(\frac{3\theta}{2}\right)\right] +$$
$$+ \frac{K_{II}}{\sqrt{2\pi r}} \sin\left(\frac{\theta}{2}\right) \cos\left(\frac{\theta}{2}\right) \cos\left(\frac{3\theta}{2}\right), \qquad (5.2)$$

$$\tau_{12} = 2\nu \left[\frac{K_I}{\sqrt{2\pi r}} \cos\left(\frac{\theta}{2}\right) - \frac{K_{II}}{\sqrt{2\pi r}} \sin\left(\frac{\theta}{2}\right)\right], \qquad (5.3)$$

where $K_I$ and $K_{II}$ are, respectively, the Mode I and Mode II stress intensity factors (SIF), and $r$ and $\theta$ are the coordinates of the local crack-front polar coordinate system centred at the crack tip as shown in Fig.(5.4).

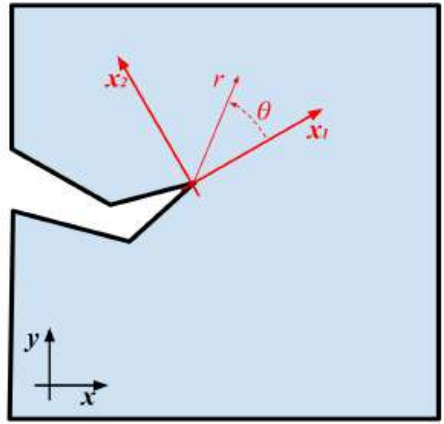

Figure 5.4: Local crack-front coordinate system.

There are several approaches for numerically evaluating the SIFs. The *stress interpretation method* (SIM) is one of the simplest techniques and it is based on the evaluation of the local normal stress $\sigma_{22}$ and shear stress $\tau_{12}$ on the ligament in front the crack ($\theta = 0$). From the near field solutions SIFs can be computed as:

$$K_I = \lim_{r \to 0} \sigma_{22} \sqrt{2\pi r}, \qquad (5.4)$$

$$K_{II} = \lim_{r \to 0} \tau_{12} \sqrt{2\pi r}. \qquad (5.5)$$

The maximum circumferential stress criterion (MCSC) [71] has been used to predict the angle $-\pi < \theta_c < \pi$ by which, for each crack growth increment, the new crack surface deviates from the original crack tip direction. $\theta_c$ is given

by

$$\theta_c = 2\arctan\left(\frac{1}{4}\frac{K_I}{K_{II}} - \frac{1}{4}\sqrt{\left(\frac{K_I}{K_{II}}\right)^2 + 8}\right). \tag{5.6}$$

A new crack tip is added at a distance $\Delta a$ in the direction identified by the crack growth direction $\theta_c$. The magnitude of increment $\Delta a$ can be set arbitrarily in quasi-static loading conditions. Smaller values of the increment lead to more accurate, stable and time-consuming simulations. A constant value $\Delta a = D/10$ has been used.

The computed crack propagation path for the unstable growth of the two kinked cracks is shown in Fig.(5.5). The obtained results agree well with previous results in [143] and, due to the relative orientation of the kinked cracks with respect to the applied load, the crack growth is, as expected, dominated by Mode I.

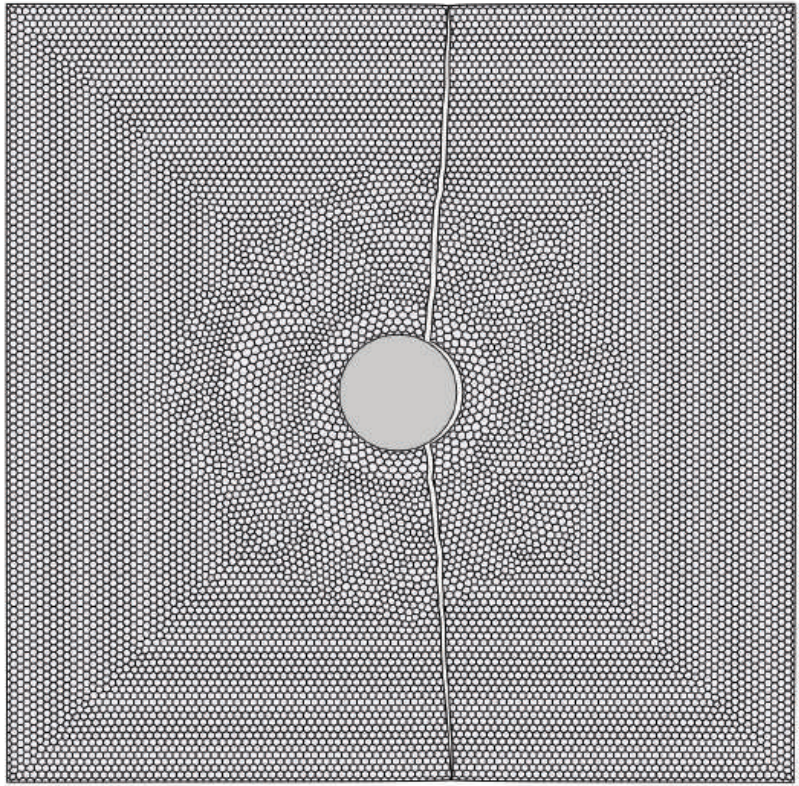

Figure 5.5: Simulated crack propagation in the considered test case.

## 5.2 Applications of an isotropic damage model

In this Section, an extension of the hybrid virtual-boundary element formulation presented in Chapter (4) for modelling regions exhibiting isotropic damage is described. Within the continuum damage context, some recent applications of BEM can be found, for instance, in Refs. [123, 121], while VEM has been recently applied in its lowest-order VEM formulation for modelling the strain-softening response of concrete-like materials as reported in the literature [68]. The present application, albeit being of general application, is ultimately aimed to model the degradation of the matrix phase of unidirectional fibre-reinforced composite material.

### 5.2.1 VEM for domains exhibiting isotropic damage

Continuum damage mechanics [100, 151, 92] describes the progressive loss of material integrity due to the propagation and coalescence of microscopic defects. These microstructure changes lead to irreversible material degradation, characterized by a loss of stiffness observed on the macroscopic scale. Different approaches have been proposed to model the growth and effects of distributed microscopic defects at the macroscopic scale.

*Isotropic damage models* [109, 157, 158, 60] are the simplest damage mechanics models. They are based on the simplifying assumption that the loss of integrity of the material is caused by an equal degradation of the bulk and shear moduli, governed by a single internal scalar *damage variable*, $\omega$. This variable is used to track and measure the loss of the material's stiffness and grows monotonically within its admissible range $0 \leq \omega \leq 1$ where 0 represents the undamaged material and 1 a fully degraded material. Under such assumptions, the constitutive equations for an isotropic damage model are defined by [110]

$$\sigma = (1 - \omega)\, \mathbf{C}^0\, \varepsilon = (1 - \omega)\tilde{\sigma}, \tag{5.7}$$

where, in Voigt notation, $\sigma$ and $\varepsilon$ collect, respectively, the stress and strain components, $\mathbf{C}^0$ is the elasticity matrix for the pristine elastic material, and $\tilde{\sigma}$ represents the stress components that would be associated to the strains $\varepsilon$ in the undamaged material.

The evolution of damage is triggered upon fulfilment of the activation threshold condition

$$F(\varepsilon) = \tau(\varepsilon) - r = 0, \qquad r = \max_{\lambda \in \mathcal{H}} \{\tau(\lambda)\}, \tag{5.8}$$

where $\tau(\varepsilon)$ is a suitably chosen norm of the strains, used to determine if the considered stress state belongs to the elastic domain, when $F(\varepsilon) < 0$, or if it induces damage initiation or evolution, $F(\varepsilon) = 0$, and the monotonically increasing internal variable $r$ represents the damage threshold at the current loading step $\lambda$ and it is a function of the loading history $\mathcal{H}$.

Different choices for the threshold function $\tau(\varepsilon)$ are available in the literature, defining different shapes of the elastic domain in the strains space. An expression proposed by Mazars [126] and frequently used in the modelling of quasi-brittle materials, e.g. concrete, defines $\tau(\varepsilon)$ as

$$\tau(\varepsilon) = \sqrt{\sum_{i}^{3} \langle \varepsilon_i \rangle^2}, \qquad (5.9)$$

where $\varepsilon_i$ are the principal strains and $\langle \cdot \rangle$ are the Macaulay brackets such that $\langle \varepsilon_i \rangle = (\varepsilon_i + |\varepsilon_i|)/2$.
To model the damage onset and progress of materials having different degradation behaviours in tension and compression, different expressions for the threshold function are used, such as the one introduced in Refs.[119, 140], which reads

$$\tau(\varepsilon) = \beta \sqrt{2\Psi^0(\varepsilon)}, \qquad (5.10)$$

where

$$\Psi^0(\varepsilon) = \frac{1}{2} \varepsilon^{\mathrm{T}} \tilde{\sigma}, \qquad (5.11)$$

is the initial elastic stored energy function of the undamaged material. The parameter $\beta$ in Eq.5.10 allows modelling materials having different degradation behaviours in tension and compression and is given as

$$\beta = m + \frac{1-m}{n}, \qquad (5.12)$$

where $n = \frac{f_c}{f_t}$ is the ratio between the compressive strength $f_c$ and the tensile strength $f_t$ of the material and $\theta$ is a weighting factor defined as

$$m = \frac{\sum_{i=1}^{3} \langle \tilde{s}_i \rangle}{\sum_{i=1}^{3} |\tilde{s}_i|}, \qquad (5.13)$$

where $\tilde{s}_i$ are the components of the effective principal stress tensor and $\langle \cdot \rangle$ are the Macaulay brackets. The values of the weighting factor are in the range $0 \leq m \leq 1$, where 0 represents a state of triaxial compression ($0 \geq \tilde{s}_1 \geq \tilde{s}_2 \geq \tilde{s}_3$) and 1 represents a state of triaxial tension ($\tilde{s}_1 \geq \tilde{s}_2 \geq \tilde{s}_3 \geq 0$).

Epoxy resins, often used as the matrix in fibre-reinforced composite materials, exhibit different tension and compression behaviour. To model the onset and evolution of damage in such materials, Melro et al.[130] proposed the following law

$$\tau(\varepsilon) = \frac{3\tilde{J}_2}{X_m^c X_m^t} + \frac{\tilde{I}_1(X_m^c - X_m^t)}{X_m^c X_m^t}, \qquad (5.14)$$

where $X_m^t$ and $X_m^c$ are, respectively, the tensile and compressive strengths of the epoxy resin and $\tilde{I}_1$ and $\tilde{J}_2$ are, respectively, the first stress invariant and the second deviatoric stress invariant; both quantities are defined using the effective stress components $\tilde{\sigma}$ that would be active in the undamaged material.

The evolution of damage is governed by the Kuhn-Tucker flow rules, which read

$$F \leq 0, \quad \dot{r} \geq 0, \quad \dot{r} F = 0, \qquad (5.15)$$

and allow distinguishing between loading and unloading conditions. Unloading occurs when $\dot{\tau} \leq 0$; otherwise, damage evolves, and the following consistency condition must be satisfied

$$\dot{F} = \dot{\tau} - \dot{r} = 0. \qquad (5.16)$$

While the chosen definition for the threshold function defines the shape of the elastic domain in the strain space, the shape of the stress-strain diagram after damage onset is guided by *damage evolution law* that expresses the dependence of the damage variable $\omega$ on the internal variable $r$.

Different choices can be found in the literature for the function $\omega(r)$. A rather simple one is the linear softening law which reads

$$\omega(r) = \left[\frac{r_f}{r_f - r_0}\left(1 - \frac{r_0}{r}\right)\right] \cdot H(r - r_0), \qquad r = \max_{\lambda \in \mathcal{H}}\{\tau(\lambda), r_f\}, \qquad (5.17)$$

where $H(\cdot)$ denotes the Heaviside step function, the parameter $r_0$ identifies the damage initiation condition and $r_f$ limits the maximum admissible value of the state variable $r$. The value of the damage threshold $r_0$ can be inferred from the stress-strain diagram under uniaxial tension and depends on the expression chosen for the threshold function.

An exponential softening can be instead modelled by adopting a damage law defined as in Ref.[97] as

$$\omega(r) = \left[1 - \frac{r_0}{r} \exp\left(-\frac{r - r_0}{r_f - r_0}\right)\right] \cdot H(r - r_0), \qquad r = \max_{\lambda \in \mathcal{H}} \{\tau(\lambda)\}, \quad (5.18)$$

where $r_f$ controls the exponential softening response behaviour.

Under uniaxial tension, the stress-strain relation has the form

$$\sigma(\varepsilon) = \begin{cases} E\varepsilon & \text{if } r \leq \varepsilon_0 \\ [1 - \omega(r)]\, E\varepsilon & \text{if } r > \varepsilon_0 \end{cases} \quad (5.19)$$

which is valid for all the previously defined damage laws, if the threshold function is defined according to Eq.(5.9), thus leading to the equivalence $r_0 = \varepsilon_0$ and $r_f = \varepsilon_f$. A graphical representation of the elastic and damage behaviour for the simple case of a bar under uniaxial tension with linear (Eq.(5.17)) and exponential (Eq.(5.18)) softening, is shown in Fig.5.6.

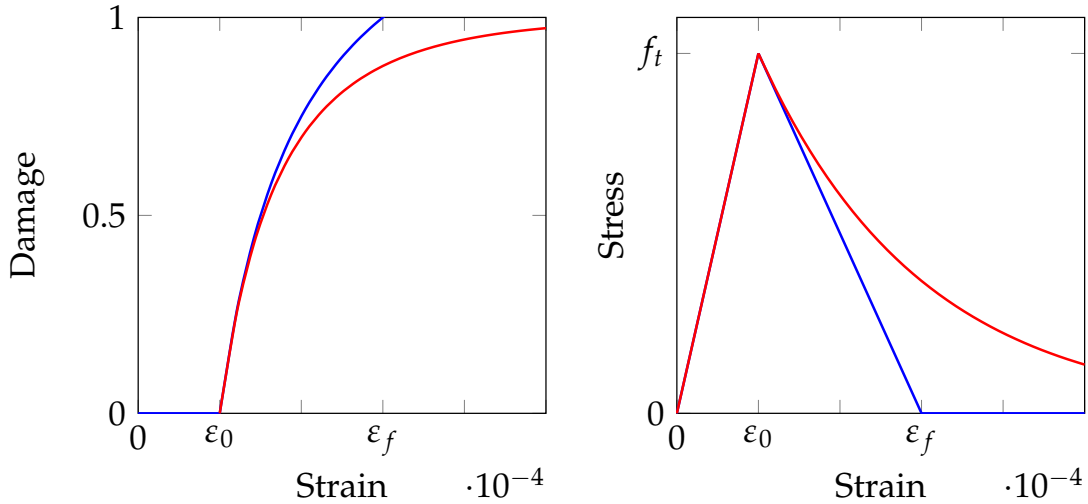

Figure 5.6: Damage-strain diagram (left) and tress-strain diagram (right) for the case of a bar under uniaxial tension. Blue curves refer to the linear damage law in Eq.(5.17); red curves refer to the exponential damage law in Eq.(5.18).

The VEM formulation described in Section 2.3 can be readily extended to problems involving non-linear material behaviours such as degradation and damage evolution, as described in Refs.[37, 14]. As in non-linear finite element formulations, the non-linear constitutive laws appearing in Eq.(5.7) can be treated using standard incremental-iterative algorithms. The stress at a generic point $x$ and at a generic loading step $\lambda$ is given by the expression

$$\sigma = \sigma(\lambda, x, \varepsilon_\Pi, \mathcal{H}), \tag{5.20}$$

where $\varepsilon_\Pi$ is the approximated virtual strain computed as in Eq.(2.61), using the matrix projector operator $\mathbf{\Pi}$. The tangent material stiffness matrix $\mathbf{C}_{tan}$ is consistently computed from the constitutive law in Eq.(5.20) as

$$\mathbf{C}_{tan}(t, x, \varepsilon_\Pi, \mathcal{H}) = \frac{\partial \sigma}{\partial \varepsilon_\Pi}. \tag{5.21}$$

### 5.2.2 Regularisation techniques

In the previous Section, the essential components of isotropic damage models have been presented. Although these models are relatively simple to implement, their numerical application may lead to physically unrealistic results because the damage process localises in a zone of the discretisation whose size depends on the mesh elements' size. As a consequence, the computed force-displacement curves are mesh-dependent. Thus, damage models require regularisation techniques to correct localised zones' thickness and avoid the numerical results' sensitivity to the mesh size. Two alternative regularisation techniques are commonly used, namely, the so-called *crack band theory* [28, 139, 140] and the *integral-type non-local damage models* [146, 96, 27, 54, 98]. In this Section, the formulations of both regularisation approaches are reviewed.

**Crack band theory**

In numerical simulation involving a numerical discretisation of the analysis domain, the damage growth localises into a band of width $h_b$ whose value depends on the size, shape and orientation of finite elements. To overcome this mesh-dependency issue, local regularisation methods based on the crack band theory introduce an explicit dependence of specific parameters regulating the damage-related softening process depending on the mesh elements' size.

Considering the uniaxial stress-diagram in Fig.(5.6) and assuming the linear softening damage law in Eq.(5.17), the area under the stress-strain diagram represents the energy dissipated per unit volume, defined as

$$g_f = \int_0^\infty \sigma(\varepsilon) \, d\varepsilon = \frac{1}{2} f_t \varepsilon_f \tag{5.22}$$

If the damage localises into a band of size $h_b$, the energy dissipated per unit area at complete failure is $g_f h_b$. This quantity must be equal to the fracture energy $G_f$, which is a material property, so that

$$g_f = \frac{G_f}{h_b} \qquad (5.23)$$

Accordingly, to Eq.(5.23), the stress-strain diagram is no longer considered a unique curve characterising the material response but must be adjusted according to the size of the localised damage band $h_b$, which in turn depends on the finite element mesh. Since the tensile strength of the material must remain independent of the finite element discretisation, to verify Eq.(5.23) for every value of $h_b$, the only parameter that can be adjusted is $\varepsilon_f$. Substitution of Eq.(5.22) in Eq.(5.23), one obtains

$$\varepsilon_f = \frac{2G_f}{h_b f_t} \qquad (5.24)$$

Eq.(5.24) states that the value of the parameter $\varepsilon_f$ increases with decreasing size of the local mesh elements' size and, in turn, of the localised damage band $h_b$, leading to a more ductile behaviour of the stress-strain curve. Conversely, as the local mesh elements' size increases, the value of $\varepsilon_f$ must decrease. A condition on the minimum value of $\varepsilon_f$, motivated by physical evidence of the damage process, prescribes that $\varepsilon_f$ must not be smaller than the limit elastic strain under uniaxial tension $\varepsilon_0$. This condition leads to a restriction on the maximum size of the localised damage band which can be stated as

$$h_b \leq h_{b,\max} = \frac{2G_f}{\varepsilon_0 f_t} \qquad (5.25)$$

In a two-dimensional modelling framework, the crack band size can be estimated as the element area's square root. However, this approach can induce significant errors when the crack band is not aligned with the mesh or when elongated, or more generally, distorted mesh elements are used. For mesh elements of arbitrary shape, a more general approach consists in estimating the crack band size by projecting the element onto the direction perpendicular to the assumed crack band direction [98].

As noted in Ref.[98], the advantage of such an approach is that the algorithmic structure of the finite element code requires only minor modifications, limited to the part of the code responsible for the evaluation of the state variables corresponding to a given strain increment.

**Integral-type non-local damage model**

Non-local damage models assume that damage parameters at a point do not depend only on the strain state at the point under consideration. In general, the integral-type non-local approach consists in replacing the value of a certain variable at a certain material point with its non-local counterpart obtained by weighted averaging over a spatial neighbourhood of each point under consideration. Given some local field $f(x)$ defined in a domain $V$, the corresponding non-local field is defined as

$$\bar{f}(x) = \int_V \alpha(x, \xi) f(\xi) \, dV, \tag{5.26}$$

where $x$ is a certain material point, $\xi$ is one of its neighbour points and $\alpha(x, \xi)$ is a given non-local weight function that depends on the relative position between the two points.

Different integral-type non-local regularisation models are known in the literature [96] which differ on the variable on which the non-local regularisation is based. The non-local approach herein used consists in replacing the local value of the equivalent strain $\tau(x)$ with its weighted average $\bar{\tau}(x)$ over a region surrounding each material point $x_p$

$$\bar{\tau}(x_p) = \int_\Omega \alpha(x_p, x_q) \tau(x_q) \, d\Omega(x_q), \tag{5.27}$$

where $\Omega$ is the analysis domain. Eq.(5.27) embodies the assumption that strains (and stresses) at a certain point depend not only on the state variables at that point but also on the distribution of the state variables over the whole body or over a finite neighbourhood of the point under consideration. A required property of the non-local operator $\alpha$ consists of not altering a uniform field, which means that the weighting function must satisfy the normalizing condition:

$$\int_\Omega \alpha(x_p, x_q) \, d\Omega(x_q) = 1 \quad \forall x_p \in \Omega. \tag{5.28}$$

This is achieved by adopting the following scaled expression for the non-local weight function

$$\alpha(x_p, x_q) = \frac{\alpha_0(d)}{\int_\Omega \alpha_0(d) \, d\Omega(x_q)}, \tag{5.29}$$

where $\alpha_0(d)$ is a non-negative weight function of the distance $d = ||x_p - x_q||$ between two considered material points, monotonically decreasing for $d \geq 0$.

The weight function $\alpha_0$ is often chosen as the Gauss distribution function

$$\alpha_0(d) = \exp\left(-\frac{d^2}{2l_c^2}\right), \tag{5.30}$$

where $l_c$ is known as the *internal length* or *characteristic length* of the non-local continuum, a parameter that depends on the heterogeneous material properties. Another common choice is the truncated quadratic polynomial function

$$\alpha_0(d) = \left\langle 1 - \frac{d^2}{R^2} \right\rangle^2, \tag{5.31}$$

where $R$ is known as the *interaction radius* [96] and it is a parameter related to the characteristic length $l_c$.

The implementation of the non-local damage model based on averaging of equivalent strain requires the computation of the point-wise values of the non-local equivalent strain. Afterwards, before damage is evaluated, the local equivalent strains are replaced by their non-local counterpart. The integral defined in Eq.(5.27) is evaluated numerically at each point $x_p$ as

$$\bar{\tau}(x_p) = \sum_q w_q \, \alpha_{pq} \, \tau(x_q), \tag{5.32}$$

where $x_q$ are the coordinates of the integration points, $w_q$ are coefficients representing the weights of the chosen integration rule and $\alpha_{pq}$ are the non-local interaction weights between points $p$ and $q$, defined as

$$\alpha_{pq} = \frac{\alpha_0(d_{pq})}{\sum_d w_d \, \alpha_0(d_{pr})}. \tag{5.33}$$

When a weight function $\alpha_0$ with a bounded support is chosen, as in Eq.(5.31), $\alpha_{pq}$ vanishes when the distance $d_{pq}$ between points $p$ and $q$ is greater than the interaction radius $R$.

It is worth noting that in order for the non-local model to be effective, the mesh elements' size within the zone where the damage process occurs must be smaller than the interaction radius $R$. In the lowest-order VEM formulation, the computed strain field is constant over a generic mesh element, and no integration weight actually exist. The approach herein adopted consist in considering the centroid of each element as the evaluation point $p$ (resp. $q$). The corresponding weight is taken as the product of the area of the element $p$ (resp. $q$) and its thickness.

Depending on the number of elements in the finite element model, the process of searching neighbour integration points within the interaction domain centred at a certain integration point, and subsequent computation of the non-local interaction weights might be a computationally demanding task.

In order to decrease the computational cost, an efficient numerical implementation for non-local averaging can be built according to the following procedure whose steps are executed once for all points $p$:

- Find all points $q$ whose distance from point $p$ is smaller than $R$, and for each of them evaluate $a_{pq} = w_{pq}\,\alpha_0(d_{pq})$;

- Compute the sum $a_p = \sum_q a_{pq}$;

- Divide each $a_{pq}$ by $a_p$ and store the results $\bar{a}_{pq}$ in a table where its position is associated with points $p$ and $q$.

The procedure outlined above is performed only once as a part of the initialization tasks before the beginning of the incremental-iterative solution of the non-linear problem.

At each iteration, the non-local equivalent strain at a generic point $p$ can be straightforwardly computed as

$$\bar{\tau}(x_p) = \sum_q \tau(x_q)\,\bar{a}_{pq} \tag{5.34}$$

where $\tau(x_q)$ is the local equivalent strain at a generic point $q$ within the interaction radius $R$ and $\bar{a}_{pq}$ is the corresponding entry in the non-local averaging table.

### 5.2.3 Numerical examples

In this Section, the developed hybrid VE-BE formulation is assessed in the non-linear framework of damage modelling. Three numerical applications are considered. The first application consists in the so-called *tension specimen test* [139] and aims to validate the non-linear virtual element formulation by adopting the local regularisation technique recalled in Section 5.2.2.

In the second application, where a three-point bending test involving a quasi-brittle concrete notched beam is considered, is meant to assess the accuracy of the non-linear virtual element formulation against cases available in the literature by implementing an isotropic damage and the non-local regularisation approach recalled in Section 5.2.2. The last application validates the same damage model by employing the hybrid VE-BE formulation. This test considers a unit cell consisting of a circular elastic fibre in epoxy matrix where partial debonding between fibre and matrix triggers damage onset and evolution. All the numerical experiments have been performed using an in-house developed `MATLAB` code which addresses all the stages of the computations, as well as pre- and post-processing tasks.

**Tension specimen test**

This numerical example simulates a case where a localisation is generated by a non-uniform stress field. The specimen, whose geometry and boundary conditions are shown in Fig.(5.7), has a central neck inducing localisation in the zone with minor cross-section. The thickness of the specimen is $t = 20$ cm. Material parameters are Young's modulus $E = 20000$ kp/cm$^2$, Poisson's ratio $\nu = 0.2$, tensile strength $f_t = 10$ kp/cm$^2$ and fracture energy per unit area $G_f = 0.125$ kp/cm. The threshold function $\tau$ is computed following the definition in Eq.(5.10). The adopted damage law is the linear softening law defined in Eq.(5.17). The initial damage threshold is set as $r_0 = \frac{f_t}{E}$ and the maximum admissible value $r_f$ is obtained, depending on the mesh size, accordingly to Eq.(5.24), where $h_b$ is computed considering the projection of mesh elements' geometry on the horizontal axis.

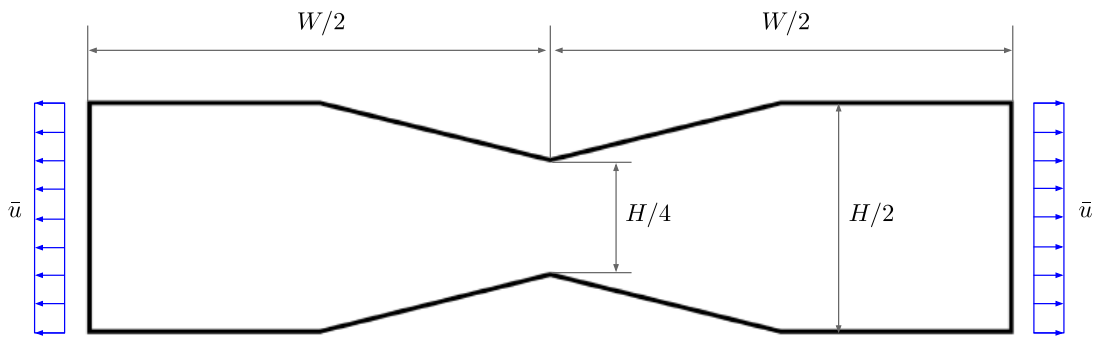

Figure 5.7: Geometry and boundary conditions of the tension specimen test.

Numerical tests have been performed with different meshes to determine convergence with mesh refinement and the regular/irregular mesh element shapes' influence. Specifically, two types of mesh are adopted. The first mesh type consists of 8-node convex polygonal virtual element discretisations with different refinement levels, referred to as mesh C1, C2, C3 and C4. The second mesh type, referred to as mesh NC4, consists of non-convex 8-node polygonal virtual elements obtained by randomly perturbing the position of the nodes of mesh C4. Both types of meshes are shown in Fig.(5.8).

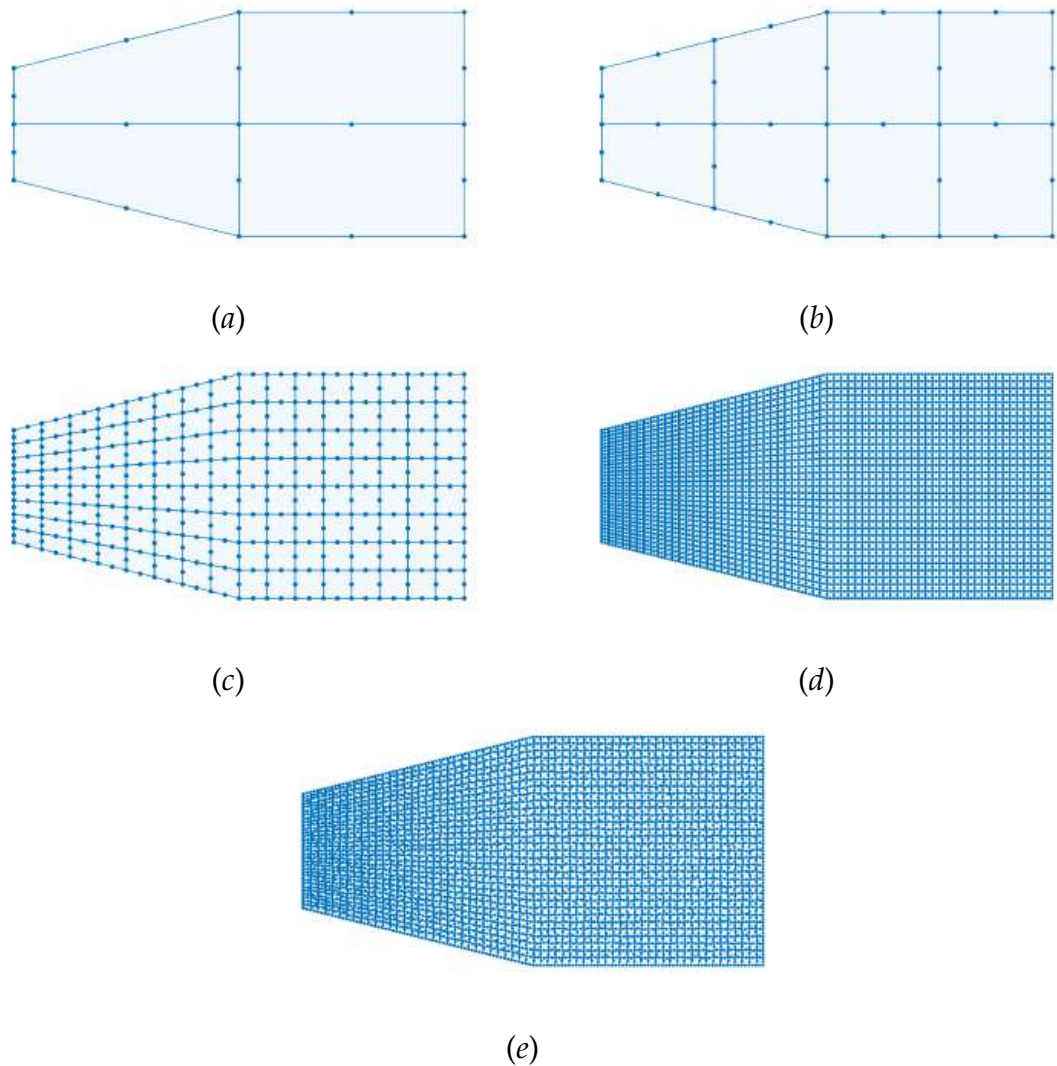

(a)　　　　　　　　　　　　　(b)

(c)　　　　　　　　　　　　　(d)

(e)

Figure 5.8: Convex meshes of 8-nodes VEM elements for the three-point bending test. (a) mesh C1, (b) mesh C2, (c) mesh C3, (d) mesh C4 and (e) NC4.

Results are observed in terms of horizontal imposed displacements and reaction forces. As reported in Fig.(5.9), convergence for progressive mesh-refining is achieved for both types of meshes and both the original results in Ref.[139] and more recent results in Ref.[68] are correctly replicated. As reported in Ref.[139], damage localises within the first rows of elements that are closer to the central neck, as shown in Fig.(5.10) for mesh C4 and mesh NC4.

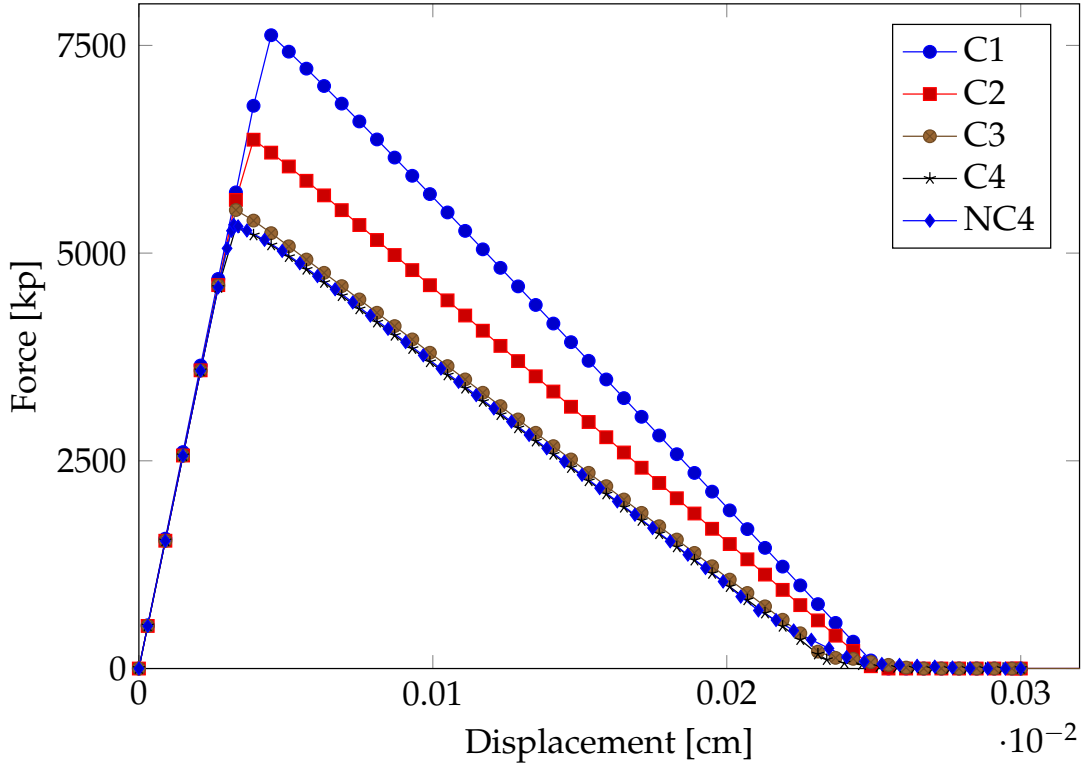

Figure 5.9: Force-displacement diagram for the tension specimen test.

**Three-point bending test**

The three-point bending (TPB) test is considered a benchmark to validate the implemented VEM for isotropic damage combined with an integral-type, non-local regularisation technique. This numerical example investigates damage initiation and evolution up to a notched concrete beam's failure, where the damage growth is dominated by mode $I$ loading.

Geometry and boundary conditions for this problem are shown in Fig.(5.11). The beam has square cross-section of side $H = 100$ mm and spans $W = 450$ mm. The notch is $A = 5$ mm wide and extends up to one half of the beam height. These dimensions correspond to the experiments performed in Ref.[105]. The material's Young's modulus is $E = 20\,000$ MPa and the Poisson's ratio is $\nu = 0.2$. The law with exponential softening defined in Eq.(5.18)

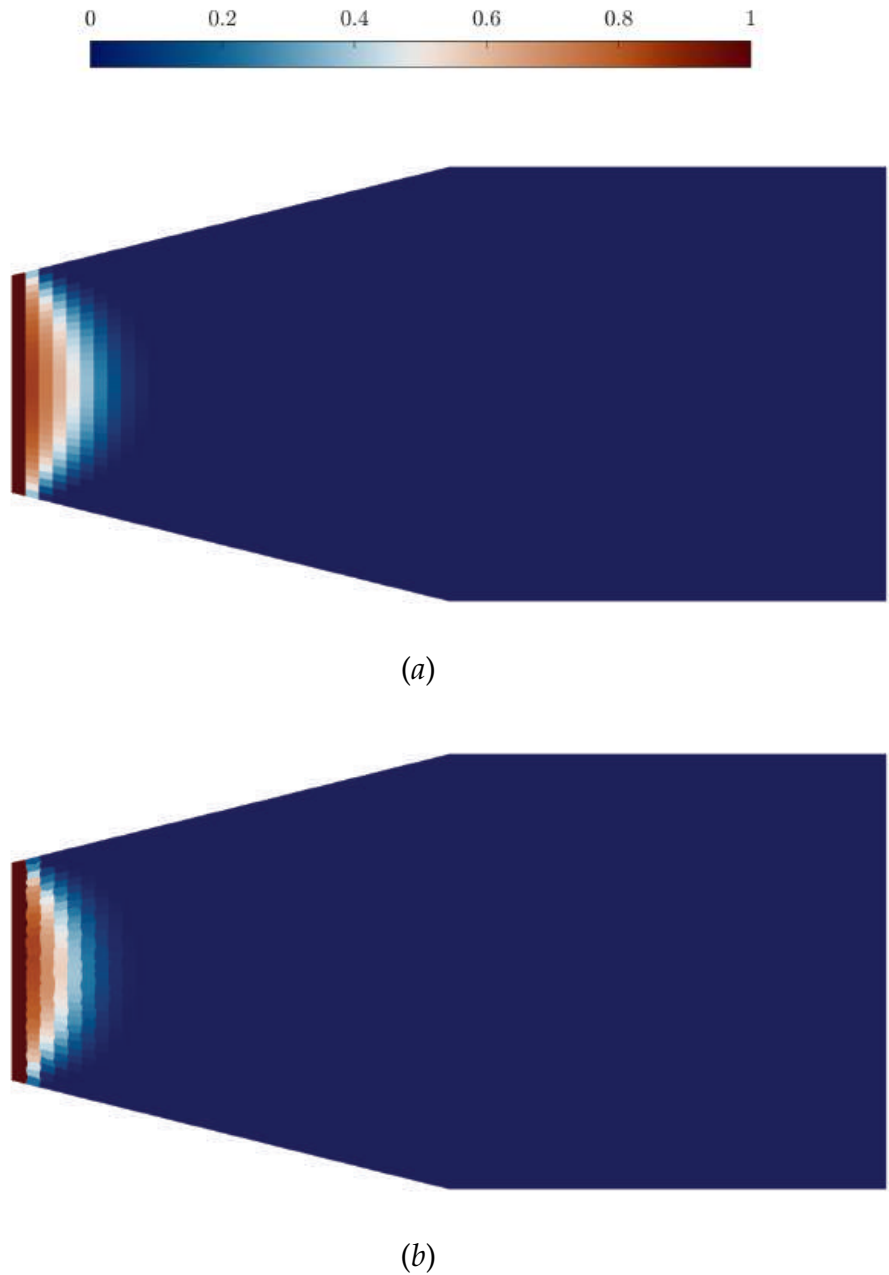

Figure 5.10: Damage profile of the tension specimen test: (*a*) mesh C4; (*b*) mesh NC4.

is adopted to model damage evolution. The damage parameters are chosen as in Ref.[97] as $r_0 = 9.0e{-}5$ and $r_f = 7.0e{-}3$. The threshold function $\tau$ is computed following the definition of Mazars in Eq.(5.9) and the non-local interaction radius is set to $R = 4\,\text{mm}$. The tests have been performed, under plane strain assumptions, using three different meshes of polygonal elements to determine convergence with mesh refinement. The two coarser meshes, referred to as V1 and V2, contains 6072 and 8493 elements and are shown in Fig.5.12. A further finer mesh, referred to as V3, is considered an over-kill discretisation, and it contains 18546 polygonal elements.

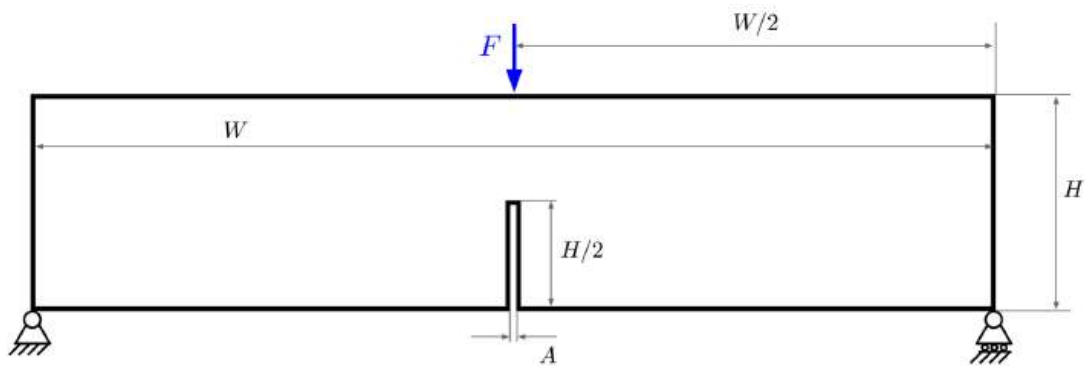

Figure 5.11: Geometry and boundary conditions of the three-point bending test.

The simulations are performed in displacement control using a Newton-Raphson scheme. Results are observed in terms of force versus displacement at the point where the vertical displacement is applied and compared with experimental results from Ref.[105] and with numerical results from Ref.[97]. The computed force-displacement curves, the reference numerical solution from Ref.[97] and the experimental bounds from Ref.[105] are depicted in Fig.(5.13). They reveal good agreement with the experimental bounds for the most part of the force-displacement diagram for all the considered discretisations. Comparison with the reference numerical solution obtained with full integrated 4-node bilinear isoparametric elements and mesh size of 1.67 mm, shows a better reproduction of the experimental data in the first part of the softening branch. A slight underestimation of the computed load can be noticed in the last part of the curve's softening branch. This difference, already noted in Ref.[68], where a similar numerical test with polygonal virtual elements has been performed, is likely due to the unstructured character of the virtual element mesh with respect to the finite element reference mesh. The evolution of the damage profile is shown in Fig.(5.14): damage originates at the bottom of the notch and grows in a limited zone along the structure's vertical axis of symmetry.

The presented results validate the implemented VE isotropic damage model,

which will be used in the next computational test.

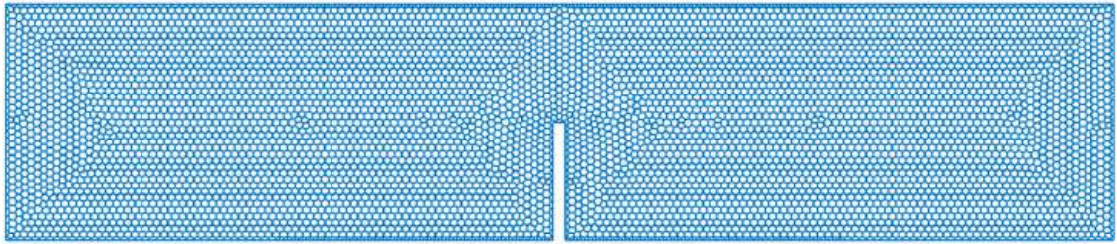

(a)

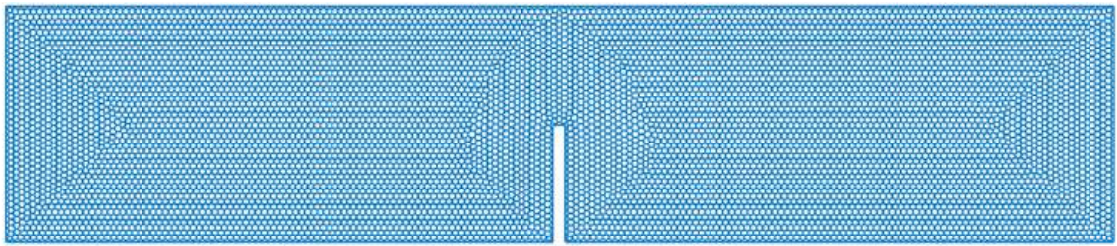

(b)

Figure 5.12: Polygonal meshes used in the numerical simulation of the three-point bending test: (a) mesh V1, (b) mesh V2.

**Transverse failure of a composite fibre-reinforced unit cell**

In the present Section, the hybrid virtual-boundary element formulation, combined with an isotropic damage model for the regions modelled with virtual elements, is used in the computational simulation of the damage evolution under transverse tensile loading of a unit cell comprising a single fibre embedded in an epoxy matrix, with initial partial debonding between fibre and matrix. The study of such fibre-matrix system has been the subject of a considerable number of studies [171, 143, 122, 169, 78].

The test case is shown in Fig.(5.15). In the initial configuration, it is assumed that the circular fibre is debonded from the matrix in the interface region identified by $|\theta_d| \leq 70°$, see Ref.[143]. Outside the debonded region, the inclusion is perfectly bonded to the matrix. This test aims to simulate the progression into the matrix of the two kinked cracks that start from both ends of the debonded zone, and this initial condition is assumed as no cohesive inter-faces have been included so far in the model, which identifies a direction of further development.

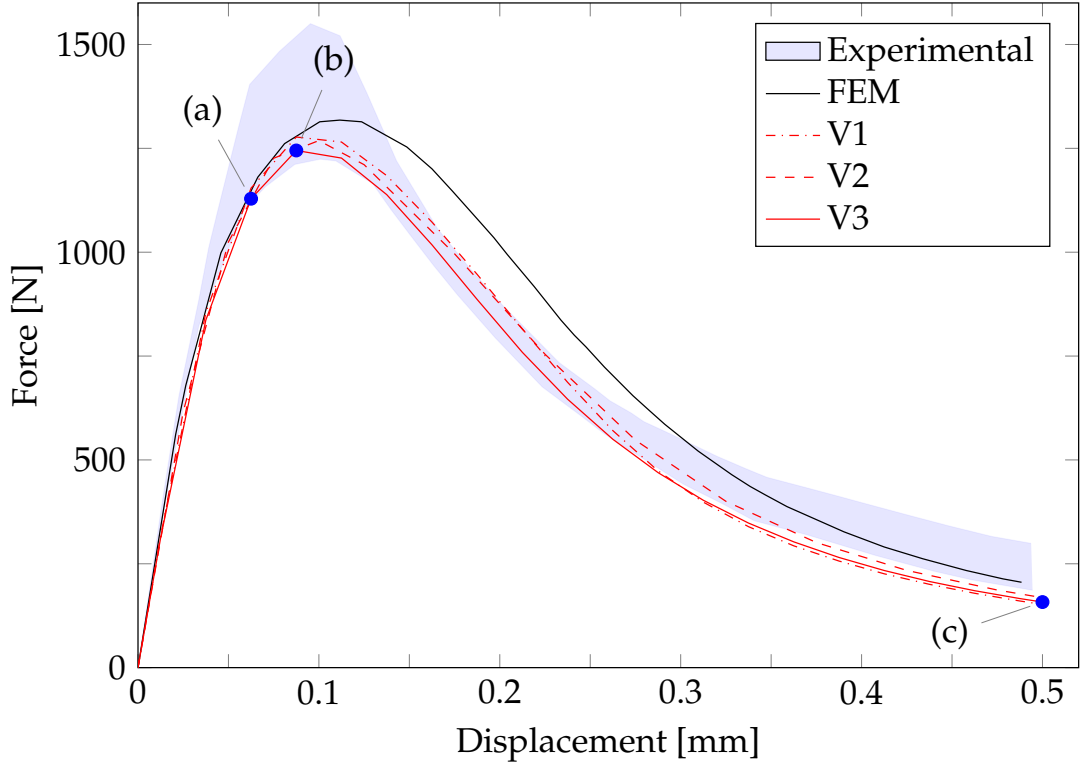

Figure 5.13: Force-displacement diagram for the three-point bending test. Comparison between the virtual element discretisations, the reference FEM solution in Ref.[97] and experimental data in Ref.[105].

The fibre diameter is $D = 0.025$ mm, and the side length of the unit cell is $L = 0.2$ mm, giving a corresponding volume fraction $V_f = 0.0123$. The centre of the circle coincides with the centre of the square. The tensile loading is applied by prescribing uniform displacements $\bar{u}$ at the sample left and right edges. Plane strain conditions are assumed. The fibre material is assumed linear elastic, and it does not develop damage. The matrix material is treated as linear elastic until the damage onset, governed by the loading function in Eq.(5.14). The exponential damage evolution law in Eq.(5.18) is assumed, with $r_0 = 1$, $r_f = 234$, according to strength and fracture toughness data about epoxy, and $R = D/3$. The transverse elastic material parameters are $E_F = 201$ GPa and $\nu_F = 0.22$ for the fibre and $E_M = 2.8$ GPa and $\nu_M = 0.33$ for the matrix. The fracture toughness of the epoxy matrix is $G_{fr} = 0.09$ N/mm.

To make the mesh consistent with the parameter assumed in the non-local continuum damage model, the matrix region is discretised with 8047 2D lowest-order virtual polygon elements, which induce 256 1D linear boundary elements on the fibre-matrix interface, where conformal meshes are employed. The overall mesh is shown in Fig.(5.16). The simulations are performed under displacement control using a Newton-Raphson with adaptive load step to track the steep softening branch. The simulation is arrested at a nominal macro-strain $\varepsilon_x = 0.05$. For each load increment, the plotted reaction force is computed as the sum of the right edge's nodal reaction forces.

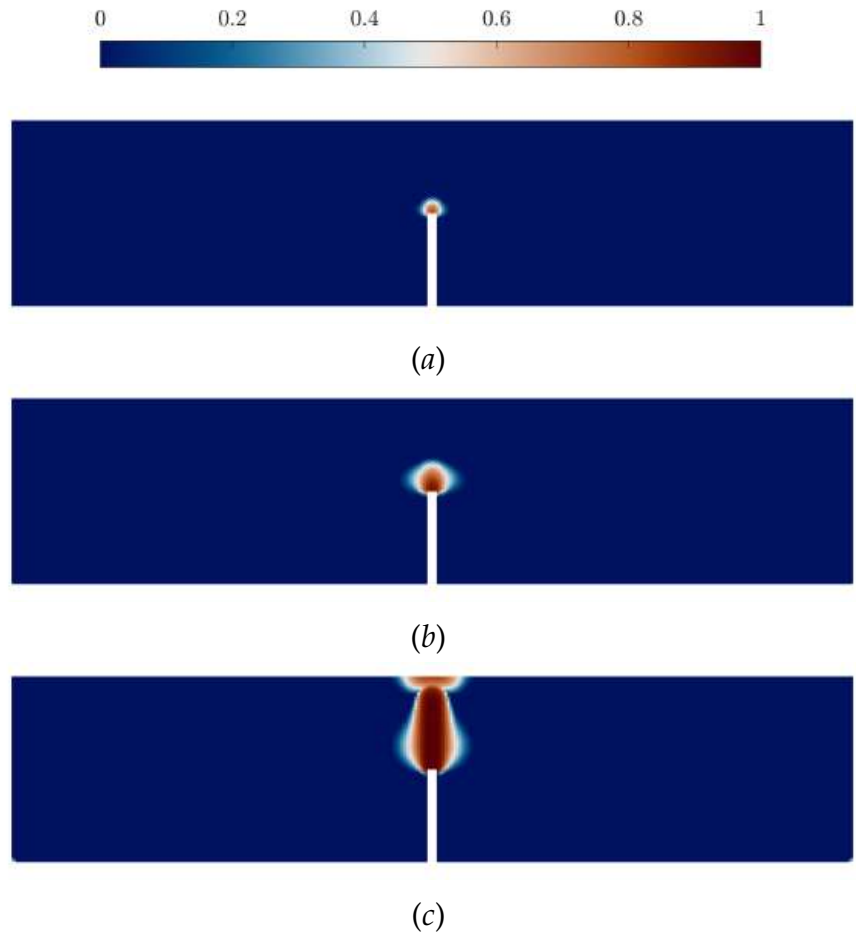

(a)

(b)

(c)

Figure 5.14: Damage profile evolution for the three-point bending test (mesh V3).

Fig.(5.17) shows the load-displacement diagram; the identified labels correspond to the damage profiles shown in Fig.(5.18). Linear elastic behaviour is exhibited up to slightly before the point (*a*) in the curve, which marks the initiation of damage at the ends of the debonded interface, where stress con-

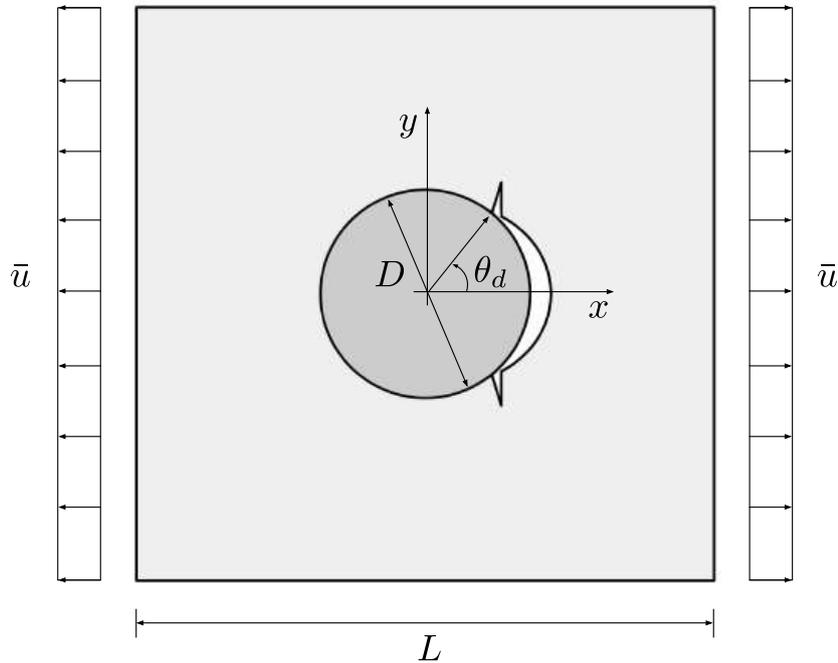

Figure 5.15: Geometry and boundary conditions of the composite unit cell containing a circular fibre partially debonded from the matrix.

centration is expected. Once damage is activated, the two symmetric damaged/failed region progress within the matrix, following a kinked path consistent with the behaviour reported in Refs.[143, 63]. As the loading increases, the material failure evolves, affecting regions oriented perpendicularly with respect to the load direction up to the unit cell boundary, which causes a progressive decrease of the load-carrying capability identified by the softening branch of the load-displacement diagram.

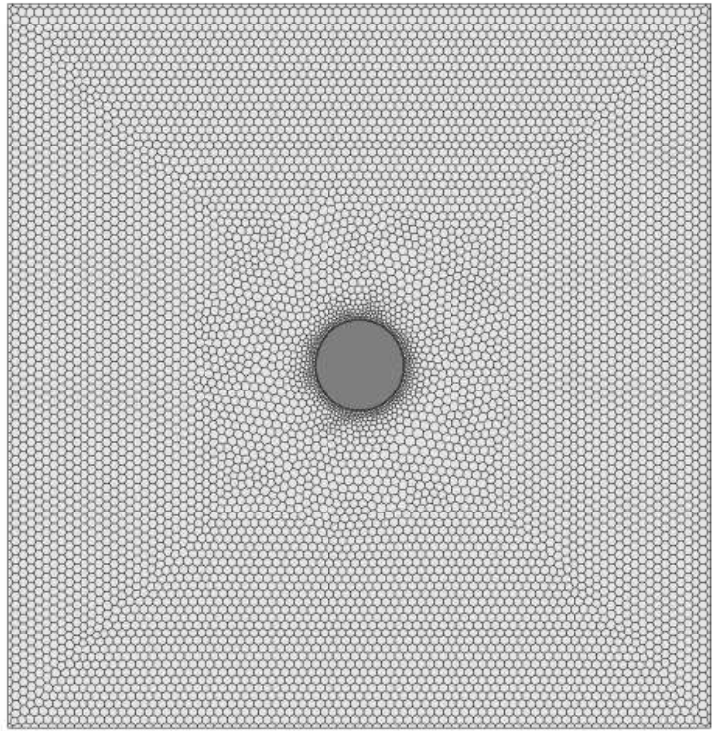

Figure 5.16: The mesh adopted to simulate the transverse failure behaviour of a composite unit cell with partial debonding.

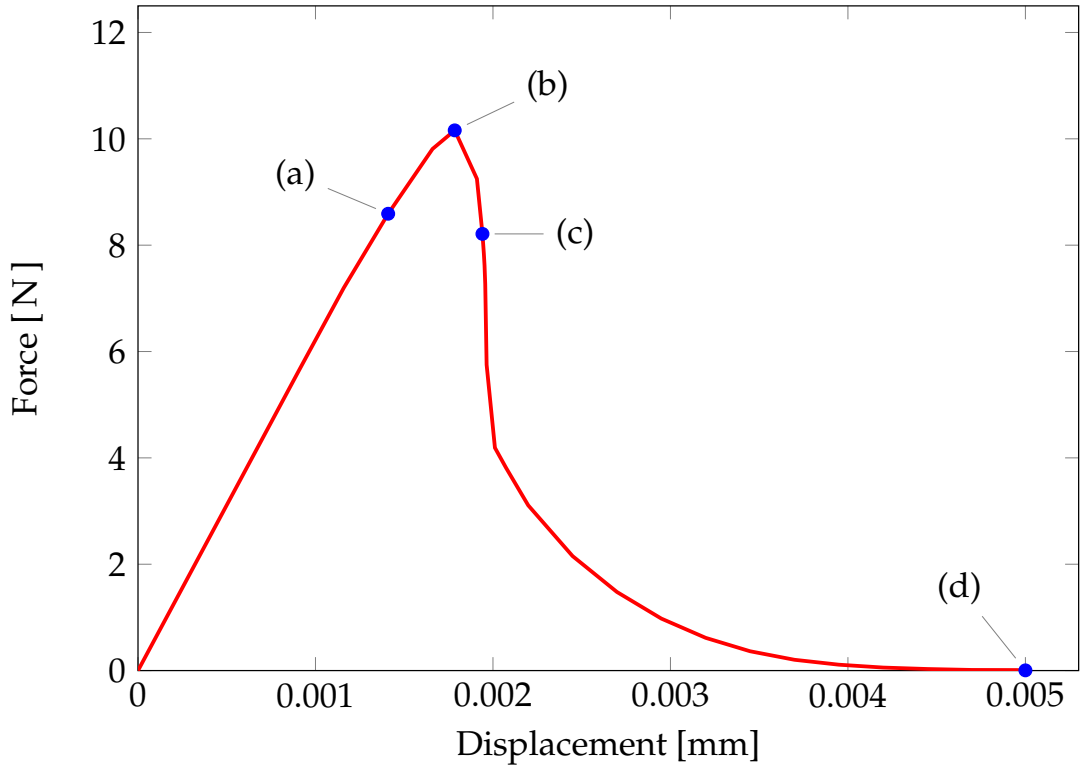

Figure 5.17: Force-displacement diagram for the composite unit cell test under tensile loading.

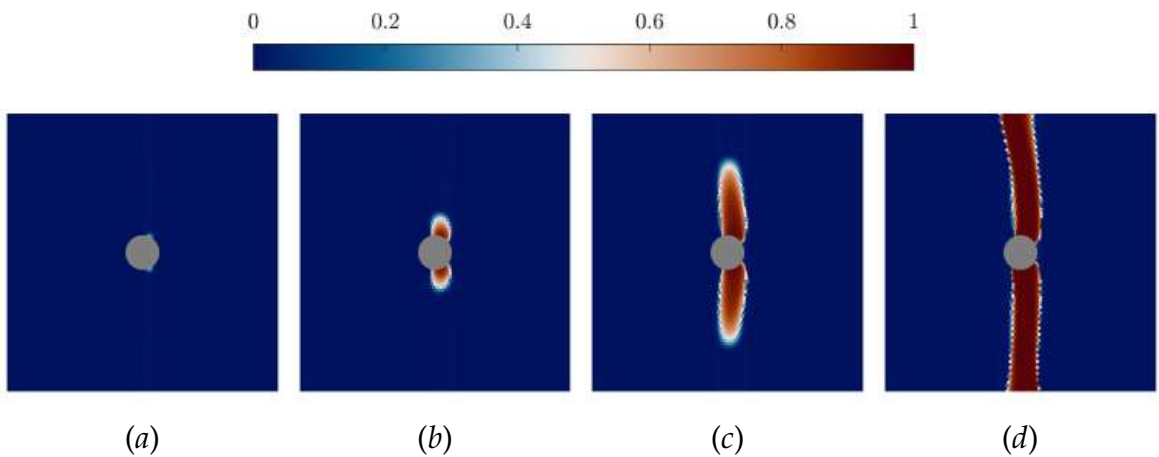

Figure 5.18: Damage profile evolution for the composite unit cell under tensile loading.

# Conclusions

A computational framework for microstructural modelling of transverse behaviour of heterogeneous materials has been developed in this thesis. The framework has been based on the lowest-order formulation of the Virtual Element Method (VEM) and applied to computational homogenisation problems and failures analysis at the microscale of polycrystalline and unidirectional fibre-reinforced composite materials. Several aspects of applying the VEM formulation to microstructural material modelling have been investigated and addressed in this work.

In the introductive Chapter (1), the still relevant interest within the field of computation micromechanics for the development of computational techniques capable of dealing with complex and evolving geometries and meshes with accuracy, effectiveness, efficiency, and robustness has been reviewed introducing VEM as the numerical technique chosen for this research work for its peculiar features, the robust treatment of complex mesh features and simplification of the analysis's pre-processing stage, being among them.

The core features of VEM have been discussed in Chapter (2), where the two-dimensional lowest order formulation for linear elasticity problems has been reviewed. Stemming from the work performed for this thesis, some practical aspects of VEM implementation have been discussed, highlighting both its common feature and difference with a standard FEM implementation with respect to mesh generation and handling and computational cost associated to the assembly of the system of equations.

In Chapter (3) an application of VEM for computational homogenisation of composite and heterogeneous materials has been presented. The selected applications have been focused on modelling the transverse mechanical behaviour of polycrystalline and unidirectional fibre-reinforced composite materials. The study has shown that VEM's capability to deal with very general polygonal mesh elements, including non-convex and highly distorted ele-ments, can be profitably exploited to relax the mesh quality requirements that may hinder the automatic analysis of micro-morphologies presenting complex or highly statistically varying features. This research activity has also driven the development of a polygonal mesh generator for general multi-region domains bounded with convex or non-convex boundaries, whose robustness has been proven in the discretisation of polycrystalline and composite microstructures.

In Chapter (4), a novel two-dimensional hybrid virtual-boundary element formulation for the analysis of multi-region two-dimensional elastic problems has been presented. The numerical tests performed on composite materials with inclusions of complex shape have assessed the advantages of the proposed formulation. The proposed method's accuracy has been tested against pure FEM and VEM solution of the same problem. The application of such a novel formulation to the computational homogenisation problem of a composite material with randomly distributed inclusions of complex shape is also reported. With respect to standard FEM or VEM discretisation, beside a further simplification pre-processing stage, the inherent reduction of the system's degrees of freedom has led to an appreciable decrease in the analysis's computational cost associated with the global system matrix assembly and the system solution.

In Chapter (5), further applications of the proposed hybrid VEM-BEM formulation for modelling damage phenomena in heterogeneous materials have been presented. In the first part of this Chapter, it has been demonstrated how the use of VEM within the framework of Linear Elastic Fracture Mechanics can be extremely useful for simulation of crack propagation, as its inherent flexibility concerning the shapes of the admissible elements can be fully exploited to avoid mesh-dependency of the computed crack propagation path. Simultaneously, mesh topology modification can be restricted only to the elements containing the crack tip node. In the second part of Chapter (5), the proposed hybrid formulation has been extended to a non-linear framework by adopting a constitutive law based on an isotropic damage model for the VEM subdomain. The proposed application has concerned the analysis of matrix degradation in a fibre-reinforced composite unit cell under progressive loading by implementing a damage model combined with a non-local integral regularisation technique for the matrix phase modelled with VEM. The FEM-like, straightforward VEM's capability to include non-linear constitutive models, combined with the use of BEM to model the linear elastic behaving inclusion, has allowed an appreciable reduction of the degrees of freedom of the problem, thus reducing the computational cost associated with the global system matrix assembly and the problem solution.

To conclude, it is worth highlighting some further investigation directions that may be identified for the proposed framework.

Lowest-order VEM, $k = 1$, has been employed in this thesis. However, higher-order virtual element formulations have been proposed in the literature [31, 34, 35]. Higher-order formulations are based on: *i)* the definition of a local virtual element space, for trial and test functions, which contains the set of all polynomial functions up to the selected degree *k plus* a set of additional functions, whose explicit knowledge is never required for the construction of the method; *ii)* the selection of a suitable set of degrees of freedom, grouped into a set *boundary* degrees of freedom, associated to the element vertices and to points lying on their edges, which maintain the physical meaning of displacements, *plus* a set of *internal* degrees of freedom, which represents suitably defined integrals, or moments, over the elements, of the functions belonging to the local virtual element space. If the virtual element space and the degrees of freedom are properly chosen, the projection operator, the stabilization term and, therefore, the local stiffness matrix entries can be still computed without the explicit knowledge of the unknown additional functions. However, as shown in Ref. [13], the procedure for determining the above quantities within higher-order formulations requires the numerical evaluation of integrals of polynomial functions that require the use of suitable quadrature rules specific to polygons [162, 134].

As an example, a VEM of order $k = 2$ would imply a quadratic approximation of the displacements over the edges of the virtual elements, expressed in terms of nodal displacements associated with the vertices and to the midpoints of the edges, which could be readily coupled with a quadratic formulation of the boundary element model of the inclusions. The coupling between higher-order virtual elements and higher-order boundary elements could be a direction of further research and could lead to remarkable benefits in solution accuracy.

An important caveat about the use of BEM is related to the fact that the method induces non-symmetric and non-definite fully populated solving matrices, see, e.g. Ref.[20]. As long as the number of elements used for modelling each inclusion is limited, this does not require additional consideration, and the potential of BEM in reducing the computational burden is preserved. However, should an inclusion need several hundred boundary elements, the presence of fully populated blocks in the solving systems could reduce the computation's effectiveness and increase the computational costs. These aspects could be mitigated and effectively addressed by using fast iterative solvers in conjunction with special matrix representations, e.g. fast multipoles [112], or hierarchical matrices [29, 43, 51, 44].

Another aspect that needs to be addressed in the framework of BEM and its coupling with VEM is the inclusions with sharp corners. The consideration of such geometrical entities is generally known to be problematic in BEM due to the non-unique definition of the normal at the corners. Some strategies to address such an issue have been proposed in the literature [7, 82], consisting in the employment of semi-discontinuous elements or hyper-singular traction boundary integral equations, and their inclusion in the present framework could be investigated in future studies.

Eventually, it is worth noting that the present technique has been hitherto developed only for two-dimensional problems. Although the considered test cases allow highlighting the potential benefits of the proposed method, 2D models generally present strong limitations in the computation of the effective properties of real materials, as they often neglect important inherent three-dimensional morphological or physical material features. In fact, while in this work the scheme has been successfully employed to compute the transverse elastic constants of composite laminae reinforced by unidirectional fibres, it would not be possible to employ it to compute the in-plane properties, or even the transverse properties, of laminates with general lay-ups, due to the impossibility of rendering in a 2D scheme the inherent 3D morphological features related to the mutual orientation of the fibres belonging to different contiguous laminae.

For such reasons, an interesting direction of further research could be related to the extension of the proposed scheme to three-dimensional problems. In the literature, three-dimensional formulations have been developed both for VEM and BEM, see, e.g. Refs.[77, 7]. The coupling between the two techniques in the 3D case could be readily applied, for example, to the computational homogenisation of polycrystalline materials, which has been successfully addressed separately both with VEM [124] and BEM [46, 45, 83, 48], both in the case of linear and non-linear material behaviour. Polycrystals represent another class of materials for which three-dimensional effects, related to the mutual orientation of the crystallographic lattices of different grains in the 3D space, play an essential role in determining the macroscopic effective properties.

Eventually, the extension of the present virtual-boundary element frame-work to the analysis of multi-phase microstructures exhibiting general non-linear behaviours, along the lines discussed above, and the analysis of three-dimensional micro-morphologies and a comprehensive investigation about the computational advantages offered by the framework, can form the object of further investigations.